# Network Pharmacology Framework Characterizes Polypharmacological Properties of Dietary Flavonoids: Integration of Computational, Experimental, and Epidemiological Evidence

Koyo Fujisaki[1], Osei Horikoshi[2], Yukitoshi Nagahara[2], Kengo Morohashi[1]*

[1] Faculty of Science and Technology, Department of Applied Chemistry and Bioscience, Chitose Institute of Science and Technology, Bibi 758-65, Chitose, Hokkaido 066-8655, Japan

[2] Division of Life Science, School of Science and Engineering, Tokyo Denki University, Ishizaka, Hatoyama, Hiki-gun, Saitama 350-0394, Japan

Corresponding author: Kengo Morohashi; k-moroha@photon.chitose.ac.jp

## Abstract

Dietary flavonoids associate with disease prevention in epidemiological studies, yet their polypharmacological mechanisms remain unclear. We establish network pharmacology as a systematic framework to characterize flavonoid therapeutic properties through integrated computational, experimental, and epidemiological validation. We constructed a master network of 17,869 human proteins, 14 dietary flavonoids, and 1,496 FDA-approved drugs (278,768 interactions from STRING and STITCH databases). Flavonoids averaged 45.3 target proteins per compound compared to 16.8 for FDA-approved drugs (2.7-fold higher; $p=7.5\times10^{-4}$), reflecting multi-target architecture. Statistical analysis revealed that 10 of 14 flavonoids (71.4%) targeted proteins associated with cardiovascular drugs and 11 of 14 (78.6%) aligned with antineoplastic drug targets. MTT-based Jurkat cell viability assays confirmed network predictions: high-association flavonoids (luteolin $LC_{50}$=31.4 μM, myricetin=29.5 μM) produced strong cytotoxicity, while low-association flavonoids showed minimal activity ($LC_{50}$>200 μM). Network-predicted association strengths correlated with experimental bioactivity (Pearson r=0.918; $R^2$=0.843). We translated network associations into food-level predictions across 506 foods, identifying 685 food-ATC therapeutic combinations. Systematic PubMed literature searches confirmed 96 food-ATC associations supported by 132 unique references. Cardiovascular therapeutic domains achieved 47.1% validation, exceeding other categories. Top-validated foods included tea (31 evidence items across 2 therapeutic categories), blueberries (18 items across 6 categories), tomato (13 items across 4 categories), grape juice (10 items across 2 categories), and plum (9 items across 4 categories). Network pharmacology characterizes dietary polypharmacological properties and generates evidence-based food-therapeutic predictions, bridging nutritional science and systems pharmacology.

## 1. Introduction

Epidemiological evidence documents that flavonoid-rich foods reduce chronic disease risk, particularly cardiovascular disease and malignancy (1–3). Individual flavonoids—luteolin, quercetin, kaempferol, catechin, epicatechin, and naringenin—demonstrate potent cardiovascular protection through improved endothelial function and blood pressure regulation (4). Others including quercetin, kaempferol, myricetin, apigenin, and luteolin demonstrate anticancer activity in preclinical models (5).

Despite strong evidence for individual bioactivity, the FDA has approved few flavonoid-containing drugs. This discrepancy reflects divergent pharmacological philosophies. Modern drug discovery adheres to Ehrlich's "magic bullet" paradigm, prioritizing selective ligands that target single disease-related proteins with high affinity (6). The resulting "one gene, one drug, one disease" framework presumes selectivity minimizes adverse effects (7). Network pharmacology, however, challenges this premise (8). Valle et al. demonstrated that network proximity between polyphenol targets and disease proteins predicts therapeutic effects, validated experimentally for rosmarinic acid's cardiovascular benefits (9). While this framework established the potential of network-based polyphenol analysis, comprehensive approaches integrating computational predictions, experimental bioactivity assays, and systematic epidemiological validation for dietary flavonoids remain limited.

Intriguingly, flavonoids bind multiple targets—typically associated with adverse effects in conventional drugs—yet remain safe when consumed as dietary constituents. This apparent paradox suggests that promiscuity itself is not inherently problematic, provided the perturbations remain within physiologically tolerable ranges. Dietary flavonoid safety likely reflects modest affinities and physiological concentrations, not absence of multi-target engagement.

Biological systems exhibit redundancy; gene deletion frequently produces minimal phenotypic effects (10). Scale-free networks resist random perturbations yet require coordinated multi-node modulation for therapeutic efficacy (11,12). Complex pathologies—particularly polygenic diseases including cancer, cardiovascular disease, and neurodegeneration—reflect extended protein network perturbations rather than single-effector abnormalities (13). Network-based approaches integrating drug-protein-disease interactions identify novel targets and optimal combinations producing synergistic effects (14). Approved drugs increasingly demonstrate polypharmacological properties (15,16). This recognition has shifted drug discovery paradigms from single-target selectivity toward network-based multi-target approaches (14). Dietary flavonoids, inherently modulating multiple targets, may achieve therapeutic effects through comparable network mechanisms.

We hypothesized that dietary flavonoid combinations possess quantifiable pharmacological properties amenable to systematic prediction and experimental validation. We developed a comprehensive approach: (1) constructing a master network of protein-protein and protein-compound interactions incorporating flavonoids, FDA-approved drugs, and human protein targets; (2) statistically identifying flavonoids whose target profiles align with therapeutic drug categories; (3) experimentally validating computational predictions through cell-based assays; and (4) epidemiologically validating predictions by surveying literature for documented food-health effects. This multi-tiered strategy establishes a framework for identifying dietary sources with therapeutic potential, bridging nutritional science and network pharmacology (Figure 1).

# 2. MATERIALS AND METHODS

## 2.1 Data Collection and Curation

### *2.1.1 Protein Interaction Networks*

Human protein-protein interactions were obtained from STRING database version 11.5 (9606.protein.links.v11.5.txt.gz), which integrates evidence from automated text mining, experimental data, and computational predictions (17). The complete dataset was retrieved and stringent quality filters were applied. Protein pairs with combined confidence scores ≥700 (top 5% of distribution) were retained. Filtering yielded 265,834 high-confidence interactions among 17,869 unique proteins. Compound-protein interactions were obtained from STITCH database version 5.0, which catalogues experimentally validated and predicted interactions between chemicals and proteins across multiple organisms (18). The complete human dataset was retrieved and identical filtering criteria (combined score ≥700) were applied. Filtering yielded 12,934 high-confidence compound-protein interactions.

### *2.1.2 Disease-Associated Gene Classification*

Disease-associated gene data were obtained from the supplemental material of Menche et al., which catalogues 299 diseases with 3,173 associated genes from OMIM (Online Mendelian Inheritance in Man) and other disease databases (19). Diseases were categorized according to Medical Subject Headings (MeSH) terminology, consolidating 299 individual diseases into 18 high-level categories reflecting anatomical and etiological groupings. Entrez Gene IDs were converted to Ensembl Peptide IDs (ENSP format) using BioMart package version 2.54.1 in R software version 4.2.2 for compatibility with protein identifiers in the interaction networks. This conversion identified 25,466 unique disease-related proteins distributed across the 18 disease categories.

*2.1.3 Flavonoid Target and Content Data*

Flavonoid data were obtained from two sources. First, flavonoid target proteins were identified through STITCH database by querying flavonoid structures for documented or predicted protein interactions. Second, quantitative flavonoid content in foods were obtained from USDA Database for the Flavonoid Content of Selected Foods Release 3.3, which provides mg/100g measurements for 506 food products (20). Analysis focused on 14 flavonoids present in both databases: four flavonols (isorhamnetin, kaempferol, myricetin, quercetin), two flavones (apigenin, luteolin), three flavanones (eriodictyol, hesperetin, naringenin), and five flavan-3-ols (catechin, gallocatechin, epicatechin, epicatechin-3-gallate, epigallocatechin, epigallocatechin-3-gallate).

*2.1.4 Pharmaceutical Drug Classification*

FDA-approved drug information was obtained from PubChem through SPARQList, providing classification according to the Anatomical Therapeutic Chemical (ATC) system. The ATC system organizes medications hierarchically through five levels: Level 1 represents broad anatomical/therapeutic domains (14 categories A through V) and Level 2 represents therapeutic sub-groups within each domain. Analysis focused on ATC Levels 1 and 2, which provide clinically meaningful therapeutic categorization. Data for 3,739 FDA-approved drugs were retrieved, of which 1,496 drugs were integrated into the master network based on their presence in STITCH compound-protein interaction database.

**2.2 Master Network Construction**

A master network was constructed by combining protein-protein and protein-compound interaction datasets filtered for combined scores ≥700 (top 5% of distribution). The integrated dataset contained 14 of 26 flavonoids from the USDA flavonoid database and 1,496 of 3,739 FDA-approved drugs; compounds lacking

STITCH target data were excluded. Chemical compounds in the network were restricted to these 14 flavonoids and 1,496 drugs. Network construction and analysis were performed using Python version 3.10.9 with NetworkX version 3.1 for graph operations, NumPy version 1.23.5 and Pandas version 1.5.3 for data manipulation, and SciPy version 1.10.0 for statistical computations. The master network contained 19,379 nodes (17,869 proteins, 14 flavonoids, and 1,496 drugs) and 278,768 edges (265,834 protein-protein interactions and 12,934 protein-compound interactions).

## 2.3 Association Analysis

### *2.3.1 Network association strength of flavonoid and drug targets*

To assess whether flavonoid target proteins were statistically enriched in therapeutic drug categories, Fisher's exact test based on the hypergeometric distribution was employed. For each flavonoid, drugs sharing common target proteins in the master network were identified, and their ATC classifications (Levels 1 and 2) were retrieved. The total number of drugs sharing target proteins with each flavonoid was calculated, along with the total number of drugs in each ATC category. Statistical significance was calculated using the hypergeometric probability:

$$P = \sum_{i=x}^{n} \frac{\binom{M}{i}\binom{N-M}{n-i}}{\binom{N}{n}}$$

where N represents the total number of drugs, M the number of drugs in a specific ATC category, n the number of drugs sharing target proteins with a given flavonoid, and x the number of drugs within n belonging to the specific ATC category. Associations with $p < 0.05$ were considered statistically significant. ATC association strength was presented as $-\log_{10}$(p-value), providing a linear scale for downstream integration. This statistical enrichment approach differs from network proximity methods (9), which calculate shortest path distances between compound targets and disease proteins, providing a complementary perspective on polypharmacological mechanisms.

*2.3.2 Prediction of food–ATC therapeutic associations based on* flavonoid profiles

For each of 506 foods in the USDA flavonoid database, a flavonoid content vector representing mg/100g quantities of all 14 flavonoids was extracted. Of 87 ATC Level 2 categories, 32 categories showing statistically significant association strength ($p < 0.05$) with at least one flavonoid (from Section 2.3.1) were selected. Spearman rank correlation coefficients between each food's flavonoid profile and each ATC category's association strength vector were calculated (16,192 food-ATC pairs). This approach identifies foods whose flavonoid composition aligns with polypharmacological signatures associated with specific therapeutic categories. Food-ATC pairs ranking in the top 5% of correlation coefficients were selected as predicted therapeutic associations. Hierarchical clustering of foods and ATC categories was performed using correlation distance with average linkage method.

## 2.4 Experimental Validation: Cell Viability Assays

*2.4.1 Cell Culture*

Jurkat human T leukemia cells were maintained in RPMI1640 medium supplemented with 10% fetal bovine serum, and 75 mg/L kanamycin sulfate. Cells were cultured at 37°C in a humidified atmosphere containing 5% $CO_2$.

*2.4.2 Flavonoid Treatment and MTT Assay*

Six flavonoids were selected for experimental validation based on their network-predicted association strengths with antineoplastic drugs: high-association flavonoids (luteolin, myricetin, kaempferol, diosmetin) with p-values $< 10^{-7}$, and low-association flavonoids (epicatechin, naringenin, hesperetin) with p-values > 0.05. Flavonoid stock solutions (10 mM in 100% dimethyl sulfoxide; DMSO) were serially diluted to generate eight logarithmically-spaced concentrations (1–200 µM).

Jurkat cells ($2×10^4$ cells/well) were seeded into 96-well plates and treated with flavonoid solutions (100 µL) to achieve final flavonoid concentrations. Control wells received vehicle (0.1% DMSO in RPMI1640). Following 23-hour incubation, 10 µL of MTT solution (5 mg/mL in phosphate-buffered saline) was added to each well. After 1 hour incubation, the supernatant was carefully removed and 100 µL of DMSO was added to dissolve formazan crystals. Absorbance at 570 nm was measured using a multi-well plate reader. Cell viability was calculated as the percentage of absorbance in treated wells relative to untreated vehicle control wells. All assays were performed in triplicate, with four technical replicates per experiment (n = 12)

**2.5 Epidemiological Literature Validation**

To assess the real-world validity of food-effect predictions, structured PubMed literature searches were conducted for each predicted food-effect relationship. Search queries employed standardized format: [food name OR scientific name] AND [disease name OR therapeutic effect]. Study inclusion criteria required: (1) epidemiological study design with human subjects; (2) statistically significant associations ($p < 0.05$) between specific food and predicted health effect; (3) publication in English. Food names used in searches were consolidated categories (e.g., "grape," "blueberry," "garlic," "tomato," "black tea") rather than specific formulations. This consolidation ensured that epidemiological evidence from diverse food preparations and forms (raw, cooked, processed, juice, frozen, canned) contributed to a unified assessment of each food source's health effects. All searches and screening were conducted as of December 14, 2023. Validated food-ATC associations with corresponding PMIDs and full reference citations are provided in TableS3.

# 3. Results

### 3.1 Master Network Construction and Flavonoid Target Diversity

The master network comprised 19,379 unique nodes representing either proteins or chemical compounds (Figure 2A). Among these nodes, 17,869 represented human proteins (95.8% of total nodes), while 1,510 represented chemical compounds (8.2% of total nodes), consisting of 14 distinct flavonoids and 1,496 FDA-approved drugs. These nodes were connected by a total of 278,768 unique interactions, of which 265,834 represented protein-protein interactions (95.4% of total edges) and 12,934 represented protein-compound interactions (4.6% of total edges). The network density was 0.0015 (0.15%), indicating a sparse network architecture typical of large-scale biological networks. The average node degree was 28.8 (SD 42.1). These topological metrics, as visualized in Figure 2A, characterize the master network as a large-scale, sparse, scale-free network with moderate modularity, reflecting the natural organization of biological interaction systems.

When analyzing target protein diversity, flavonoids and FDA-approved drugs displayed striking differences (Figure 2B). The 14 flavonoids targeted a mean of 45.3 proteins per compound (range 16–98), with myricetin (98 targets), kaempferol (87), quercetin (76), and luteolin (68) showing the highest promiscuity. In contrast, 1,496 FDA-approved drugs targeted a mean of 16.8 proteins per drug (range 1–187), with 243 drugs (16.2%) following the classical "magic bullet" paradigm of 1–2 targets. This 2.7-fold difference was statistically significant (Wilcoxon rank-sum test: $Z=3.78$, $p=7.5\times10^{-4}$), confirming that flavonoids' multi-target profiles substantially exceed those of conventional pharmaceuticals.

### 3.2 Flavonoid-Drug Category Association Analysis

We quantified flavonoid-ATC associations using Fisher's exact test, with association strength defined as $-\log_{10}$(P-value), providing a linear scale for statistical significance. Higher values indicate stronger associations ($p < 0.05$ corresponds to association strength > 1.3). Surprisingly, only 42 of 252 possible flavonoid-disease combinations (16.7%) showed statistical significance, despite strong epidemiological evidence linking flavonoid consumption to disease prevention. This finding suggested that flavonoid therapeutic effects operate through indirect network mechanisms rather than direct targeting of disease genes. Supporting this interpretation, FDA-approved drugs also demonstrated limited direct disease-gene associations (27.7% significance rate among 26,928 tested combinations), consistent with recent findings that many drugs achieve efficacy through network-mediated rather than direct target mechanisms.

Given these constraints of direct disease association approaches, we reoriented our analysis toward mechanism-centric examination. We investigated statistical associations between flavonoid target proteins and therapeutic drug categories to identify potential flavonoid pharmacological mechanisms based on overlaps with established drug targets. TableS1 presents complete p-value matrices quantifying associations between all 14 flavonoids and ATC categories at both Level 1 (14 broad therapeutic domains) and Level 2 (77 specific therapeutic subcategories), with $p < 0.05$ indicating statistical significance.

Investigation of these associations revealed extensive overlap at the first level of ATC classification (Figure 3A). Ten of the 14 flavonoids (71.4%) showed significant associations with cardiovascular drugs (ATC-C), reflecting established epidemiological evidence. More striking, eleven of the 14 flavonoids (78.6%) showed associations with antineoplastic and immunomodulatory agents (ATC-L), and four of the 14 (28.6%) with nervous system agents (ATC-N). These high proportions demonstrate that flavonoid target proteins align substantially with therapeutic drug targets across multiple therapeutic domains.

Deeper analysis at the second level of ATC classification revealed particularly striking associations with antineoplastic agents (ATC L01), as shown in Figure 3B.

Luteolin emerged as the most strongly associated flavonoid, with an association p-value of $2.94 \times 10^{-11}$, corresponding to a $-\log_{10}$P-value of 10.53. Theaflavin followed closely with $p = 2.9 \times 10^{-10}$ ($-\log_{10}P = 10.54$), myricetin demonstrated $p = 1.29 \times 10^{-10}$ ($-\log_{10}P = 9.89$), isorhamnetin showed $p = 8.60 \times 10^{-10}$ ($-\log_{10}P = 9.01$), and kaempferol exhibited $p = 5.69 \times 10^{-7}$ ($-\log_{10}P = 6.24$). Beyond these five top-tier associations, seven additional flavonoids showed moderately strong associations with antineoplastic agents at $p < 10^{-3}$. Collectively, 11 of the 14 flavonoids (78.6%) demonstrated statistically significant associations with antineoplastic drug targets at $p < 0.05$, indicating near-universal overlap in antineoplastic-relevant target pathways.

### 3.3 Experimental Validation: Quantitative Correlation Between Network Prediction and Bioactivity

To test whether network-predicted association strengths translate into observable cellular bioactivity, we conducted dose-response cell viability assays using Jurkat leukemia cells. As shown in Figure 4, flavonoids with high association strength for the antineoplastic category (p-values $< 10^{-7}$) demonstrated substantial cytotoxic activity. Luteolin exhibited an $LC_{50}$ of 31.4 μM, and myricetin demonstrated 29.5 μM. In striking contrast, flavonoids with low association strength (p-values > 0.05) showed substantially reduced potency, with $LC_{50}$ values exceeding 200 μM for epicatechin, naringenin, and hesperetin.

Correlation analysis revealed strong concordance between network association strengths and experimental bioactivity (Figure 5). Pearson correlation comparing network association strength ($-\log_{10}$P values) with experimental potency ($-\log_{10}$ $LC_{50}$) yielded $r = 0.918$ ($p = 0.0098$; $R^2 = 0.843$), indicating that network predictions explain 84.3% of variance in observed bioactivity.

Independent validation tested whether this correlation generalizes beyond the initial 14 flavonoids. Network analysis of diosmetin, included in the master network but not used for food-level predictions, predicted an antineoplastic association strength ($-\log_{10}P$ = 5.32). Experimental measurement yielded an $LC_{50}$ of 31 μM, demonstrating close agreement with the predicted range and confirming that the predictive framework generalizes to novel compounds (Figure 5).

### 3.4 Food-Level Therapeutic Effect Predictions

To translate network associations into dietary predictions, we integrated USDA flavonoid content data with association strengths for 32 ATC Level 2 categories that demonstrated statistically significant flavonoid associations. For each of 506 foods, flavonoid profiles were compared against therapeutic category vectors using Spearman rank correlation (16,192 food-ATC pairs). Food-ATC pairs ranking in the top 5% of correlation coefficients were classified as predicted therapeutic effects, yielding 685 food-ATC associations across 216 distinct foods and 26 ATC categories (Figure 6). The 26 ATC categories represent those with at least one food achieving top-5% correlation strength among the 32 statistically significant categories. TableS2 presents the complete 216 × 26 correlation matrix, from which Figure 6 was derived.

The food-ATC correlation heatmap revealed heterogeneous associations across the 26 therapeutic categories (Figure 6). While some foods showed strong positive correlations (red) with specific ATC classes, others displayed weak or negative correlations (blue), indicating substantial variation in food-level flavonoid signatures. Among these patterns, tea varieties showed high positive correlations with C01 (cardiac therapy; 0.547–0.603), contrasting with the majority of foods that showed negative correlations (mean −0.24, SD 0.238). Among 218 foods analyzed, tea consistently ranked in the top positions for C01 associations. This pattern suggests that tea's flavonoid profile aligns with cardiac therapy drug targets.

### 3.5 Epidemiological Literature Validation

To validate whether network-predicted food-effect relationships correspond to documented health associations, systematic PubMed literature searches were conducted for each of the 685 predicted food-ATC category associations (TableS3). Our literature searches identified 241 pieces of epidemiological supporting evidence, validating 96 food-ATC category associations (14.0% overall validation rate). These 96 validated predictions were collectively supported by 132 unique references, yielding an average of 2.5 evidence pieces per validated prediction (range: 1-9 studies per prediction) (TableS3).

Validation rates varied substantially across therapeutic domains, as shown in Figure 7. Cardiovascular categories demonstrated the highest rates: cardiac therapy 75% (6/8 predictions), lipid-modifying agents 50% (2/4), calcium channel blockers 45.8% (33/72), antihypertensives 35.0% (7/20), and antineoplastic agents 33.3% (1/3). The combined cardiovascular categories achieved 47.1% (48/102). Intermediate rates appeared in endocrine therapy (33.3%), corticosteroids (20.0%), bone disease agents (18.4%), psycholeptics (14.3%), anti-inflammatory agents (15.4%), antihistamines (15.4%), immunosuppressants (15.2%), and emollients (12.3%). Lower rates were observed in topical pain products (9.1%), otologicals (2.5%), and antimycotics (1.8%). The most robustly validated food-effect relationships included: tea (31 evidence counts across 2 ATC categories: C01, C08), blueberries (18 counts across 6 categories: C02, C08, L04, M01, M05, N05), tomato (13 counts across 4 categories: C08, N05, N06, R06), grape juice (10 counts across 2 categories: C08, C02), and plum (9 counts across 4 categories: C02, C08, M05, S02) (Figure 8). Tea demonstrated the highest validation among all foods examined, with particularly strong evidence for cardiac therapy associations (C01: 9 studies for black tea, 7 studies each for green tea varieties).

# 4. Discussion

## 4.1 Network Pharmacology as a Validated Framework

Network pharmacology provides a validated framework for systematically characterizing dietary polypharmacological properties and generating evidence-based food-therapeutic predictions. The convergence of computational predictions, experimental validation, and epidemiological confirmation demonstrates the utility of integrating multiple evidence streams. Flavonoids exhibit 2.7-fold greater target diversity than conventional drugs (45.3 vs. 16.8 proteins; $p = 7.5 \times 10^{-4}$), reflecting fundamental differences in molecular design between natural products and rationally designed pharmaceuticals (7,21). The strong correlation between network-predicted association strengths and experimental bioactivity ($r = 0.918$; $R^2 = 0.843$) confirms that network predictions reliably capture flavonoid bioactivity, validating the framework's predictive foundation. This finding parallels Valle et al.'s demonstration that network proximity between polyphenol targets and disease proteins predicts therapeutic effects, as validated by rosmarinic acid's platelet inhibition (9). While their study employed network proximity metrics based on shortest path distances, our statistical enrichment approach identifies shared pharmacological neighborhoods between flavonoids and therapeutic drug categories, offering complementary mechanistic insights into polypharmacological effects of dietary bioactives.

## 4.2 Flavonoid-Drug Category Associations Reveal Network-Mediated Therapeutic Mechanisms

A striking parallel emerged from our analysis: flavonoids and FDA-approved drugs show strikingly low direct associations with disease-related genes (flavonoids: 16.7%, drugs: 27.7% of tested combinations), yet both achieve therapeutic effects through indirect network mechanisms. This systematic dissociation reveals a fundamental principle: both natural products and synthetic pharmaceuticals achieve therapeutic efficacy not through direct targeting of disease genes, but through coordinated multi-target perturbations of interconnected biological networks. The convergence between flavonoids and drugs at the therapeutic mechanism level---despite low direct disease-gene associations---validates a network-mediated model of drug action that transcends the traditional "magic bullet" paradigm. Complex diseases likely benefit from coordinated multi-target perturbations rather than single-target selectivity (16,19), consistent with emerging polypharmacological approaches to drug discovery.

The broad alignment of flavonoid targets with cardiovascular drug mechanisms provides computational support for epidemiological benefits observed in flavonoid-rich diets (4,22,23). This alignment demonstrates the framework's biological relevance. Critically, the near-universal association between flavonoids and antineoplastic agents (78.6% of flavonoids; 11 of 14 compounds) indicates a class-wide therapeutic signature rather than isolated bioactivity. This pattern indicates that flavonoid target profiles align with therapeutic networks engaged by established anticancer drugs, suggesting that dietary flavonoid combinations may exert coordinated polypharmacological effects. Furthermore, the significant engagement of 28.6% of flavonoids with nervous system drug targets highlights under-explored neuroactive mechanisms, suggesting potential for dietary strategies in neurological health beyond established cardiovascular and metabolic domains.

### 4.3 Epidemiological Validation and Domain-Specific Research Maturity

The 14.0% overall validation rate is more appropriately interpreted as reflecting uneven research maturity across therapeutic domains rather than intrinsic limitations of the framework. Multi-level analysis provides complementary perspectives: 31.6% of foods demonstrated at least one validated prediction, and 64.0% of therapeutic categories received epidemiological support, indicating that the framework preferentially recapitulates relationships in well-studied cardiovascular and metabolic domains while generating hypotheses in understudied areas. Cardiovascular domains achieved the highest validation (47.1% overall, with cardiac therapy reaching 75%), consistent with the extensive cardiovascular literature documenting flavonoid-rich foods as cardioprotective (4,22,23).

In contrast, domains such as psycholeptics (16.4%) and antimycotics (1.8%) showed lower validation rates, highlighting gaps in epidemiological investigation rather than a failure of the predictive framework. The 584 unvalidated predictions (85.4%) are therefore better viewed as a structured set of computationally derived hypotheses for future investigation, consistent with the view that network-based analyses primarily act as generators of testable hypotheses in systems pharmacology and network medicine (14).

### 4.4 Multi-Functional Foods and Precision Nutrition

Six foods demonstrated broad therapeutic associations: tea (31 evidence counts, 2 categories), blueberries (18 counts, 6 categories), tomato (13 counts, 4 categories), grape juice (10 counts, 2 categories), plum (9 counts, 4 categories), and cranberry (10 counts, 4 categories). Tea showed the highest overall validation, with particularly concentrated evidence for cardiac therapy (C01). Both black tea and green tea varieties demonstrated cardiovascular protective effects through multiple mechanisms, including improved endothelial function and reduced blood pressure, consistent with tea's catechin-rich flavonoid profile (22–26). Blueberries reduced blood pressure (systolic −4.8 mmHg, diastolic −3.1 mmHg) and improved vascular function in randomized controlled trials (27,28). Tomatoes showed therapeutic associations spanning cardiovascular protection and neurological health. Cranberry demonstrated endothelial-dependent vasodilation and reduced arterial stiffness, while grape juice improved endothelial-dependent vasodilation and reduced arterial stiffness (29,30). Plum showed broad antioxidant and anti-inflammatory effects supporting cardiovascular health (31).

These multi-domain food profiles exemplify network-predicted polypharmacological effects aligned with precision nutrition principles, demonstrating that dietary combinations modulate interconnected biological systems through coordinated multi-target mechanisms. This network-level modulation is comparable to the actions of multi-targeted pharmaceuticals (32) and is consistent with contemporary precision nutrition frameworks (33,34).

### 4.5 Limitations and Future Directions

Our analysis has several constraints. We examined only 14 of the 26 flavonoids in the USDA database because the remaining compounds lacked sufficient STITCH target coverage, which may bias the network toward well-characterized flavonoids. Moreover, low oral bioavailability and unmodeled inter-individual differences in flavonoid metabolism and gut microbiota complicate direct translation of *in vitro* potency to *in vivo* effects (5,35).

These limitations highlight clear priorities for future work rather than undermining the framework. Pharmacokinetic modeling, pathway-level transcriptomic and proteomic validation, systematic evaluation of flavonoid combinations, and integration of individual omics and microbiome data will be essential to refine predictions and support precision nutrition applications. The framework's scalability and modularity position it for expansion, and as network databases mature and systems biology tools advance, network pharmacology will provide an increasingly systematic basis for precision nutrition and evidence-based dietary intervention development.

**Figure 1. Overview of the Multi-Tiered Validation Strategy for Dietary Flavonoid Therapeutic Discovery**

(1) Master network of flavonoid-protein and drug-protein interactions. (2) Statistical enrichment analysis of flavonoid target overlap with therapeutic drug categories. (3) Cell-based bioassays measuring $LC_{50}$ values. (4) Epidemiological validation via literature survey. Framework integrates network pharmacology, experimental validation, and epidemiological evidence for identifying dietary sources with therapeutic potential.

**Figure 2. Master Network Architecture and Target Diversity**

(A) Network visualization of 19,379 nodes and 278,768 interactions. Node types: human proteins (n=17,869), flavonoids (n=14), drugs (n=1,496). Edges: protein-protein interactions (n=265,834) and protein-compound interactions (n=12,934). Scale-free topology (density 0.0015) reflects hub-and-spoke network architecture. (B) Target protein counts. Flavonoids (mean 45.3; range 16–98) exceed FDA-approved drugs (mean 16.8; range 1–187) by 2.7-fold (Wilcoxon $p=7.5\times10^{-4}$).

**Figure 3. Flavonoid-ATC Association Analysis**

Heatmaps quantifying statistical association strength between 14 flavonoids and therapeutic drug categories via Fisher's exact test. (A) ATC Level 1 analysis across 14 broad therapeutic domains ($-\log_{10}$ p-values). (B) ATC Level 2 analysis across 20 specific therapeutic categories from ATC Level 1 categories L, C, and N.

**Figure 4. Dose-Response Cytotoxicity Curves**

Dose-response curves (1–200 μM) for six flavonoids in Jurkat leukemia cells. Orange lines: high-association flavonoids (luteolin, myricetin, diosmetin). Gray lines: low-association flavonoids (epicatechin, naringenin, hesperetin). Cell viability expressed as percentage of vehicle control (mean±SD, n=12) determined via MTT assay.

**Figure 5. Network Prediction vs. Experimental Bioactivity Correlation**

Correlation between network-predicted association strength ($-\log_{10}$ p-value) and experimental potency ($1/LC_{50}$, $\mu M^{-1}$). High-association flavonoids (red circles: luteolin, myricetin, diosmetin) show strong cytotoxicity; low-association flavonoids (gray circles: epicatechin, hesperetin, naringenin) show minimal activity. Linear regression (Pearson $r=0.918$; $p=0.0098$; $R^2=0.843$) demonstrates that network predictions explain 84.3% of bioactivity variance.

**Figure 6. Food-ATC Correlation Heatmap**

Heatmap of Spearman rank correlations between 506 foods and 26 ATC therapeutic categories. Red indicates positive correlations; gray indicates no correlation; blue indicates negative correlations. Rows and columns are organized by hierarchical clustering.

**Figure 7. Epidemiological Validation Rates by Therapeutic Domain**

Bars represent validation rates (%) for food-ATC associations across 26 therapeutic categories, color-coded by domain: cardiovascular (C, dark red), cancer/immune (L, medium red), nervous system (N, light red), and other (gray). Cardiovascular categories achieved highest rates (47.1% combined).

**Figure 8. Top-Validated Foods: Evidence Counts**

Top six foods ranked by literature-supported therapeutic evidence. Tea leads with 31 publications (combining all tea types), followed by blueberries (18), tomato (13), grape juice (10), plum (9), and cranberry (10).

## Acknowledgments

This work was supported by a grant from the Akiyama Life Science Foundation (THE AKIYAMA LIFE SCIENCE FOUNDATION). We utilized STRING, STITCH, USDA Flavonoid Content Database, and PubMed. The manuscript was edited for clarity and grammar using artificial intelligence (Perplexity AI). All substantive scientific content, analysis, and interpretations were conducted and validated by the authors.

## Conflict of Interest

The authors declare no conflict of interest.

## Data Availability

All data derive from publicly available databases: STRING (https://string-db.org), STITCH (http://stitch.embl.de), USDA Flavonoid Content Database (https://www.ars.usda.gov/), PubMed (https://pubmed.ncbi.nlm.nih.gov/). Master network construction files, statistical scripts (Python 3.10.9, R 4.2.2), and supplementary tables available upon request.

## Supplementary Information

**Table S1**: Flavonoid-ATC Association Strength

**Table S2:** Correlation of Food Nutrient Profiling and Flavonoid-ATC association strength

**Table S3:** Validation of Food-ATC associations

Fig. 1

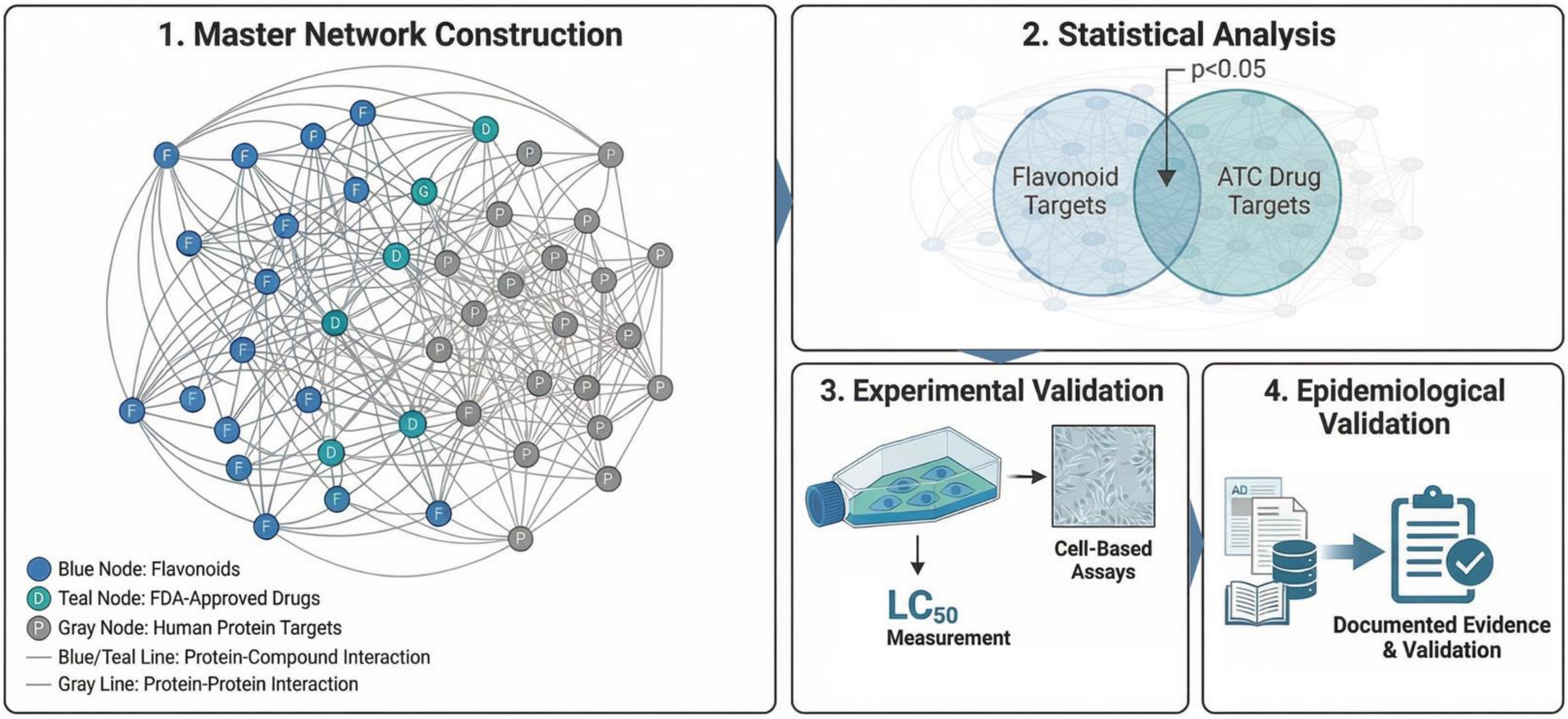

Figure 1: Overview of the Multi-Tiered Validation Strategy for Dietary Flavonoid Therapeutic Discovery

(1) Master network of flavonoid-protein and drug-protein interactions. (2) Statistical enrichment analysis of flavonoid target overlap with therapeutic drug categories. (3) Cell-based bioassays measuring $LC_{50}$ values. (4) Epidemiological validation via literature survey. Framework integrates network pharmacology, experimental validation, and epidemiological evidence for identifying dietary sources with therapeutic potential.

Fig. 2

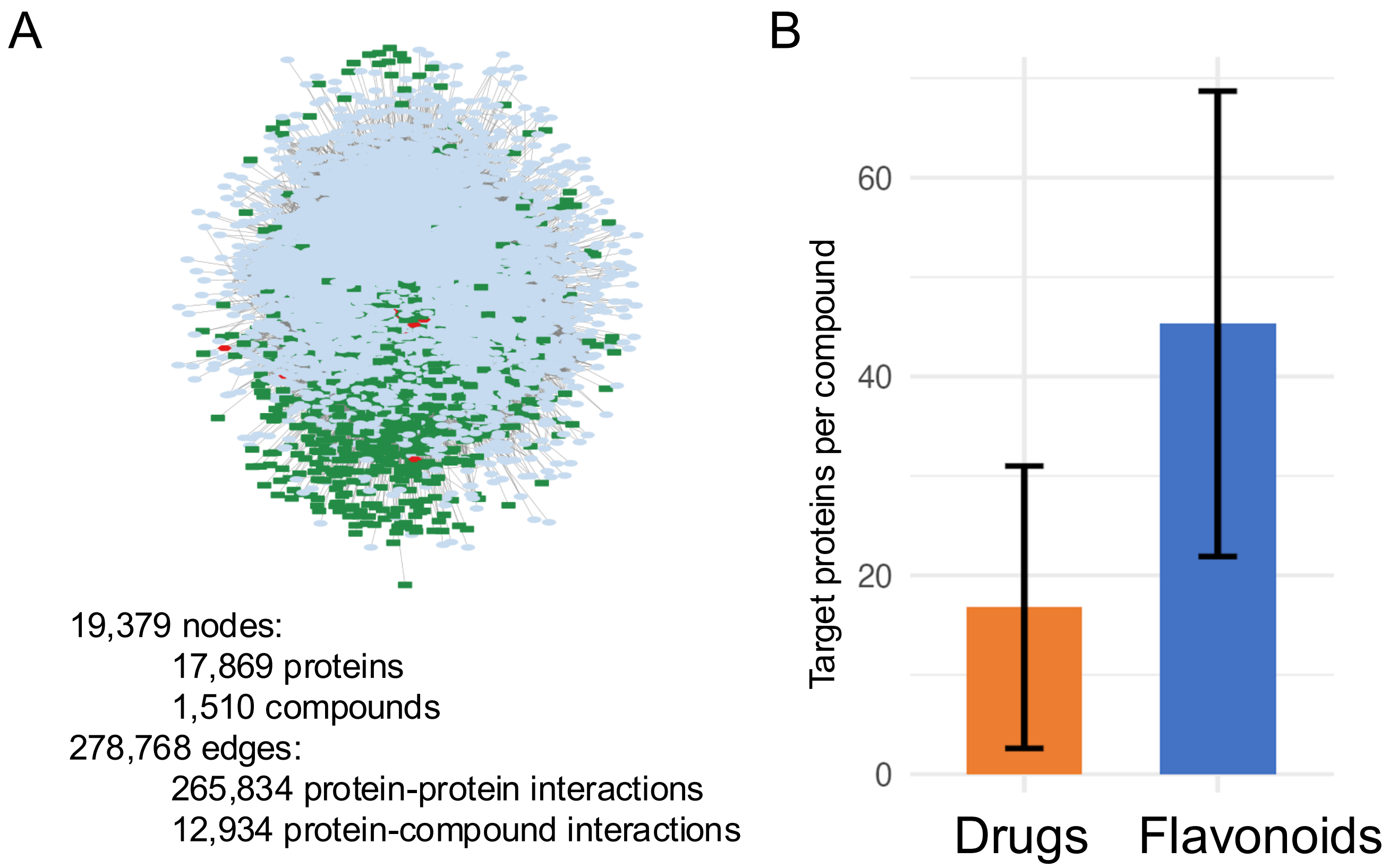

# Figure 2: Master Network Architecture and Target Diversity

(A) Network visualization of 19,379 nodes and 278,768 interactions. Node types: human proteins (n=17,869), flavonoids (n=14), drugs (n=1,496). Edges: protein-protein interactions (n=265,834) and protein-compound interactions (n=12,934). Scale-free topology (density 0.0015) reflects hub-and-spoke network architecture. (B) Target protein counts. Flavonoids (mean 45.3; range 16–98) exceed FDA-approved drugs (mean 16.8; range 1–187) by 2.7-fold (Wilcoxon $p=7.5 \times 10^{-4}$).

Fig. 3

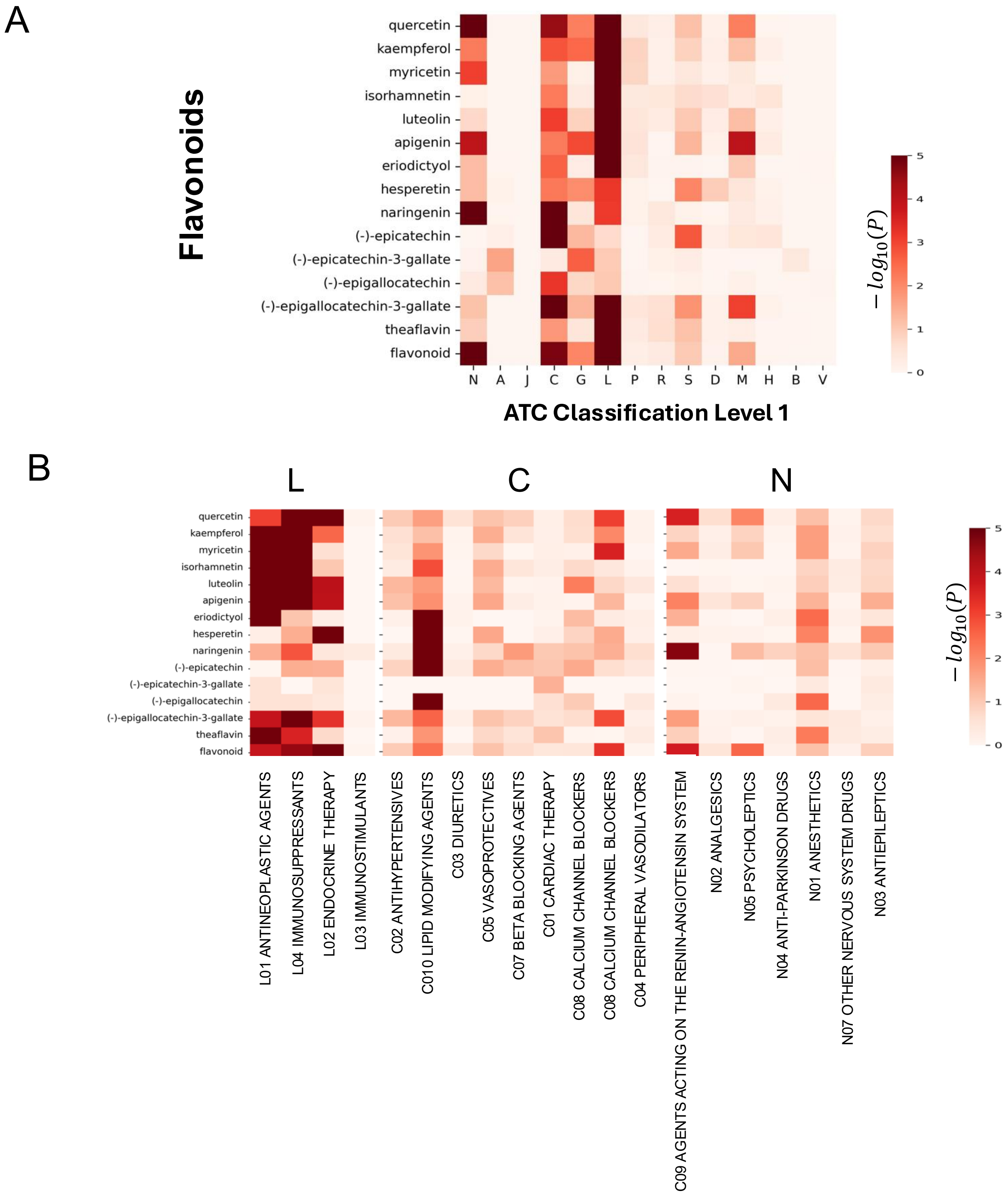

**Figure 3. Flavonoid-ATC Association Analysis**

Heatmaps quantifying statistical association strength between 14 flavonoids and therapeutic drug categories via Fisher's exact test. (A) ATC Level 1 analysis across 14 broad therapeutic domains ($-\log_{10}$ p-values). (B) ATC Level 2 analysis across 20 specific therapeutic categories from ATC Level 1 categories L, C, and N.

Fig. 4

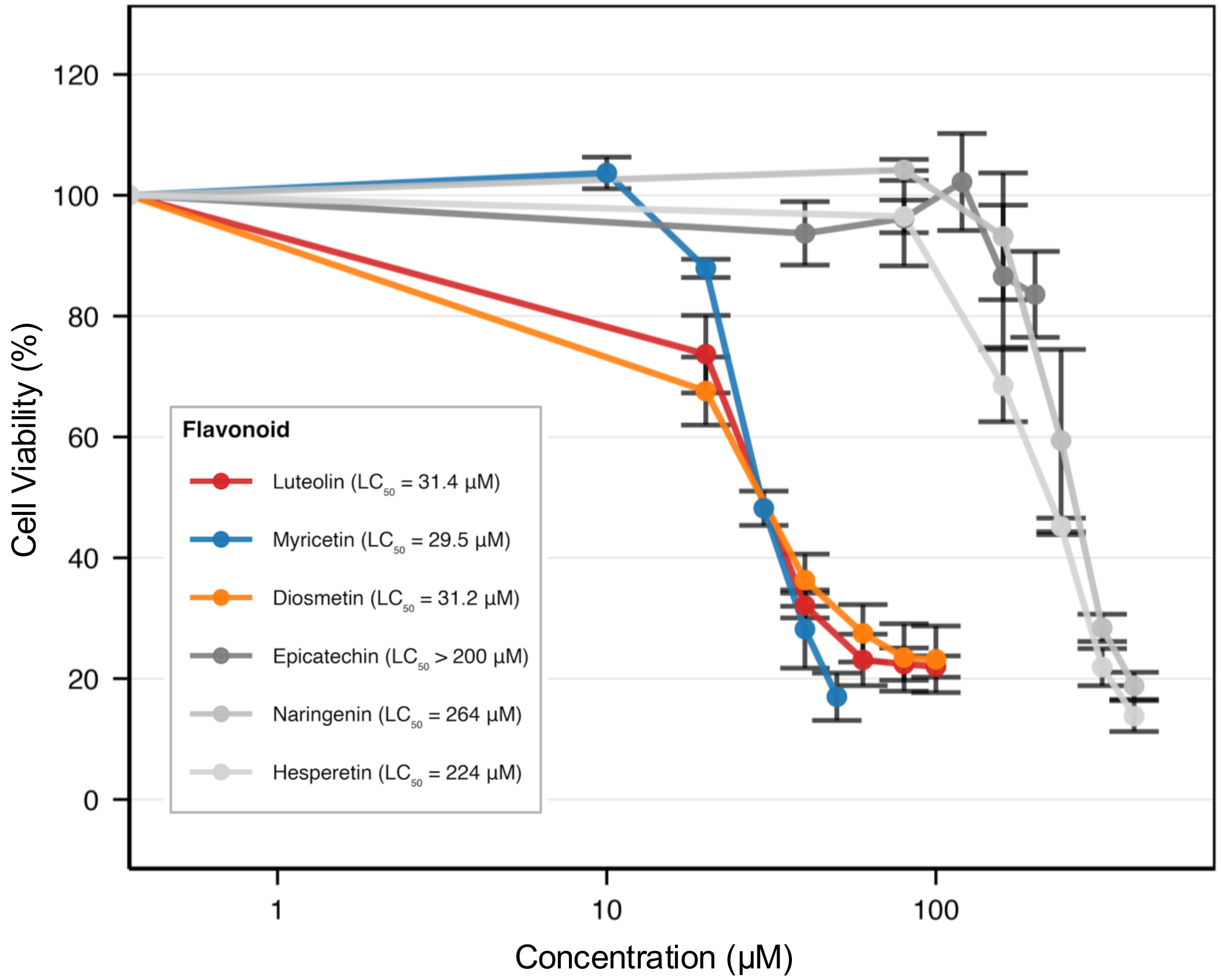

## Figure 4. Dose-Response Cytotoxicity Curves

Dose-response curves (1–200 µM) for six flavonoids in Jurkat leukemia cells. Orange lines: high-association flavonoids (luteolin, myricetin, diosmetin). Gray lines: low-association flavonoids (epicatechin, naringenin, hesperetin). Cell viability expressed as percentage of vehicle control (mean ± SD, n=12) determined via MTT assay.

Fig.5

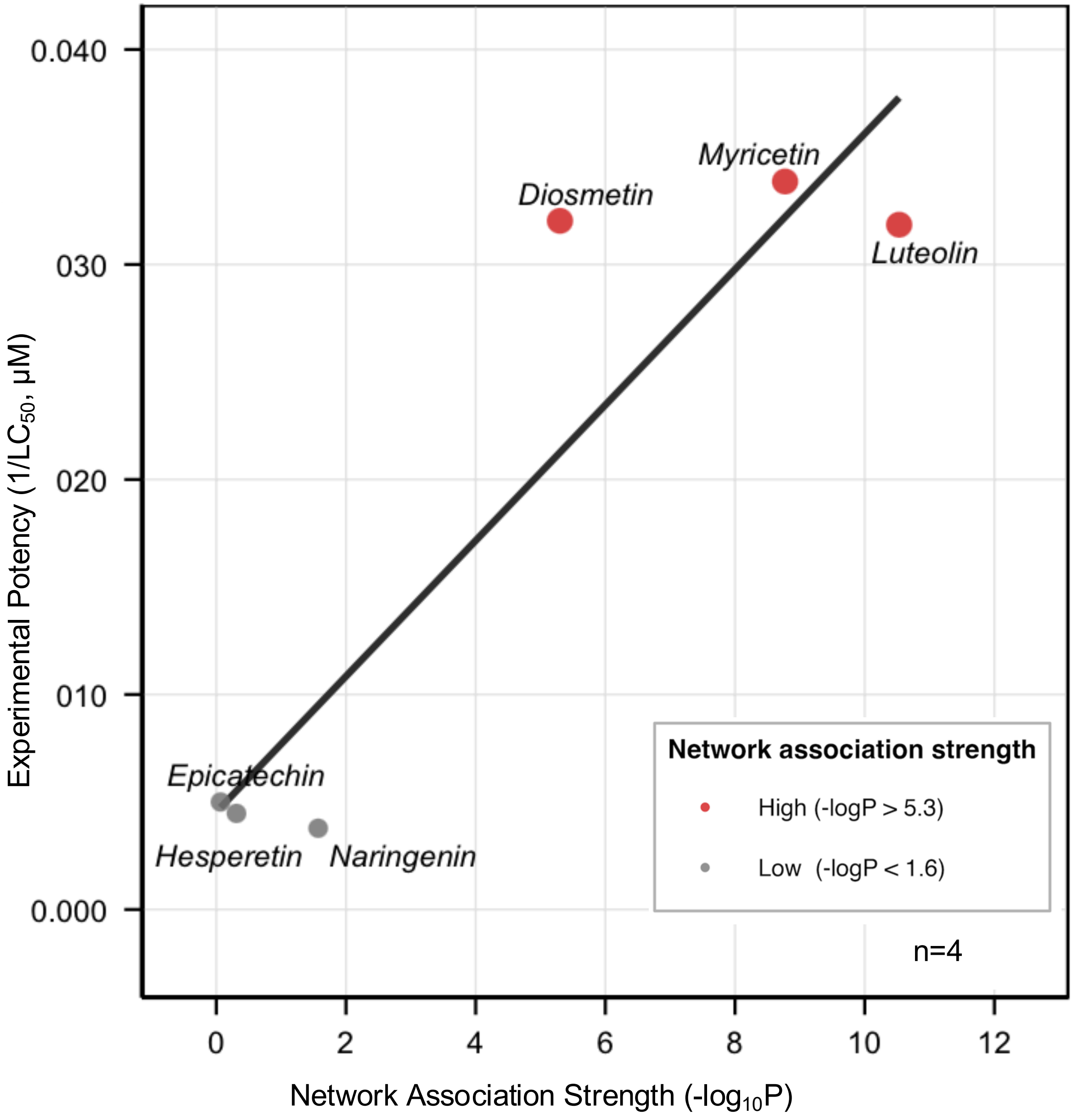

## Figure 5. Network Prediction vs. Experimental Bioactivity Correlation

Correlation between network-predicted association strength ($-\log_{10}$ p-value) and experimental potency ($1/LC_{50}$, $\mu M^{-1}$). High-association flavonoids (red circles: luteolin, myricetin, diosmetin) show strong cytotoxicity; low-association flavonoids (gray circles: epicatechin, hesperetin, naringenin) show minimal activity. Linear regression (Pearson r=0.918; p=0.0098; $R^2$=0.843) demonstrates that network predictions explain 84.3% of bioactivity variance.

Fig.6

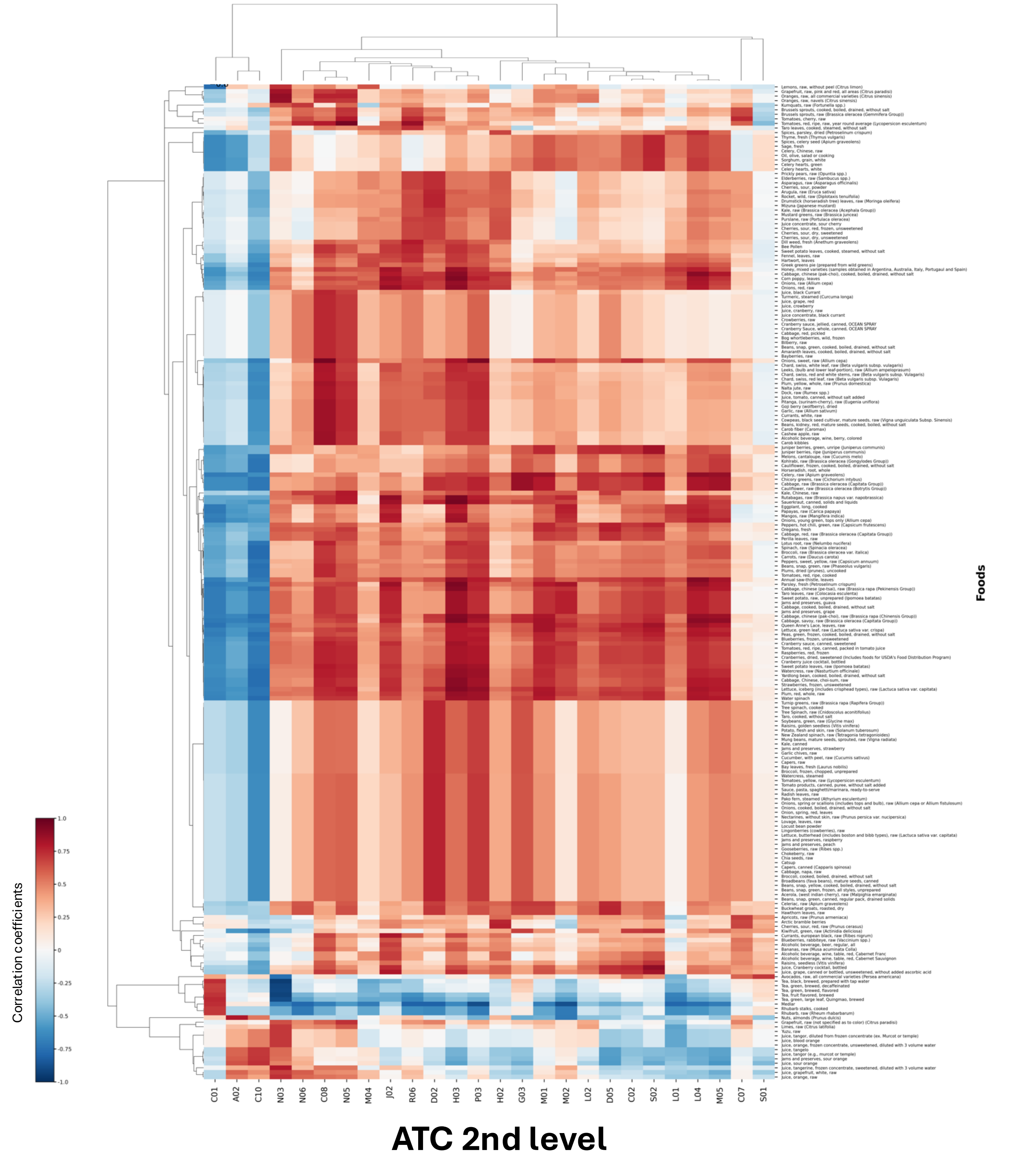

# Figure 6. Food-ATC Correlation Heatmap

Heatmap of Spearman rank correlations between 506 foods and 26 ATC therapeutic categories. Red indicates positive correlations; white indicates no correlation; blue indicates negative correlations. Rows and columns are organized by hierarchical clustering.

Fig. 7

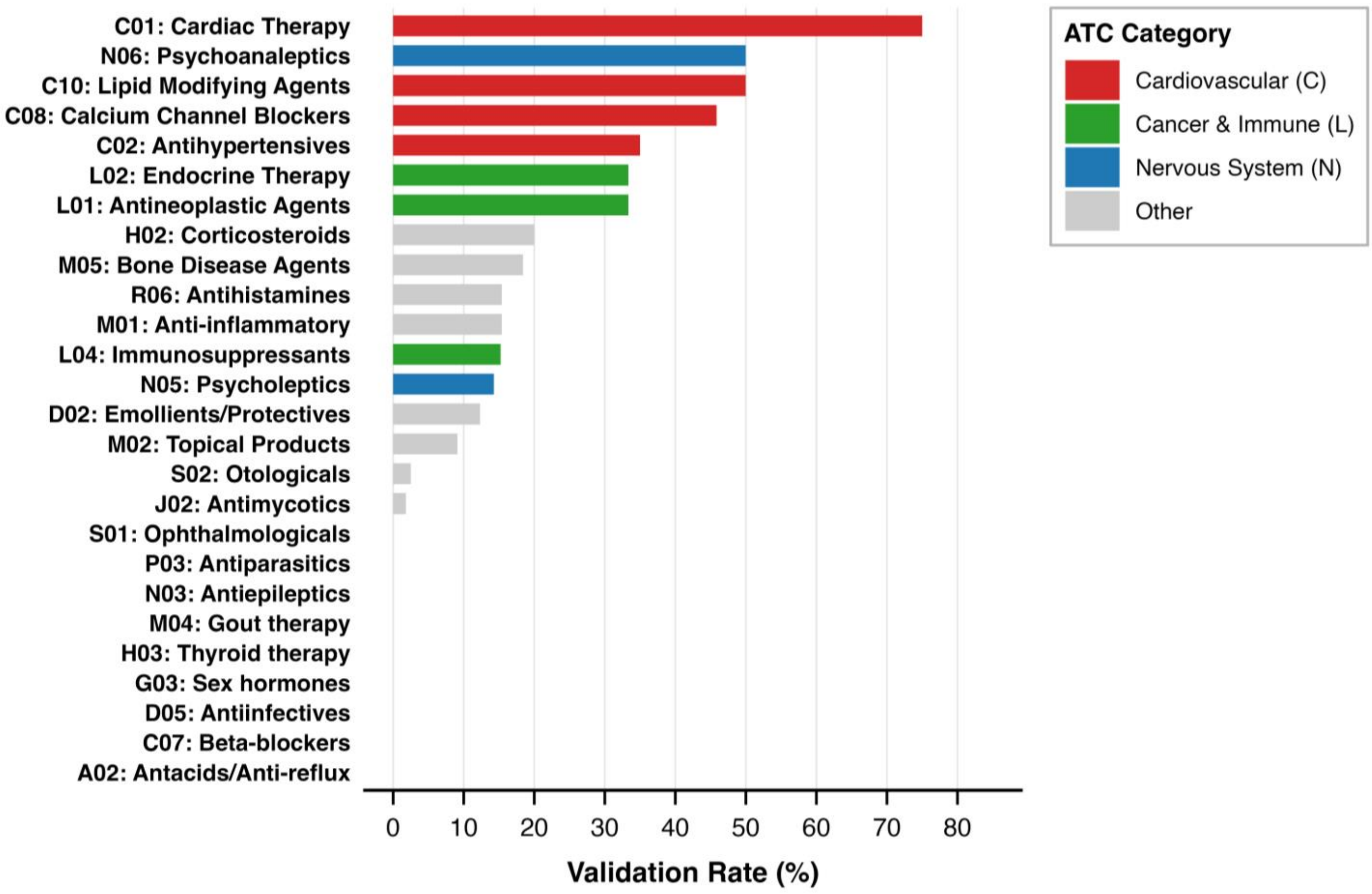

## Figure 7. Epidemiological Validation Rates by Therapeutic Domain

Bars represent validation rates (%) for food-ATC associations across 26 therapeutic categories, color-coded by domain: cardiovascular (C, dark red), cancer/immune (L, medium red), nervous system (N, light red), and other (gray). Cardiovascular categories achieved highest rates (47.1% combined).

Fig. 8

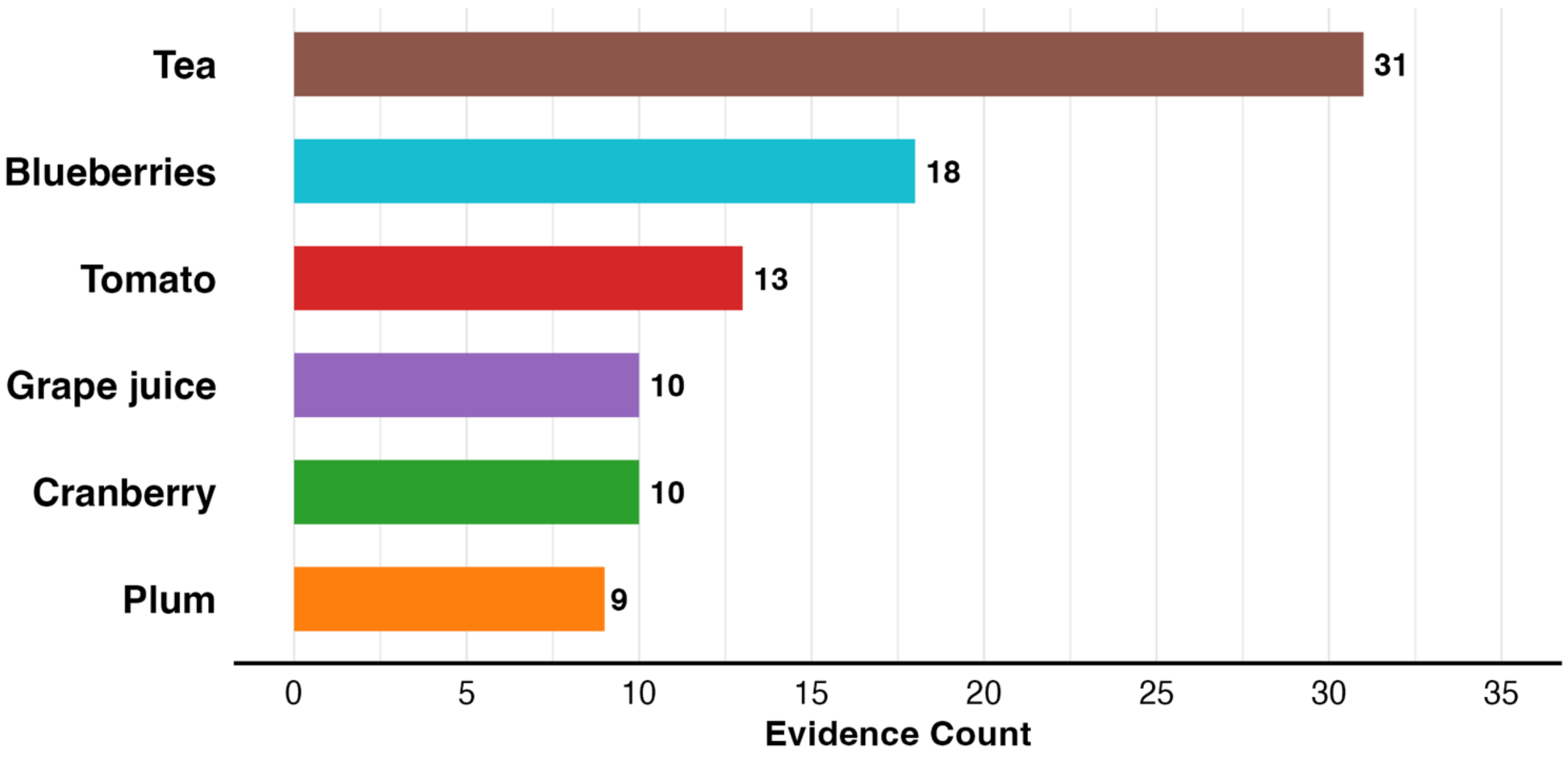

## Figure 8. Top-Validated Foods: Evidence Counts

Top six foods ranked by literature-supported therapeutic evidence. Tea leads with 31 publications (combining all tea types), followed by blueberries (18), tomato (13), grape juice (10), plum (9), and cranberry (10).

# Supplementary Tables

## Table S1

| Chemical Name | A | B | C | D | G | H | J | L | M | N | P | R | S | V |
|---|---|---|---|---|---|---|---|---|---|---|---|---|---|---|
| quercetin | 0.999999547 | 0.999968 | 0.001499742 | 0.931713106 | 0.041309943 | 0.962511539 | 1 | 6.10117981287119e-08 | 0.036225713 | 0.000124427 | 0.640742196 | 0.860902672 | 0.218095006 | 1 |
| kaempferol | 0.999999755 | 0.999992162 | 0.02623688 | 0.880768354 | 0.022199982 | 0.79479258 | 1 | 5.60964872500301e-10 | 0.209261116 | 0.079539989 | 0.318417012 | 0.916635513 | 0.33642635 | 1 |
| myricetin | 0.999968247 | 0.999850221 | 0.105747429 | 0.920776204 | 0.864360875 | 0.942270701 | 0.999988185 | 1.17440572525974e-10 | 0.638106124 | 0.014732143 | 0.314573755 | 0.893620078 | 0.62188606 | 0.999999942 |
| isorhamnetin | 0.997845119 | 0.999032694 | 0.037174655 | 0.492237926 | 0.675885827 | 0.44432387 | 1 | 5.0625915138553995e-12 | 0.663156573 | 0.891897832 | 0.603854207 | 0.580630974 | 0.368350914 | 0.999981605 |
| luteolin | 0.999863069 | 0.999799437 | 0.016305278 | 0.857529286 | 0.345410216 | 0.791862538 | 1 | 7.66946184800389e-16 | 0.191723375 | 0.568080819 | 0.571501687 | 0.786743312 | 0.26755789 | 0.999999251 |
| apigenin | 0.999999673 | 0.999999836 | 0.071563779 | 0.926203625 | 0.011617053 | 0.684828147 | 1 | 9.076245298532511e-10 | 0.001264316 | 0.005659608 | 0.537214378 | 0.989013172 | 0.167533601 | 1 |
| eriodictyol | 0.919314463 | 0.983226521 | 0.005062422 | 0.997784936 | 0.591991526 | 1 | 0.999999883 | 1.03548921459896e-05 | 0.128791944 | 0.094662915 | 0.464167634 | 0.88356378 | 0.869932433 | 0.933776788 |
| hesperetin | 0.943243195 | 0.996195736 | 0.03875455 | 0.305509019 | 0.043896076 | 0.872693872 | 1 | 0.005578689 | 0.537011269 | 0.239749989 | 0.890431923 | 0.988238707 | 0.033905168 | 0.99999943 |
| naringenin | 0.969467935 | 0.999944082 | 9.88372155040807e-07 | 0.967739438 | 0.510132859 | 0.812741176 | 1 | 0.003942512 | 0.604723895 | 1.01971946248293e-05 | 0.85944827 | 0.594926723 | 0.873695631 | 0.999999995 |

(-)-Epicatechin 0.86342788 0.99677041 1.2921259598679e-07
0.808085623 0.129789621 0.471171076 0.999999991 0.37747287
0.528094695 0.994090084 0.997645057 0.698587581 0.008003839
0.999803565

(-)-Epicatechin-3-gallate 0.067265481 0.523966332 0.605716168 0.974336737
0.006393504 1 0.999044796 0.181521895 1 0.888805634 1
0.773275787 0.634093767 1

(-)-Epigallocatechin 0.083617486 0.910003225 0.000913907 0.992911644
0.186694102 1 0.999917017 0.109818107 0.721935456 0.511823203
0.879608492 0.895154404 0.788522122 0.863293097

(-)-Epigallocatechin-3-gallate 0.999999873 0.999993307 0.00020266
0.94592795 0.174865323 0.889235634 1 8.64433706861018e-08
0.007257384 0.38487457 0.589810901 0.547838219 0.069535329 1

Theaflavin 0.99977624 0.965737919 0.109135462 0.91836823 0.587135254
0.959943164 1 1.72398001245585e-11 0.785674247 0.419187789
0.649192666 0.475432604 0.205100474 0.999954337

flavonoid 0.999998422 0.999995846 0.000830188 0.968093274 0.049791355
0.96512289 1 1.56358098844137e-08 0.113277743
4.5604455034087e-05 0.797399056 0.758087517 0.272157024 1

## Table S2

Food A02 C01 C02 C07 C08 C10 D02 D05 G03 H02 H03
J02 L01 L02 L04 M01 M02 M04 M05 N03 N05 N06
P03 R06 S01 S02

Acerola, (west indian cherry), raw (Malpighia emarginata) -0.238009825 -
0.191129102 0.259932879 0.362216623 0.462102897 -0.501263116
0.616259102 0.328526278 0.501263116 0.300639797 0.44190428
0.240669235 0.036062095 0.367833366 0.33898369 0.360620947
0.097367656 0.166619646 0.387571786 0.057699352 0.469323254
0.27407192 0.565057931 0.361017888 0.057699352 0.259932879

Alcoholic beverage, beer, regular, all -0.107549894 -0.044123034
0.306440478 0.412713881 0.49969123 -0.352984268 0.218897464
0.270551053 0.195795961 0.385014963 0.476421393 0.552123539 -
0.085488377 0.190280582 0.220615168 0.220615168 0.082730688 0.038778485

0.235440805 -0.085488377 0.295397578 0.055153792 0.282528966
0.237422352 0.212342099 0.356133528

Alcoholic beverage, wine, berry, colored -0.162082542 -0.223245766
0.229614551 0.162799725 0.682720597 -0.46789866 0.411280107
0.284722043 0.174315187 0.162799725 0.568263192 0.51023472
0.198780476 0.168198865 0.290525312 0.308874279 0.345572213
0.205803426 0.279524057 0.217129443 0.590874777 0.339455891
0.568263192 0.456167574 -0.076454029 0.180630113

Alcoholic beverage, wine, table, red, Cabernet Franc -0.002433926 -
0.104658812 0.445898702 0.371593717 0.389856789 -0.316410363
0.065219718 0.384983579 -0.026773185 0.466936842 0.398485368
0.55621344 0.146035552 0.321278215 0.29450503 0.180110514
0.031641036 0.36425963 0.227356682 0.017037481 0.165689135 -
0.094923109 0.180907467 0.307012221 0.236090809 0.506813826

Alcoholic beverage, wine, table, red, Cabernet Sauvignon -0.002433926 -
0.104658812 0.445898702 0.371593717 0.389856789 -0.316410363
0.065219718 0.384983579 -0.026773185 0.466936842 0.398485368
0.55621344 0.146035552 0.321278215 0.29450503 0.180110514
0.031641036 0.36425963 0.227356682 0.017037481 0.165689135 -
0.094923109 0.180907467 0.307012221 0.236090809 0.506813826

Amaranth leaves, cooked, boiled, drained, without salt-0.003606209 -
0.097367656 0.213000554 0.126775818 0.602899873 -0.313740224
0.284427278 0.375458604 0.032455885 0.065198992 0.347727958
0.335478328 0.082942818 0.180310473 0.104580075 0.266859501
0.284890548 0.260795968 0.05795466 0.245222244 0.563187905
0.414714089 0.470881609 0.407950213 -0.036062095 0.166068229

Annual saw-thistle, leaves -0.419200591 -0.553247292 0.575815319
0.1591198 0.490418979 -0.565433356 0.537249013 0.556296155
0.438698293 0.518975347 0.702575116 0.601324583 0.528875165
0.53374959 0.733601035 0.575182207 0.470382059 0.443087442
0.665855162 0.448447144 0.509938142 0.27540504 0.631583205
0.46358013 0.058493106 0.602654168

Apricots, raw (Prunus armeniaca) -0.064221385 0.070337707 0.327583426
0.55597642 0.290845098 -0.272176345 0.238109536 0.33370648
0.321106923 0.457682246 0.273380671 0.312325495 -0.290525312
0.314990601 0.143733575 0.211013121 -0.143733575 0.009215079
0.181229883 -0.223245766 0.15001484 -0.003058161
0.175086497 0.113276512 0.36392118 0.425552301

Arctic bramble berries 0.068823271 0.074329133 0.259060338 0.53366662
0.212209 -0.220234468 0.321516288 0.297643793 0.118376027
0.608324644 0.306927434 0.392500404 -0.101858441 0.305575324
0.222987399 0.06331741 -0.242257915 0.243329858 0.182497393 -
0.154164128 0.10472652 -0.140399473 0.160376497 0.338983209
0.269787223 0.305911676

Arugula, raw (Eruca sativa) -0.113151964 -0.174315187
0.168384004 0.334814529 0.315337316 -0.406735436 0.609189332
0.198999277 0.2599437 0.48225579 0.48225579 0.374172128
0.223245766 0.290525312 0.425084403 0.149849898 -0.033639773
0.328671143 0.402391774 0.058105062 0.309214262 0.009174484
0.445395475 0.517398121 -0.015290806 0.131645676

Asparagus, raw (Asparagus officinalis) -0.039756095 -0.131500931
0.162260949 0.334814529 0.290845098 -0.357804858 0.590635342
0.235737605 0.186547832 0.488399176 0.426965317 0.343248811
0.211013121 0.302757956 0.382270147 0.119268286 -0.070337707
0.390105002 0.334814529 0.076454029 0.303091207 0.015290806
0.420821931 0.548013394 -0.009174484 0.119399566

Avocados, raw, all commercial varieties (Persea americana) 0.06491177
0.313740224 0.440441823 0.463637277 0.12996644 0.158673217 -
0.107571855 0.245492164 0.082942818 0.315128462 0.083309823 0.018232518 -
0.375045785 0.144248379 0.028849676 0.176704264 -0.032455885 -
0.018110831 0.02897733 -0.602236981 -0.173288586 -0.079336608
-0.039843828 -0.119135903 0.562568677 0.361017888

Bananas, raw (Musa acuminata Colla) -0.126634819 -0.068823271
0.319691481 0.425827251 0.534656443 -0.399174973 0.304814144
0.314179559 0.253269638 0.384350571 0.49218994 0.523333872 -
0.071576202 0.239504984 0.245010846 0.264281362 0.09635258 0.077423137
0.262685642 -0.052305686 0.361030897 0.118376027 0.359464563
0.283863988 0.19545809 0.361030897

Bay leaves, fresh (Laurus nobilis) -0.266859501 -0.212766359
0.245492164 0.347727958 0.447662181 -0.508475535 0.608966095
0.277983774 0.508475535 0.307884129 0.470881609 0.269841264
0.06491177 0.33898369 0.367833366 0.346196109 0.104580075
0.137642317 0.423793448 0.043274514 0.440441823 0.230797406
0.550569266 0.346577173 0.043274514 0.245492164

Bayberries, raw -0.003606209 -0.097367656 0.213000554
0.126775818 0.602899873 -0.313740224 0.284427278 0.375458604

0.032455885 0.065198992 0.347727958 0.335478328 0.082942818
0.180310473 0.104580075 0.266859501 0.284890548 0.260795968
0.05795466 0.245222244 0.563187905 0.414714089 0.470881609
0.407950213 -0.036062095 0.166068229

Beans, kidney, red, mature seeds, cooked, boiled, without salt -0.168198865
-0.217129443 0.247983715 0.205803426 0.676597543 -0.486247627
0.460757414 0.321460371 0.229362088 0.187373269 0.55597642
0.47312674 0.168198865 0.217129443 0.296641634 0.333339568
0.314990601 0.218090198 0.291810828 0.204896799 0.603120886
0.357804858 0.592836735 0.462290629 -0.05198874 0.205122332

Beans, snap, green, canned, regular pack, drained solids -0.238009825 -
0.191129102 0.259932879 0.362216623 0.462102897 -0.501263116
0.616259102 0.328526278 0.501263116 0.300639797 0.44190428
0.240669235 0.036062095 0.367833366 0.33898369 0.360620947
0.097367656 0.166619646 0.387571786 0.057699352 0.469323254
0.27407192 0.565057931 0.361017888 0.057699352 0.259932879

Beans, snap, green, cooked, boiled, drained, without salt -0.003606209 -
0.097367656 0.213000554 0.126775818 0.602899873 -0.313740224
0.284427278 0.375458604 0.032455885 0.065198992 0.347727958
0.335478328 0.082942818 0.180310473 0.104580075 0.266859501
0.284890548 0.260795968 0.05795466 0.245222244 0.563187905
0.414714089 0.470881609 0.407950213 -0.036062095 0.166068229

Beans, snap, green, frozen, all styles, unprepared -0.238009825 -
0.191129102 0.259932879 0.362216623 0.462102897 -0.501263116
0.616259102 0.328526278 0.501263116 0.300639797 0.44190428
0.240669235 0.036062095 0.367833366 0.33898369 0.360620947
0.097367656 0.166619646 0.387571786 0.057699352 0.469323254
0.27407192 0.565057931 0.361017888 0.057699352 0.259932879

Beans, snap, green, raw (Phaseolus vulgaris) -0.308861235 -0.399864991
0.425151654 0.18281286 0.601838056 -0.603934021 0.404332895
0.383740779 0.33919582 0.196662319 0.553978364 0.412698403
0.364015026 0.386076543 0.463291852 0.446745714 0.350226578
0.343466586 0.473651501 0.248192063 0.507973405 0.27025358
0.592756849 0.380980054 -0.016546138 0.425151654

Beans, snap, yellow, cooked, boiled, drained, without salt -0.238009825 -
0.191129102 0.259932879 0.362216623 0.462102897 -0.501263116
0.616259102 0.328526278 0.501263116 0.300639797 0.44190428
0.240669235 0.036062095 0.367833366 0.33898369 0.360620947

0.097367656 0.166619646 0.387571786 0.057699352 0.469323254
0.27407192 0.565057931 0.361017888 0.057699352 0.259932879

Bee Pollen -0.057811548 -0.19545809 0.192917273 0.212913626
0.575995858 -0.423951351 0.489929582 0.261816299 0.055058617
0.348404115 0.583438636 0.565089234 0.300069463 0.19270516
0.363386872 0.184446367 0.198211021 0.376055235 0.298632098
0.211975675 0.504340871 0.176187574 0.525371284 0.611823352 -
0.09635258 0.121262286

Bilberry, raw -0.003606209 -0.097367656 0.213000554 0.126775818
0.602899873 -0.313740224 0.284427278 0.375458604 0.032455885
0.065198992 0.347727958 0.335478328 0.082942818 0.180310473
0.104580075 0.266859501 0.284890548 0.260795968 0.05795466
0.245222244 0.563187905 0.414714089 0.470881609 0.407950213 -
0.036062095 0.166068229

Blueberries, frozen, unsweetened -0.427748376 -0.525080342
0.566685296 0.12091695 0.587198791 -0.586553162 0.432527836
0.546171801 0.443116581 0.31644223 0.640602564 0.520587395
0.437993846 0.484098461 0.607044102 0.612166837 0.535325812
0.347314643 0.578857739 0.422625641 0.566685296 0.381643761
0.640602564 0.361550347 0.048665983 0.597455538

Blueberries, rabbiteye, raw (Vaccinium spp.) -0.022023447 0.030282239
0.325203403 0.431357475 0.496072988 -0.289057739 0.148927458
0.34173917 0.121128957 0.367759899 0.392645907 0.512199109 -
0.165175851 0.206469814 0.12938775 0.209222745 0.052305686 0.066362688
0.124430041 -0.085340856 0.294887832 0.090846718 0.237799634
0.223232845 0.261528431 0.383078585

Bog whortleberries, wild, frozen -0.003606209 -0.097367656
0.213000554 0.126775818 0.602899873 -0.313740224 0.284427278
0.375458604 0.032455885 0.065198992 0.347727958 0.335478328
0.082942818 0.180310473 0.104580075 0.266859501 0.284890548
0.260795968 0.05795466 0.245222244 0.563187905 0.414714089
0.470881609 0.407950213 -0.036062095 0.166068229

Broadbeans (fava beans), mature seeds, canned -0.238009825 -
0.191129102 0.259932879 0.362216623 0.462102897 -0.501263116
0.616259102 0.328526278 0.501263116 0.300639797 0.44190428
0.240669235 0.036062095 0.367833366 0.33898369 0.360620947
0.097367656 0.166619646 0.387571786 0.057699352 0.469323254
0.27407192 0.565057931 0.361017888 0.057699352 0.259932879

Broccoli, cooked, boiled, drained, without salt -0.238009825 -0.191129102
0.259932879 0.362216623 0.462102897 -0.501263116 0.616259102
0.328526278 0.501263116 0.300639797 0.44190428 0.240669235
0.036062095 0.367833366 0.33898369 0.360620947 0.097367656
0.166619646 0.387571786 0.057699352 0.469323254 0.27407192
0.565057931 0.361017888 0.057699352 0.259932879

Broccoli, frozen, chopped, unprepared -0.266859501 -0.212766359
0.245492164 0.347727958 0.447662181 -0.508475535 0.608966095
0.277983774 0.508475535 0.307884129 0.470881609 0.269841264
0.06491177 0.33898369 0.367833366 0.346196109 0.104580075
0.137642317 0.423793448 0.043274514 0.440441823 0.230797406
0.550569266 0.346577173 0.043274514 0.245492164

Broccoli, raw (Brassica oleracea var. italica) -0.341363425 -0.426704282
0.427173962 0.174202057 0.570483936 -0.61390358 0.395284095
0.350007053 0.360633941 0.204618289 0.564082852 0.414769931
0.393669112 0.382657388 0.495527553 0.443221867 0.346869287
0.32904833 0.514310836 0.233999122 0.468513378 0.22574033
0.5751433 0.355518975 -0.019270516 0.432685884

Brussels sprouts, cooked, boiled, drained, without salt -0.104611372 -
0.170681713 0.363786858 0.525371284 0.429929923 -0.327598771
0.409202549 0.217720923 0.349622218 0.248860082 0.221208962
0.233830028 0.112870165 0.261528431 0.341363425 0.385410319
0.090846718 0.138255601 0.295866986 0.313834117 0.575995858
0.462492383 0.378820347 0.523632599 -0.200963952 0.270084182

Brussels sprouts, raw (Brassica oleracea (Gemmifera Group)) -0.077082064
-0.148658266 0.34173917 0.536431732 0.407882235 -0.278046016
0.370230878 0.162601702 0.311081186 0.237799634 0.188027617
0.23939741 0.107364303 0.206469814 0.32484584 0.357881011
0.079834995 0.099544033 0.268215866 0.330351702 0.570483936
0.467998245 0.32904833 0.534656443 -0.245010846 0.231500728

Buckwheat groats, roasted, dry -0.284890548 -0.238009825
0.400729856 0.22095214 0.32130592 -0.266859501 0.568854556
0.610120231 0.548143839 0.44190428 0.394816119 0.288073781 -
0.010818628 0.461594812 0.385864413 0.548143839 0.331771271 0.119531485
0.340483625 0.33898369 0.51625558 0.508475535 0.51796977
0.267153237 0.198341521 0.447662181

Cabbage, Chinese, choi-sum, raw -0.458484786 -0.519957607
0.515401559 0.151789363 0.617969033 -0.612166837 0.510227447

0.479502942 0.489221196 0.36275085 0.722928998 0.593107032
0.417502906 0.432871111 0.637780513 0.591675897 0.530203077
0.275279013 0.625166358 0.381643761 0.587198791 0.36115282
0.671474977 0.392320589 0.028175043 0.535915053

Cabbage, chinese (pak-choi), cooked, boiled, drained, without salt -0.333898145
-0.485005335 0.456260443 0.200735747 0.453820548 -0.545935654
0.579144578 0.417222117 0.336335358 0.575279276 0.722159091
0.640755703 0.545935654 0.446009931 0.721414971 0.419200591
0.333898145 0.467567412 0.648719184 0.35583306 0.441621071
0.126735062 0.587519261 0.55385626 0 0.451380652

Cabbage, chinese (pak-choi), raw (Brassica rapa (Chinensis Group)) -
0.519957607 -0.550694017 0.556428548 0.172370971 0.556428548 -
0.601921367 0.572387136 0.53078668 0.601921367 0.445077283 0.733219802
0.582747084 0.397011966 0.494343931 0.699253333 0.653148718
0.540448547 0.244406601 0.686911183 0.402134701 0.576942043
0.381643761 0.681765781 0.361550347 0.079402393 0.597455538

Cabbage, chinese (pe-tsai), raw (Brassica rapa (Pekinensis Group)) -
0.482257419 -0.561778589 0.513604518 0.072143387 0.580373106 -
0.582300181 0.41761168 0.42629175 0.433518637 0.337527989 0.731740067
0.632902174 0.492518215 0.382214657 0.661821352 0.572039385
0.582300181 0.257654953 0.641560833 0.407866647 0.529012654
0.307823884 0.613218788 0.34668305 0.005130398 0.536716722

Cabbage, cooked, boiled, drained, without salt -0.445785982 -0.559165076
0.544302754 0.023293795 0.575258361 -0.538550695 0.341332226
0.482391541 0.363328459 0.297642931 0.680696442 0.612313536
0.507629124 0.394250031 0.626161813 0.584933052 0.621008218
0.30281933 0.582344865 0.466400363 0.533984219 0.347867674
0.587521264 0.332772774 0.012883988 0.570099093

Cabbage, napa, raw -0.238009825 -0.191129102 0.259932879
0.362216623 0.462102897 -0.501263116 0.616259102 0.328526278
0.501263116 0.300639797 0.44190428 0.240669235 0.036062095
0.367833366 0.33898369 0.360620947 0.097367656 0.166619646
0.387571786 0.057699352 0.469323254 0.27407192 0.565057931
0.361017888 0.057699352 0.259932879

Cabbage, raw (Brassica oleracea (Capitata Group)) -0.501033415 -
0.501033415 0.559460092 0.24609497 0.399614352 -0.578115479
0.58457507 0.540168365 0.688232713 0.442417923 0.572378188
0.361879805 0.311081186 0.586374271 0.655197542 0.635927026

0.371645665	0.25992053	0.669157109	0.305575324	0.460245495
0.319339979	0.61661998	0.275596105	0.143152404	0.625603157

Cabbage, red, pickled	-0.003606209	-0.097367656	0.213000554
0.126775818	0.602899873	-0.313740224	0.284427278	0.375458604
0.032455885	0.065198992	0.347727958	0.335478328	0.082942818
0.180310473	0.104580075	0.266859501	0.284890548	0.260795968
0.05795466	0.245222244	0.563187905	0.414714089	0.470881609
0.407950213	-0.036062095	0.166068229

Cabbage, red, raw (Brassica oleracea (Capitata Group))	-0.294563601	-
0.418445489	0.518120676	0.038711568	0.523632599	-0.412939628
0.292287535	0.581507781	0.275293085	0.235034522	0.489424828
0.445390529	0.352375149	0.42119842	0.448727729	0.553339101
0.536821516	0.34287389	0.378820347	0.470751175	0.542924326
0.443221867	0.53366662	0.31693552	0.060564479	0.542924326

Cabbage, savoy, raw (Brassica oleracea (Capitata Group))	-0.545571282	-
0.571184957	0.561556922	0.14664396	0.53078668	-0.581430427
0.562027188	0.525658306	0.612166837	0.465658892	0.748656009
0.608646954	0.412380171	0.484098461	0.724867008	0.663394188
0.571184957	0.21867959	0.707492792	0.422625641	0.561556922
0.376521026	0.666329575	0.341036852	0.084525128	0.607712285

Capers, canned (Capparis spinosa)	-0.238009825	-0.191129102
0.259932879	0.362216623	0.462102897	-0.501263116	0.616259102
0.328526278	0.501263116	0.300639797	0.44190428	0.240669235
0.036062095	0.367833366	0.33898369	0.360620947	0.097367656
0.166619646	0.387571786	0.057699352	0.469323254	0.27407192
0.565057931	0.361017888	0.057699352	0.259932879

Capers, raw	-0.266859501	-0.212766359	0.245492164	0.347727958
0.447662181	-0.508475535	0.608966095	0.277983774	0.508475535
0.307884129	0.470881609	0.269841264	0.06491177	0.33898369
0.367833366	0.346196109	0.104580075	0.137642317	0.423793448
0.043274514	0.440441823	0.230797406	0.550569266	0.346577173
0.043274514	0.245492164

Carob fiber (Caromax)	-0.168198865	-0.217129443	0.247983715
0.205803426	0.676597543	-0.486247627	0.460757414	0.321460371
0.229362088	0.187373269	0.55597642	0.47312674	0.168198865
0.217129443	0.296641634	0.333339568	0.314990601	0.218090198
0.291810828	0.204896799	0.603120886	0.357804858	0.592836735
0.462290629	-0.05198874	0.205122332

Carob kibbles -0.162082542 -0.223245766 0.229614551
0.162799725 0.682720597 -0.46789866 0.411280107 0.284722043
0.174315187 0.162799725 0.568263192 0.51023472 0.198780476
0.168198865 0.290525312 0.308874279 0.345572213 0.205803426
0.279524057 0.217129443 0.590874777 0.339455891 0.568263192
0.456167574 -0.076454029 0.180630113

Carrots, raw (Daucus carota) -0.341363425 -0.426704282
0.427173962 0.174202057 0.570483936 -0.61390358 0.395284095
0.350007053 0.360633941 0.204618289 0.564082852 0.414769931
0.393669112 0.382657388 0.495527553 0.443221867 0.346869287
0.32904833 0.514310836 0.233999122 0.468513378 0.22574033
0.5751433 0.355518975 -0.019270516 0.432685884

Cashew apple, raw -0.162082542 -0.223245766 0.229614551
0.162799725 0.682720597 -0.46789866 0.411280107 0.284722043
0.174315187 0.162799725 0.568263192 0.51023472 0.198780476
0.168198865 0.290525312 0.308874279 0.345572213 0.205803426
0.279524057 0.217129443 0.590874777 0.339455891 0.568263192
0.456167574 -0.076454029 0.180630113

Catsup -0.238009825 -0.191129102 0.259932879 0.362216623
0.462102897 -0.501263116 0.616259102 0.328526278 0.501263116
0.300639797 0.44190428 0.240669235 0.036062095 0.367833366
0.33898369 0.360620947 0.097367656 0.166619646 0.387571786
0.057699352 0.469323254 0.27407192 0.565057931 0.361017888
0.057699352 0.259932879

Cauliflower, frozen, cooked, boiled, drained, without salt -0.388386469 -
0.425084403 0.43779841 0.224233584 0.352075644 -0.584108785
0.405095444 0.321460371 0.498480272 0.242663741 0.433108703
0.219555546 0.345572213 0.46789866 0.516829239 0.437317048
0.198780476 0.297954214 0.568263192 0.137617253 0.296968152
0.113151964 0.48225579 0.223491496 0.05198874 0.480659793

Cauliflower, raw (Brassica oleracea (Botrytis Group)) -0.501033415 -
0.501033415 0.559460092 0.24609497 0.399614352 -0.578115479
0.58457507 0.540168365 0.688232713 0.442417923 0.572378188
0.361879805 0.311081186 0.586374271 0.655197542 0.635927026
0.371645665 0.25992053 0.669157109 0.305575324 0.460245495
0.319339979 0.61661998 0.275596105 0.143152404 0.625603157

Celeriac, raw (Apium graveolens) -0.320952643 -0.266859501
0.407950213 0.184730478 0.285204132 -0.238009825 0.554268542

0.602899873 0.562568677 0.470881609 0.416549116 0.324538817
0.010818628 0.447169974 0.421926508 0.562568677 0.375045785
0.083309823 0.369460955 0.367833366 0.494594507 0.501263116
0.496236773 0.238271806 0.20555394 0.462102897

Celery hearts, green -0.461594812 -0.501263116 0.501814864 -0.097798488 0.003610179 -0.191129102 0.127627625 0.415170571
0.421926508 0.329617127 0.322372794 0.277134271 0.385864413
0.400289251 0.562568677 0.515687954 0.468807231 0.130397984
0.510725438 0.414714089 0.119135903 0.173098055 0.213707807 -0.043322147 0.158673217 0.602899873

Celery hearts, white -0.461594812 -0.501263116 0.501814864 -0.097798488 0.003610179 -0.191129102 0.127627625 0.415170571
0.421926508 0.329617127 0.322372794 0.277134271 0.385864413
0.400289251 0.562568677 0.515687954 0.468807231 0.130397984
0.510725438 0.414714089 0.119135903 0.173098055 0.213707807 -0.043322147 0.158673217 0.602899873

Celery, Chinese, raw -0.447169974 -0.508475535 0.509035222 -0.105042821 -0.003610179 -0.212766359 0.076576575 0.393509498
0.385864413 0.278906799 0.286151132 0.233376228 0.421926508
0.40750167 0.548143839 0.494050697 0.439957555 0.173863979
0.503481105 0.393076832 0.083034114 0.129823541 0.19197481 -0.057762862 0.144248379 0.610120231

Celery, raw (Apium graveolens) -0.556092032 -0.561597894
0.609067391 0.152081161 0.305911676 -0.495527553 0.512199109
0.578751819 0.693738574 0.481129491 0.566847964 0.395284095
0.360633941 0.591880133 0.710256159 0.685479782 0.470751175
0.226739186 0.696808229 0.399174973 0.410638196 0.330351702
0.555787516 0.198429195 0.176187574 0.697258144

Chard, swiss, red and white stems, raw (Beta vulgaris subsp. Vulagaris) -0.223245766 -0.247711055 0.24186066 0.224233584 0.645982269 -0.516829239 0.497865393 0.272475933 0.296641634 0.224233584 0.592836735
0.485496067 0.186547832 0.217129443 0.351688535 0.333339568
0.296641634 0.181229883 0.365531459 0.168198865 0.566382558
0.302757956 0.592836735 0.443921465 -0.05198874 0.205122332

Chard, swiss, red leaf, raw (Beta vulgaris subsp. Vulagaris) -0.223245766 -0.247711055 0.24186066 0.224233584 0.645982269 -0.516829239
0.497865393 0.272475933 0.296641634 0.224233584 0.592836735
0.485496067 0.186547832 0.217129443 0.351688535 0.333339568

0.296641634 0.181229883 0.365531459 0.168198865 0.566382558
0.302757956 0.592836735 0.443921465 -0.05198874 0.205122332

Chard, swiss, white leaf, raw (Beta vulgaris subsp. vulagaris) -0.217129443 -0.253827378 0.223491496 0.181229883 0.652105324 -0.498480272
0.448388087 0.235737605 0.241594733 0.19966004 0.605123507
0.522604046 0.217129443 0.168198865 0.345572213 0.308874279
0.327223246 0.168943111 0.353244687 0.180431509 0.554136449
0.284408989 0.568263192 0.43779841 -0.076454029 0.180630113

Cherries, sour, dry, sweetened -0.039756095 -0.143733575
0.119399566 0.297954214 0.229614551 -0.327223246 0.553527363
0.143891785 0.131500931 0.512972719 0.445395475 0.386541454
0.272176345 0.241594733 0.412851759 0.05198874 -0.088686674
0.37781823 0.359388073 0.058105062 0.223491496 -0.088686674
0.359388073 0.54189034 -0.039756095 0.070415129

Cherries, sour, dry, unsweetened -0.039756095 -0.143733575
0.119399566 0.297954214 0.229614551 -0.327223246 0.553527363
0.143891785 0.131500931 0.512972719 0.445395475 0.386541454
0.272176345 0.241594733 0.412851759 0.05198874 -0.088686674
0.37781823 0.359388073 0.058105062 0.223491496 -0.088686674
0.359388073 0.54189034 -0.039756095 0.070415129

Cherries, sour, powder -0.039756095 -0.131500931 0.162260949
0.334814529 0.290845098 -0.357804858 0.590635342 0.235737605
0.186547832 0.488399176 0.426965317 0.343248811 0.211013121
0.302757956 0.382270147 0.119268286 -0.070337707 0.390105002
0.334814529 0.076454029 0.303091207 0.015290806 0.420821931
0.548013394 -0.009174484 0.119399566

Cherries, sour, red, frozen, unsweetened -0.039756095 -0.143733575
0.119399566 0.297954214 0.229614551 -0.327223246 0.553527363
0.143891785 0.131500931 0.512972719 0.445395475 0.386541454
0.272176345 0.241594733 0.412851759 0.05198874 -0.088686674
0.37781823 0.359388073 0.058105062 0.223491496 -0.088686674
0.359388073 0.54189034 -0.039756095 0.070415129

Cherries, sour, red, raw (Prunus cerasus) 0.085340856 0.101858441
0.253548416 0.539196844 0.173625546 -0.181693436 0.2574914
0.270084182 0.085340856 0.619385092 0.284806538 0.403635167 -0.12938775 0.278046016 0.200963952 0.030282239 -0.269787223
0.210148514 0.160376497 -0.19270516 0.044095377 -0.19545809
0.094013809 0.294887832 0.297316532 0.311423598

Chia seeds, raw -0.238009825 -0.191129102 0.259932879
0.362216623 0.462102897 -0.501263116 0.616259102 0.328526278
0.501263116 0.300639797 0.44190428 0.240669235 0.036062095
0.367833366 0.33898369 0.360620947 0.097367656 0.166619646
0.387571786 0.057699352 0.469323254 0.27407192 0.565057931
0.361017888 0.057699352 0.259932879

Chicory greens, raw (Cichorium intybus) -0.501033415 -0.501033415
0.559460092 0.24609497 0.399614352 -0.578115479 0.58457507
0.540168365 0.688232713 0.442417923 0.572378188 0.361879805
0.311081186 0.586374271 0.655197542 0.635927026 0.371645665
0.25992053 0.669157109 0.305575324 0.460245495 0.319339979
0.61661998 0.275596105 0.143152404 0.625603157

Chokeberry, raw -0.238009825 -0.191129102 0.259932879
0.362216623 0.462102897 -0.501263116 0.616259102 0.328526278
0.501263116 0.300639797 0.44190428 0.240669235 0.036062095
0.367833366 0.33898369 0.360620947 0.097367656 0.166619646
0.387571786 0.057699352 0.469323254 0.27407192 0.565057931
0.361017888 0.057699352 0.259932879

Corn poppy, leaves -0.292861555 -0.446613872 0.434889471
0.19610494 0.53750384 -0.534472339 0.58486436 0.442219068
0.290421042 0.546642519 0.740296147 0.681107862 0.50274567
0.410006178 0.668700552 0.419768229 0.373398483 0.455943984
0.585863507 0.380720022 0.522844644 0.205003089 0.622633183
0.596140623 -0.014643078 0.415343876

Cowpeas, black seed cultivar, mature seeds, raw (Vigna unguiculata Subsp. Sinensis) -0.168198865 -0.217129443 0.247983715 0.205803426
0.676597543 -0.486247627 0.460757414 0.321460371 0.229362088
0.187373269 0.55597642 0.47312674 0.168198865 0.217129443
0.296641634 0.333339568 0.314990601 0.218090198 0.291810828
0.204896799 0.603120886 0.357804858 0.592836735 0.462290629 -0.05198874 0.205122332

Cranberries, dried, sweetened (Includes foods for USDA's Food Distribution Program) -0.410431846 -0.507909409 0.549556835 0.10821508
0.616325422 -0.564343788 0.43576871 0.552124857 0.415562244
0.319492142 0.65959668 0.560274056 0.420692642 0.45404023
0.589995779 0.607952172 0.564343788 0.32979834 0.5513816
0.443779433 0.600917287 0.415562244 0.649290482 0.382635366
0.041043185 0.572669038

Cranberry Sauce, whole, canned, OCEAN SPRAY -0.003606209 -
0.097367656 0.213000554 0.126775818 0.602899873 -0.313740224
0.284427278 0.375458604 0.032455885 0.065198992 0.347727958
0.335478328 0.082942818 0.180310473 0.104580075 0.266859501
0.284890548 0.260795968 0.05795466 0.245222244 0.563187905
0.414714089 0.470881609 0.407950213 -0.036062095 0.166068229

Cranberry juice cocktail, bottled -0.410431846 -0.507909409
0.549556835 0.10821508 0.616325422 -0.564343788 0.43576871
0.552124857 0.415562244 0.319492142 0.65959668 0.560274056
0.420692642 0.45404023 0.589995779 0.607952172 0.564343788
0.32979834 0.5513816 0.443779433 0.600917287 0.415562244
0.649290482 0.382635366 0.041043185 0.572669038

Cranberry sauce, canned, sweetened -0.405301448 -0.513039807
0.534148699 0.072143387 0.621461467 -0.548952594 0.394266928
0.521308586 0.369388661 0.298879746 0.669902878 0.591400392
0.446344632 0.412997045 0.584865381 0.58743058 0.589995779
0.319492142 0.541075402 0.45404023 0.590645196 0.40017105
0.628678086 0.377499321 0.020521592 0.552124857

Cranberry sauce, jellied, canned, OCEAN SPRAY -0.003606209 -
0.097367656 0.213000554 0.126775818 0.602899873 -0.313740224
0.284427278 0.375458604 0.032455885 0.065198992 0.347727958
0.335478328 0.082942818 0.180310473 0.104580075 0.266859501
0.284890548 0.260795968 0.05795466 0.245222244 0.563187905
0.414714089 0.470881609 0.407950213 -0.036062095 0.166068229

Crowberries, raw -0.003606209 -0.097367656 0.213000554
0.126775818 0.602899873 -0.313740224 0.284427278 0.375458604
0.032455885 0.065198992 0.347727958 0.335478328 0.082942818
0.180310473 0.104580075 0.266859501 0.284890548 0.260795968
0.05795466 0.245222244 0.563187905 0.414714089 0.470881609
0.407950213 -0.036062095 0.166068229

Cucumber, with peel, raw (Cucumis sativus) -0.266859501 -0.212766359
0.245492164 0.347727958 0.447662181 -0.508475535 0.608966095
0.277983774 0.508475535 0.307884129 0.470881609 0.269841264
0.06491177 0.33898369 0.367833366 0.346196109 0.104580075
0.137642317 0.423793448 0.043274514 0.440441823 0.230797406
0.550569266 0.346577173 0.043274514 0.245492164

Currants, european black, raw (Ribes nigrum) 0.007684103 -0.064034188
0.258982873 0.342169241 0.541043427 -0.335539145 0.282308587

0.279496368 -0.012806838 0.439931881 0.537694522 0.655266721
0.105016068 0.166488889 0.26894359 0.135752479 0.099893333
0.259842807 0.208388786 0.048665983 0.361550347 0.043543248
0.337023839 0.469246195 0.089647863 0.248726125

Currants, white, raw -0.168198865 -0.217129443 0.247983715
0.205803426 0.676597543 -0.486247627 0.460757414 0.321460371
0.229362088 0.187373269 0.55597642 0.47312674 0.168198865
0.217129443 0.296641634 0.333339568 0.314990601 0.218090198
0.291810828 0.204896799 0.603120886 0.357804858 0.592836735
0.462290629 -0.05198874 0.205122332

Dill weed, fresh (Anethum graveolens) -0.024776378 -0.173434644
0.170869585 0.24609497 0.476781261 -0.390916181 0.523333872
0.239768611 0.044046894 0.425827251 0.550257292 0.531684945
0.311081186 0.214728606 0.385410319 0.12938775 0.099105511
0.420297027 0.309692546 0.178940505 0.427173962 0.088093787
0.481129491 0.63387104 -0.085340856 0.099214598

Dock, raw (Rumex spp.) -0.198780476 -0.229362088 0.254106769
0.236520356 0.658228379 -0.510712916 0.504050057 0.315337316
0.290525312 0.218090198 0.568263192 0.460757414 0.162082542
0.241594733 0.327223246 0.345572213 0.290525312 0.205803426
0.334814529 0.180431509 0.590874777 0.339455891 0.605123507
0.456167574 -0.039756095 0.217368441

Drumstick (horseradish tree) leaves, raw (Moringa oleifera) -0.088686674 -
0.15596622 0.180630113 0.347101301 0.327583426 -0.400619114
0.615373995 0.24186066 0.253827378 0.476112404 0.457682246
0.349433475 0.198780476 0.314990601 0.400619114 0.162082542 -
0.039756095 0.353244687 0.371674844 0.070337707 0.33370648 0.045872418
0.457682246 0.52964423 -0.003058161 0.143891785

Eggplant, long, cooked -0.452261093 -0.562568677 0.449998179 -
0.113565565 0.416869479 -0.413653439 0.158944655 0.292636853
0.248192063 0.240980588 0.631535335 0.621836108 0.565326367
0.234403615 0.606691711 0.466049541 0.625995538 0.191122536
0.562288039 0.432957266 0.347851353 0.187522892 0.404404206
0.201532927 -0.030334586 0.47760543

Elderberries, raw (Sambucus spp.) -0.039756095 -0.131500931
0.162260949 0.334814529 0.290845098 -0.357804858 0.590635342
0.235737605 0.186547832 0.488399176 0.426965317 0.343248811
0.211013121 0.302757956 0.382270147 0.119268286 -0.070337707

0.390105002 0.334814529 0.076454029 0.303091207 0.015290806
0.420821931 0.548013394 -0.009174484 0.119399566

Fennel, leaves, raw -0.105016068 -0.315048205 0.315394984
0.162080167 0.525658306 -0.473852991 0.401447991 0.315394984
0.033297778 0.347314643 0.568566934 0.546487266 0.478975726
0.294557265 0.463607521 0.233084444 0.233084444 0.522258315
0.383332458 0.27918906 0.433347579 0.105016068 0.501676707
0.587198791 -0.089647863 0.264111246

Garlic chives, raw -0.266859501 -0.212766359 0.245492164
0.347727958 0.447662181 -0.508475535 0.608966095 0.277983774
0.508475535 0.307884129 0.470881609 0.269841264 0.06491177
0.33898369 0.367833366 0.346196109 0.104580075 0.137642317
0.423793448 0.043274514 0.440441823 0.230797406 0.550569266
0.346577173 0.043274514 0.245492164

Garlic, raw (Allium sativum) -0.168198865 -0.217129443
0.247983715 0.205803426 0.676597543 -0.486247627 0.460757414
0.321460371 0.229362088 0.187373269 0.55597642 0.47312674
0.168198865 0.217129443 0.296641634 0.333339568 0.314990601
0.218090198 0.291810828 0.204896799 0.603120886 0.357804858
0.592836735 0.462290629 -0.05198874 0.205122332

Goji berry (wolfberry), dried -0.168198865 -0.217129443
0.247983715 0.205803426 0.676597543 -0.486247627 0.460757414
0.321460371 0.229362088 0.187373269 0.55597642 0.47312674
0.168198865 0.217129443 0.296641634 0.333339568 0.314990601
0.218090198 0.291810828 0.204896799 0.603120886 0.357804858
0.592836735 0.462290629 -0.05198874 0.205122332

Gooseberries, raw (Ribes spp.) -0.238009825 -0.191129102
0.259932879 0.362216623 0.462102897 -0.501263116 0.616259102
0.328526278 0.501263116 0.300639797 0.44190428 0.240669235
0.036062095 0.367833366 0.33898369 0.360620947 0.097367656
0.166619646 0.387571786 0.057699352 0.469323254 0.27407192
0.565057931 0.361017888 0.057699352 0.259932879

Grapefruit, raw (not specified as to color) (Citrus paradisi) 0.118376027 -
0.06331741 -0.038583455 0.398176131 0.476781261 0.027529309
0.400851477 -0.049607299 0.231246191 0.116134705 0.102309145
0.164237758 -0.32209291 0.300069463 0.005505862 0.275293085
0.137646543 -0.033181344 0.024886008 0.517551 0.587019703
0.382657388 0.31522277 0.355518975 -0.038541032 0.02755961

Grapefruit, raw, pink and red, all areas (Citrus paradisi) 0.004874425 -
0.294902742 0.200071424 0.25459168 0.558736051 -0.121860637
0.197155601 0.056117595 0.131609488 0.039167951 0.188495763
0.280946731 0.051181468 0.365581911 0.182790955 0.392391251
0.348521422 0.178703775 0.164015794 0.670233503 0.588014796
0.370456336 0.337823575 0.363544417 -0.104800148 0.26106881

Greek greens pie (prepared from wild greens) -0.207470769 -0.397011966
0.40770571 0.187807177 0.4487327 -0.540448547 0.406627965
0.3512936 0.181857094 0.352460045 0.53254912 0.442887784
0.514834872 0.412380171 0.540448547 0.315048205 0.212593504
0.537694522 0.491385903 0.25869812 0.366678721 0.069156923
0.501676707 0.494888064 -0.038420513 0.392320589

Hartwort, leaves -0.115261538 -0.315048205 0.31026661
0.203243384 0.464117821 -0.478975726 0.453247732 0.300009862
0.079402393 0.409059469 0.563421532 0.520587395 0.478975726
0.325293675 0.499466667 0.217716239 0.166488889 0.53254912
0.424495675 0.243329914 0.387192216 0.048665983 0.491385903
0.592327164 -0.074279658 0.264111246

Hawthorn leaves, raw -0.284890548 -0.238009825 0.400729856
0.22095214 0.32130592 -0.266859501 0.568854556 0.610120231
0.548143839 0.44190428 0.394816119 0.288073781 -0.010818628
0.461594812 0.385864413 0.548143839 0.331771271 0.119531485
0.340483625 0.33898369 0.51625558 0.508475535 0.51796977
0.267153237 0.198341521 0.447662181

Honey, mixed varieties (samples obtained in Argentina, Australia, Italy, Portugaul and Spain) -0.27540504 -0.477693697 0.46845992 0.093023883
0.551416365 -0.511814675 0.416491207 0.446500862 0.175479317
0.423503467 0.687887134 0.665400153 0.575182207 0.3826424
0.626363674 0.421637804 0.450884357 0.494495378 0.531215332
0.43626108 0.500178561 0.212037508 0.570383282 0.539216888 -
0.038995404 0.451380652

Horseradish, root, whole -0.388386469 -0.425084403 0.43779841
0.224233584 0.352075644 -0.584108785 0.405095444 0.321460371
0.498480272 0.242663741 0.433108703 0.219555546 0.345572213
0.46789866 0.516829239 0.437317048 0.198780476 0.297954214
0.568263192 0.137617253 0.296968152 0.113151964 0.48225579
0.223491496 0.05198874 0.480659793

Jams and preserves, grape	-0.445785982	-0.559165076
0.544302754	0.023293795	0.575258361	-0.538550695	0.341332226
0.482391541	0.363328459	0.297642931	0.680696442	0.612313536
0.507629124	0.394250031	0.626161813	0.584933052	0.621008218
0.30281933	0.582344865	0.466400363	0.533984219	0.347867674
0.587521264	0.332772774	0.012883988	0.570099093

Jams and preserves, guava	-0.445785982	-0.559165076
0.544302754	0.023293795	0.575258361	-0.538550695	0.341332226
0.482391541	0.363328459	0.297642931	0.680696442	0.612313536
0.507629124	0.394250031	0.626161813	0.584933052	0.621008218
0.30281933	0.582344865	0.466400363	0.533984219	0.347867674
0.587521264	0.332772774	0.012883988	0.570099093

Jams and preserves, peach	-0.238009825	-0.191129102
0.259932879	0.362216623	0.462102897	-0.501263116	0.616259102
0.328526278	0.501263116	0.300639797	0.44190428	0.240669235
0.036062095	0.367833366	0.33898369	0.360620947	0.097367656
0.166619646	0.387571786	0.057699352	0.469323254	0.27407192
0.565057931	0.361017888	0.057699352	0.259932879

Jams and preserves, raspberry	-0.238009825	-0.191129102
0.259932879	0.362216623	0.462102897	-0.501263116	0.616259102
0.328526278	0.501263116	0.300639797	0.44190428	0.240669235
0.036062095	0.367833366	0.33898369	0.360620947	0.097367656
0.166619646	0.387571786	0.057699352	0.469323254	0.27407192
0.565057931	0.361017888	0.057699352	0.259932879

Jams and preserves, sour orange	0.486247627	-0.174315187	-0.309214262
-0.033788622	0.052045965	0.57187614	-0.18863223	-0.443921465	-
0.339455891	-0.285667442	-0.396248388	-0.176262903	-
0.339455891	-0.180431509	-0.388386469	-0.05198874	0.058105062
0.089079095	-0.439252089	0.443433371	0.070415129	0.266060022	-
0.187373269	0.033676801	-0.284408989	-0.370444808

Jams and preserves, strawberry	-0.266859501	-0.212766359
0.245492164	0.347727958	0.447662181	-0.508475535	0.608966095
0.277983774	0.508475535	0.307884129	0.470881609	0.269841264
0.06491177	0.33898369	0.367833366	0.346196109	0.104580075
0.137642317	0.423793448	0.043274514	0.440441823	0.230797406
0.550569266	0.346577173	0.043274514	0.245492164

Juice concentrate, black currant	-0.003606209	-0.097367656
0.213000554	0.126775818	0.602899873	-0.313740224	0.284427278

0.375458604	0.032455885 0.065198992 0.347727958 0.335478328
0.082942818	0.180310473 0.104580075 0.266859501 0.284890548
0.260795968	0.05795466 0.245222244 0.563187905 0.414714089
0.470881609	0.407950213 -0.036062095 0.166068229

Juice concentrate, sour cherry -0.039756095 -0.143733575
0.119399566 0.297954214 0.229614551 -0.327223246 0.553527363
0.143891785 0.131500931 0.512972719 0.445395475 0.386541454
0.272176345 0.241594733 0.412851759 0.05198874 -0.088686674
0.37781823 0.359388073 0.058105062 0.223491496 -0.088686674
0.359388073 0.54189034 -0.039756095 0.070415129

Juice, Cranberry cocktail, bottled -0.255907338 -0.260781763
0.605094064 0.347615563 0.512378038 -0.416763379 0.205781159
0.583135005 0.321712082 0.477359399 0.528767335 0.598860138
0.092614084 0.448447144 0.419200591 0.494754186 0.311963231
0.208079738 0.384335516 0.170604892 0.37818379 0.209600296
0.376991526 0.217150692 0.311963231 0.707569671

Juice, black Currant 0.003606209 -0.104580075 0.191339481 0.076065491
0.610120231 -0.292102967 0.226083221 0.332136457 -0.032455885
0.036221662 0.362216623 0.379236371 0.119004912 0.122611122
0.097367656 0.238009825 0.320952643 0.246307303 0.043465995
0.259647082 0.54874719 0.393076832 0.44190428 0.400729856 -
0.06491177 0.137186797

Juice, blood orange 0.292102967 0.028849676 -0.267153237 0.115909319
0.205780196 0.400289251 0.003646504 -0.299644847 -0.090155237
-0.112287153 -0.19197481 -0.010939511 -0.414714089
0.140642169 -0.259647082 0.057699352 0.119004912 -0.137642317
-0.246307303 0.562568677 0.285204132 0.169491845 -0.036221662
0.061373041 -0.007212419 -0.137186797

Juice, cranberry, raw -0.003606209 -0.097367656 0.213000554
0.126775818 0.602899873 -0.313740224 0.284427278 0.375458604
0.032455885 0.065198992 0.347727958 0.335478328 0.082942818
0.180310473 0.104580075 0.266859501 0.284890548 0.260795968
0.05795466 0.245222244 0.563187905 0.414714089 0.470881609
0.407950213 -0.036062095 0.166068229

Juice, crowberry -0.003606209 -0.097367656 0.213000554
0.126775818 0.602899873 -0.313740224 0.284427278 0.375458604
0.032455885 0.065198992 0.347727958 0.335478328 0.082942818
0.180310473 0.104580075 0.266859501 0.284890548 0.260795968

0.05795466 0.245222244 0.563187905 0.414714089 0.470881609
0.407950213 -0.036062095 0.166068229

Juice, grape, canned or bottled, unsweetened, without added ascorbic acid -
0.274081228 -0.283869844 0.58551354 0.321995641 0.56591476 -
0.435593381 0.261059452 0.580613845 0.330365766 0.476848507 0.587457696
0.64831826 0.119910537 0.425804766 0.442934842 0.51145515
0.369520228 0.201554523 0.403109047 0.215349537 0.445872235
0.261845459 0.435062813 0.267033372 0.271634075 0.6712582

Juice, grape, red -0.003606209 -0.097367656 0.213000554
0.126775818 0.602899873 -0.313740224 0.284427278 0.375458604
0.032455885 0.065198992 0.347727958 0.335478328 0.082942818
0.180310473 0.104580075 0.266859501 0.284890548 0.260795968
0.05795466 0.245222244 0.563187905 0.414714089 0.470881609
0.407950213 -0.036062095 0.166068229

Juice, grapefruit, white, raw 0.463607521 -0.227961709 -0.176928893
0.033445114 0.382063842 0.37139829 -0.020719896 -0.212827509
-0.304802735 -0.239261199 -0.187807177 0.010359948 -
0.27918906 -0.069156923 -0.315048205 0.099893333 0.212593504
0.239261199 -0.38847786 0.555816752 0.374371281 0.484098461
0.084899135 0.253854499 -0.28943453 -0.26154706

Juice, orange, frozen concentrate, unsweetened, diluted with 3 volume water
0.292102967 0.028849676 -0.267153237 0.115909319 0.205780196
0.400289251 0.003646504 -0.299644847 -0.090155237 -
0.112287153 -0.19197481 -0.010939511 -0.414714089 0.140642169 -
0.259647082 0.057699352 0.119004912 -0.137642317 -0.246307303
0.562568677 0.285204132 0.169491845 -0.036221662 0.061373041 -
0.007212419 -0.137186797

Juice, orange, raw 0.417502906 -0.227961709 -0.166672146
0.033445114 0.433347579 0.315048205 0.04402978 -0.151287025 -
0.233084444 -0.213534188 -0.131207754 0.023309883 -0.304802735
0.048665983 -0.294557265 0.145997949 0.243329914 0.244406601 -
0.342169241 0.581430427 0.412834084 0.458484786 0.156934765 0.233341004 -
0.192102564 -0.192314014

Juice, sour orange 0.510712916 -0.198780476 -0.296968152 -
0.046075394 0.015307637 0.584108785 -0.222647878 -0.456167574 -
0.382270147 -0.3040976 -0.426965317 -0.197909225 -0.302757956
-0.2599437 -0.394502792 -0.076454029 0.033639773

0.12593941 -0.463825632 0.394502792 0.030615273 0.290525312 -0.218090198 0.04592291 -0.351688535 -0.410244664

Juice, tangelo 0.46789866 -0.211013121 -0.33370648 -0.095222481 0.05816902 0.565759818 -0.182447566 -0.43779841 -0.321106923 -0.3040976 -0.37781823 -0.194816893 -0.357804858 -0.113151964 -0.400619114 -0.039756095 0.088686674 0.113652638 -0.426965317 0.455666015 0.039799855 0.211013121 -0.162799725 -0.033676801 -0.211013121 -0.34595259

Juice, tangerine, frozen concentrate, sweetened, diluted with 3 volume water 0.284408989 -0.027523451 -0.223491496 0.015358465 0.425552301 0.308874279 -0.052569638 -0.254106769 -0.247711055 -0.156656339 -0.009215079 0.231924873 -0.247711055 0.033639773 -0.217129443 0.070337707 0.308874279 -0.070648937 -0.242663741 0.639155686 0.425552301 0.247711055 0.058362166 0.186753168 -0.100919319 -0.143891785

Juice, tangor (e.g., murcot or temple) 0.46789866 -0.211013121 -0.33370648 -0.095222481 0.05816902 0.565759818 -0.182447566 -0.43779841 -0.321106923 -0.3040976 -0.37781823 -0.194816893 -0.357804858 -0.113151964 -0.400619114 -0.039756095 0.088686674 0.113652638 -0.426965317 0.455666015 0.039799855 0.211013121 -0.162799725 -0.033676801 -0.211013121 -0.34595259

Juice, tangor, diluted from frozen concentrate (ex. Murcot or temple) 0.292102967 0.028849676 -0.267153237 0.115909319 0.205780196 0.400289251 0.003646504 -0.299644847 -0.090155237 -0.112287153 -0.19197481 -0.010939511 -0.414714089 0.140642169 -0.259647082 0.057699352 0.119004912 -0.137642317 -0.246307303 0.562568677 0.285204132 0.169491845 -0.036221662 0.061373041 -0.007212419 -0.137186797

Juice, tomato, canned, without salt added -0.198780476 -0.229362088 0.254106769 0.236520356 0.658228379 -0.510712916 0.504050057 0.315337316 0.290525312 0.218090198 0.568263192 0.460757414 0.162082542 0.241594733 0.327223246 0.345572213 0.290525312 0.205803426 0.334814529 0.180431509 0.590874777 0.339455891 0.605123507 0.456167574 -0.039756095 0.217368441

Juniper berries, green, unripe (Juniperus communis) -0.394502792 -0.455666015 0.584751722 0.101365867 0.229614551 -0.382270147 0.324694822 0.603120886 0.504596594 0.322527758 0.334814529

0.201001556	0.308874279	0.584108785	0.522945561	0.602457752
0.382270147	0.316384372	0.488399176	0.388386469	0.33370648
0.327223246	0.433108703	0.13776873	0.180431509	0.670474488

Juniper berries, ripe (Juniperus communis)	-0.394502792	-0.455666015
0.584751722	0.101365867	0.229614551	-0.382270147	0.324694822
0.603120886	0.504596594	0.322527758	0.334814529	0.201001556
0.308874279	0.584108785	0.522945561	0.602457752	0.382270147
0.316384372	0.488399176	0.388386469	0.33370648	0.327223246
0.433108703	0.13776873	0.180431509	0.670474488

Kale, Chinese, raw	-0.223722278	-0.263563232	0.371235513
0.101582229	0.549183115	-0.300339497	0.443147296	0.558387301
0.300339497	0.323216185	0.523302394	0.50822487	0.125652238
0.309533563	0.352439205	0.499544265	0.502608954	0.172381965
0.273964195	0.432121113	0.638156915	0.548579285	0.560241387
0.3804397	0.07355253	0.371235513

Kale, canned	-0.266859501	-0.212766359	0.245492164	0.347727958
0.447662181	-0.508475535	0.608966095	0.277983774	0.508475535
0.307884129	0.470881609	0.269841264	0.06491177	0.33898369
0.367833366	0.346196109	0.104580075	0.137642317	0.423793448
0.043274514	0.440441823	0.230797406	0.550569266	0.346577173
0.043274514	0.245492164

Kale, raw (Brassica oleracea (Acephala Group))	-0.088686674	-
0.168198865	0.13776873	0.310240986	0.266352879	-0.370037502
0.578266016	0.15001484	0.198780476	0.500685947	0.476112404
0.392726118	0.2599437	0.253827378	0.431200726	0.094802996	-
0.058105062	0.340957915	0.396248388	0.05198874	0.254106769	-0.058105062
0.396248388	0.523521176	-0.033639773	0.094907348

Kiwifruit, green, raw (Actinidia deliciosa)	-0.464879841	-0.048678517
0.556764227	0.366704326	0.244878796	-0.554935098	0.145206164
0.288737684	0.547633321	0.300697547	0.325144502	0.10459766
0.014603555	0.423503101	0.357787103	0.357787103	0.021905333
0.124679471	0.427821714	-0.340749622	0.099900802	-0.133865923
0.256693028	-0.037767376	0.340749622	0.577475369

Kohlrabi, raw (Brassica oleracea (Gongylodes Group))	-0.388386469	-
0.425084403	0.43779841	0.224233584	0.352075644	-0.584108785
0.405095444	0.321460371	0.498480272	0.242663741	0.433108703
0.219555546	0.345572213	0.46789866	0.516829239	0.437317048

0.198780476 0.297954214 0.568263192 0.137617253 0.296968152
0.113151964 0.48225579 0.223491496 0.05198874 0.480659793

Kumquats, raw (Fortunella spp.) -0.014424838 -0.021637257
0.184119123 0.329617127 0.104695188 0.219978778 0.185971682
0.068593399 0.169491845 0.333239293 0.068821158 0.342771335 -
0.06491177 -0.079336608 0.223584987 0.277678129 0.223584987 -
0.260795968 0.068821158 0.501263116 0.462102897 0.562568677 0.02897733
0.375458604 -0.234403615 0.068593399

Leeks, (bulb and lower leaf-portion), raw (Allium ampeloprasum) -0.217129443
-0.253827378 0.223491496 0.181229883 0.652105324 -0.498480272
0.448388087 0.235737605 0.241594733 0.19966004 0.605123507
0.522604046 0.217129443 0.168198865 0.345572213 0.308874279
0.327223246 0.168943111 0.353244687 0.180431509 0.554136449
0.284408989 0.568263192 0.43779841 -0.076454029 0.180630113

Lemons, raw, without peel (Citrus limon) 0.208880482 -0.617253558
0.030544216 -0.134369325 0.392375694 0.105613727 -0.014239174
-0.096331757 -0.119695557 -0.242807728 -0.025930922 -
0.004746391 0.016428802 0.222962312 -0.063368236 0.28867752
0.368474558 0.511546379 -0.073078054 0.584395955 0.234955505
0.30275935 0.228663589 0.086933537 -0.157247104 0.046991101

Lettuce, butterhead (includes boston and bibb types), raw (Lactuca sativa var. capitata) -0.238009825 -0.191129102 0.259932879 0.362216623
0.462102897 -0.501263116 0.616259102 0.328526278 0.501263116
0.300639797 0.44190428 0.240669235 0.036062095 0.367833366
0.33898369 0.360620947 0.097367656 0.166619646 0.387571786
0.057699352 0.469323254 0.27407192 0.565057931 0.361017888
0.057699352 0.259932879

Lettuce, green leaf, raw (Lactuca sativa var. crispa) -0.463607521 -
0.545571282 0.602583912 0.115771548 0.541043427 -0.555816752
0.458427706 0.602583912 0.504589402 0.373041654 0.640602564
0.520587395 0.422625641 0.525080342 0.642903248 0.663394188
0.566062222 0.326733035 0.604584749 0.463607521 0.566685296
0.417502906 0.640602564 0.335908478 0.089647863 0.648739275

Lettuce, iceberg (includes crisphead types), raw (Lactuca sativa var. capitata) -
0.478975726 -0.514834872 0.541043427 0.177516373 0.602583912 -
0.586553162 0.582747084 0.556428548 0.566062222 0.434786479 0.733219802
0.598287006 0.36115282 0.478975726 0.658271453 0.648025983

0.550694017 0.249552003 0.635457162 0.417502906 0.62822578
0.432871111 0.70234739 0.397448963 0.074279658 0.571813669

Limes, raw (Citrus latifolia) 0.247711055 0.009174484 -0.113276512
0.273380671 0.388813972 0.198780476 0.244294199 -0.033676801
0.082570352 -0.009215079 -0.082935709 0.009276995 -
0.431200726 0.327223246 -0.180431509 0.217129443 0.125384608
0.003071693 -0.168943111 0.565759818 0.499028957 0.357804858
0.205803426 0.223491496 0.045872418 0.003061527

Lingonberries (cowberries), raw -0.238009825 -0.191129102
0.259932879 0.362216623 0.462102897 -0.501263116 0.616259102
0.328526278 0.501263116 0.300639797 0.44190428 0.240669235
0.036062095 0.367833366 0.33898369 0.360620947 0.097367656
0.166619646 0.387571786 0.057699352 0.469323254 0.27407192
0.565057931 0.361017888 0.057699352 0.259932879

Locust bean powder -0.238009825 -0.191129102 0.259932879
0.362216623 0.462102897 -0.501263116 0.616259102 0.328526278
0.501263116 0.300639797 0.44190428 0.240669235 0.036062095
0.367833366 0.33898369 0.360620947 0.097367656 0.166619646
0.387571786 0.057699352 0.469323254 0.27407192 0.565057931
0.361017888 0.057699352 0.259932879

Lotus root, raw (Nelumbo nucifera) -0.313834117 -0.410186697
0.383078585 0.129960265 0.614579313 -0.586374271 0.356312424
0.305911676 0.278046016 0.176967169 0.597264196 0.487145892
0.404680835 0.300069463 0.467998245 0.404680835 0.390916181
0.30139721 0.475599267 0.250516707 0.490561066 0.231246191
0.564082852 0.377566663 -0.057811548 0.372054741

Lovage, leaves, raw -0.238009825 -0.191129102 0.259932879
0.362216623 0.462102897 -0.501263116 0.616259102 0.328526278
0.501263116 0.300639797 0.44190428 0.240669235 0.036062095
0.367833366 0.33898369 0.360620947 0.097367656 0.166619646
0.387571786 0.057699352 0.469323254 0.27407192 0.565057931
0.361017888 0.057699352 0.259932879

Mangos, raw (Mangifera indica) -0.45423359 -0.547833239 0.418906079 -
0.082953361 0.446465689 -0.437716005 0.20599312 0.259060338
0.267034293 0.254390306 0.666391997 0.645816268 0.545080308
0.211975675 0.608397718 0.448727729 0.605644787 0.165906721
0.5751433 0.399174973 0.366542819 0.178940505 0.428592363
0.228744767 -0.044046894 0.438197806

Medlar 0.082570352 0.590225107 -0.355137172 -0.168943111 -
0.551074922 0.082570352 -0.649389643 -0.404121609 -0.094802996
-0.218090198 -0.512972719 -0.278309847 -0.633039363 -
0.327223246 -0.602457752 -0.553527173 -0.700318909 -
0.562119806 -0.562119806 -0.541294528 -0.462290629 -
0.626923041 -0.709561067 -0.649043797 0.003058161 -0.192876223

Melons, cantaloupe, raw (Cucumis melo) -0.382270147 -0.437317048
0.462290629 0.19966004 0.327583426 -0.57187614 0.355618138
0.339829535 0.474014982 0.218090198 0.396248388 0.18863223
0.370037502 0.486247627 0.510712916 0.443433371 0.204896799
0.328671143 0.55597642 0.15596622 0.272475933 0.107035641
0.457682246 0.198999277 0.058105062 0.511275066

Mizuna (Japanese mustard) -0.088686674 -0.15596622 0.180630113
0.347101301 0.327583426 -0.400619114 0.615373995 0.24186066
0.253827378 0.476112404 0.457682246 0.349433475 0.198780476
0.314990601 0.400619114 0.162082542 -0.039756095 0.353244687
0.371674844 0.070337707 0.33370648 0.045872418 0.457682246
0.52964423 -0.003058161 0.143891785

Mung beans, mature seeds, sprouted, raw (Vigna radiata) -0.266859501 -
0.212766359 0.245492164 0.347727958 0.447662181 -0.508475535
0.608966095 0.277983774 0.508475535 0.307884129 0.470881609
0.269841264 0.06491177 0.33898369 0.367833366 0.346196109
0.104580075 0.137642317 0.423793448 0.043274514 0.440441823
0.230797406 0.550569266 0.346577173 0.043274514 0.245492164

Mustard greens, raw (Brassica juncea) -0.088686674 -0.168198865
0.13776873 0.310240986 0.266352879 -0.370037502 0.578266016
0.15001484 0.198780476 0.500685947 0.476112404 0.392726118
0.2599437 0.253827378 0.431200726 0.094802996 -0.058105062
0.340957915 0.396248388 0.05198874 0.254106769 -0.058105062
0.396248388 0.523521176 -0.033639773 0.094907348

Nalta jute, raw -0.198780476 -0.229362088 0.254106769
0.236520356 0.658228379 -0.510712916 0.504050057 0.315337316
0.290525312 0.218090198 0.568263192 0.460757414 0.162082542
0.241594733 0.327223246 0.345572213 0.290525312 0.205803426
0.334814529 0.180431509 0.590874777 0.339455891 0.605123507
0.456167574 -0.039756095 0.217368441

Nectarines, without skin, raw (Prunus persica var. nucipersica) -0.238009825
-0.191129102 0.259932879 0.362216623 0.462102897 -0.501263116

0.616259102	0.328526278	0.501263116	0.300639797	0.44190428
0.240669235	0.036062095	0.367833366	0.33898369	0.360620947
0.097367656	0.166619646	0.387571786	0.057699352	0.469323254
0.27407192	0.565057931	0.361017888	0.057699352	0.259932879

New Zealand spinach, raw (Tetragonia tetragonioides)	-0.266859501	-
0.212766359	0.245492164	0.347727958	0.447662181	-0.508475535
0.608966095	0.277983774	0.508475535	0.307884129	0.470881609
0.269841264	0.06491177	0.33898369	0.367833366	0.346196109
0.104580075	0.137642317	0.423793448	0.043274514	0.440441823
0.230797406	0.550569266	0.346577173	0.043274514	0.245492164

Nuts, almonds (Prunus dulcis)	0.645417219	0.180716821	-0.169167964
0.285240147	-0.216159065	0.429495822	-0.132898957	-
0.328937708	-0.406026105	0.266381294	-0.195660597	0.14239174	-
0.260513859	-0.335616954	-0.152553161	-0.387250331	-
0.570314124	0.00707207	-0.304099	-0.232350199	-0.249052836	-
0.218268369	-0.37953441	0.216159065	-0.312147237	-0.305442157

Oil, olive, salad or cooking	-0.447169974	-0.508475535	0.509035222	-
0.105042821	-0.003610179	-0.212766359	0.076576575	0.393509498
0.385864413	0.278906799	0.286151132	0.233376228	0.421926508
0.40750167	0.548143839	0.494050697	0.439957555	0.173863979
0.503481105	0.393076832	0.083034114	0.129823541	0.19197481	-
0.057762862	0.144248379	0.610120231

Onion, spring, red, leaves	-0.238009825	-0.191129102	0.259932879
0.362216623	0.462102897	-0.501263116	0.616259102	0.328526278
0.501263116	0.300639797	0.44190428	0.240669235	0.036062095
0.367833366	0.33898369	0.360620947	0.097367656	0.166619646
0.387571786	0.057699352	0.469323254	0.27407192	0.565057931
0.361017888	0.057699352	0.259932879

Onions, cooked, boiled, drained, without salt	-0.238009825	-0.191129102
0.259932879	0.362216623	0.462102897	-0.501263116	0.616259102
0.328526278	0.501263116	0.300639797	0.44190428	0.240669235
0.036062095	0.367833366	0.33898369	0.360620947	0.097367656
0.166619646	0.387571786	0.057699352	0.469323254	0.27407192
0.565057931	0.361017888	0.057699352	0.259932879

Onions, raw (Allium cepa)	-0.240958661	-0.423503101	0.416659443
0.19068625	0.482447776	-0.498954803	0.553752319	0.423969258
0.221487254	0.55494588	0.694293524	0.652197175	0.525727988
0.40646562	0.652292133	0.365088881	0.316410363	0.508496665

0.55494588 0.369956732 0.467828147 0.146035552 0.574503444
0.609151233 -0.021905333 0.392293394

Onions, red, raw -0.226355106 -0.408899546 0.406913023
0.171128685 0.501940616 -0.464879841 0.553752319 0.443462097
0.197147995 0.564724662 0.708961697 0.691575118 0.506256581
0.382126362 0.637688578 0.369956732 0.355353177 0.488939101
0.525609534 0.404031694 0.501940616 0.189846218 0.579392835
0.623770862 -0.021905333 0.377673764

Onions, spring or scallions (includes tops and bulb), raw (Allium cepa or Allium fistulosum) -0.238009825 -0.191129102 0.259932879 0.362216623
0.462102897 -0.501263116 0.616259102 0.328526278 0.501263116
0.300639797 0.44190428 0.240669235 0.036062095 0.367833366
0.33898369 0.360620947 0.097367656 0.166619646 0.387571786
0.057699352 0.469323254 0.27407192 0.565057931 0.361017888
0.057699352 0.259932879

Onions, sweet, raw (Allium cepa) -0.312954283 -0.310389084
0.580373106 0.278267349 0.765270733 -0.551517793 0.503209106
0.472516157 0.392475453 0.226736359 0.662173229 0.36054673
0.312954283 0.407866647 0.523300604 0.582300181 0.495083414
0.394212078 0.525616104 0.048738782 0.577805083 0.40017105
0.754929013 0.446835931 0.105173161 0.444267908

Onions, young green, tops only (Allium cepa) -0.481762899 -0.558844963
0.424418001 -0.05530224 0.429929923 -0.459739452 0.244964791
0.253548416 0.32209291 0.282041426 0.677452445 0.634681505
0.539574447 0.233999122 0.635927026 0.459739452 0.58362134
0.154846273 0.613854868 0.377151527 0.355518975 0.16242292
0.439652811 0.223232845 -0.03303517 0.44922165

Oranges, raw, all commercial varieties (Citrus sinensis) 0.04137674 -0.279901475 0.119393642 0.212688509 0.614024443 -0.087621331
0.24365102 0.058478518 0.131431997 0.017112869 0.202909727
0.270723356 -0.058414221 0.423503101 0.107092738 0.391862065
0.37725851 0.178462772 0.112455993 0.703404577 0.616461048
0.362654955 0.38626189 0.326505061 -0.021905333 0.231477468

Oranges, raw, navels (Citrus sinensis) 0.014623276 -0.309526018
0.131754353 0.195839754 0.578255214 -0.092614084 0.216871161
0.036598431 0.141358339 0.014687982 0.198287751 0.261231171 -0.021934915 0.428949442 0.134046701 0.392391251 0.377767975 0.178703775

0.144431818 0.699480056 0.573375423 0.321712082 0.362303544
0.295227346 -0.021934915 0.251309228

Oregano, fresh -0.338610495 -0.429457213 0.534656443
0.05530224 0.496072988 -0.390916181 0.364663496 0.625603157
0.363386872 0.317987882 0.522606172 0.473227438 0.313834117
0.448727729 0.492774622 0.602891856 0.564350824 0.293101874
0.417531915 0.498280484 0.564972014 0.487268761 0.555787516
0.311423598 0.099105511 0.570483936

Pako fern, steamed (Athyrium esculentum) -0.238009825 -0.191129102
0.259932879 0.362216623 0.462102897 -0.501263116 0.616259102
0.328526278 0.501263116 0.300639797 0.44190428 0.240669235
0.036062095 0.367833366 0.33898369 0.360620947 0.097367656
0.166619646 0.387571786 0.057699352 0.469323254 0.27407192
0.565057931 0.361017888 0.057699352 0.259932879

Papayas, raw (Carica papaya) -0.464095481 -0.574594405
0.464606318 -0.11098786 0.398233987 -0.419895911 0.156427022
0.29867549 0.265197418 0.244173291 0.621532014 0.603361371
0.574594405 0.254147525 0.618793975 0.475145373 0.618793975
0.199778147 0.57713687 0.430945804 0.331861656 0.176798278
0.399556295 0.188054938 -0.022099785 0.497792484

Parsley, fresh (Petroselinum crispum) -0.514834872 -0.576307692
0.541043427 0.033445114 0.541043427 -0.504589402 0.43770781
0.500016437 0.473852991 0.414204871 0.753801411 0.686346566
0.463607521 0.391889231 0.694130598 0.632657778 0.668516923
0.203243384 0.645747966 0.494343931 0.556428548 0.386766496
0.609730151 0.325651731 0.048665983 0.576942043

Peas, green, frozen, cooked, boiled, drained, without salt -0.468730256 -
0.560939487 0.582070417 0.115771548 0.541043427 -0.601921367
0.422167888 0.520529932 0.478975726 0.326733035 0.640602564
0.505047473 0.473852991 0.499466667 0.648025983 0.617289572
0.525080342 0.342169241 0.63031176 0.407257436 0.515401559
0.33041641 0.620020956 0.325651731 0.053788718 0.623097407

Peppers, hot chili, green, raw (Capsicum frutescens) -0.349622218 -
0.456986521 0.567727975 0.05530224 0.451977611 -0.418445489
0.320124443 0.620091235 0.368892734 0.284806538 0.472834155
0.400851477 0.357881011 0.492774622 0.503786346 0.602891856
0.525809792 0.34287389 0.439652811 0.476257037 0.504340871
0.432210144 0.528136396 0.272840143 0.104611372 0.614579313

Peppers, sweet, yellow, raw (Capsicum annuum) -0.286304808 -
0.410186697 0.454733572 0.135490489 0.575995858 -0.575362548
0.317340752 0.410638196 0.278046016 0.149316049 0.497720164
0.375798259 0.399174973 0.399174973 0.440468936 0.448727729
0.368892734 0.389880795 0.436887699 0.283551878 0.479537222
0.269787223 0.553022404 0.350007053 -0.013764654 0.460245495

Perilla leaves, raw -0.294563601 -0.418445489 0.518120676
0.038711568 0.523632599 -0.412939628 0.292287535 0.581507781
0.275293085 0.235034522 0.489424828 0.445390529 0.352375149
0.42119842 0.448727729 0.553339101 0.536821516 0.34287389
0.378820347 0.470751175 0.542924326 0.443221867 0.53366662
0.31693552 0.060564479 0.542924326

Pitanga, (surinam-cherry), raw (Eugenia uniflora) -0.168198865 -
0.217129443 0.247983715 0.205803426 0.676597543 -0.486247627
0.460757414 0.321460371 0.229362088 0.187373269 0.55597642
0.47312674 0.168198865 0.217129443 0.296641634 0.333339568
0.314990601 0.218090198 0.291810828 0.204896799 0.603120886
0.357804858 0.592836735 0.462290629 -0.05198874 0.205122332

Plum, red, whole, raw -0.462884484 -0.530496825 0.570130806
0.143658967 0.596164176 -0.582506317 0.512758567 0.57533748
0.514893977 0.3891852 0.694787002 0.570608252 0.403073568
0.49409018 0.644917708 0.650118657 0.564302995 0.300377839
0.613815585 0.442080687 0.609180861 0.426477839 0.679115115
0.380087204 0.070212815 0.603974187

Plum, yellow, whole, raw (Prunus domestica) -0.183881325 -0.223722278
0.251581092 0.221633955 0.668837536 -0.499544265 0.483433413
0.319078458 0.260498543 0.203164459 0.563319636 0.467938753
0.165493192 0.229851656 0.312598252 0.34018045 0.303404185
0.212399207 0.313981437 0.193075391 0.598272108 0.349374517
0.600258629 0.460209314 -0.045970331 0.211696284

Plums, dried (prunes), uncooked -0.293844219 -0.401956714
0.435701153 0.158709736 0.602211148 -0.590460552 0.364401646
0.396848821 0.302160564 0.172631642 0.534601215 0.40644799
0.379779792 0.385324023 0.449082674 0.446310558 0.365919216
0.364753954 0.451069775 0.268895181 0.505080318 0.277211527
0.579151316 0.374647488 -0.019404807 0.435701153

Potato, flesh and skin, raw (Solanum tuberosum) -0.266859501 -
0.212766359 0.245492164 0.347727958 0.447662181 -0.508475535

0.608966095 0.277983774 0.508475535 0.307884129 0.470881609
0.269841264 0.06491177 0.33898369 0.367833366 0.346196109
0.104580075 0.137642317 0.423793448 0.043274514 0.440441823
0.230797406 0.550569266 0.346577173 0.043274514 0.245492164

Prickly pears, raw (Opuntia spp.) -0.039756095 -0.131500931
0.162260949 0.334814529 0.290845098 -0.357804858 0.590635342
0.235737605 0.186547832 0.488399176 0.426965317 0.343248811
0.211013121 0.302757956 0.382270147 0.119268286 -0.070337707
0.390105002 0.334814529 0.076454029 0.303091207 0.015290806
0.420821931 0.548013394 -0.009174484 0.119399566

Purslane, raw (Portulaca oleracea) -0.015290806 -0.125384608
0.131645676 0.310240986 0.24186066 -0.321106923 0.559712026
0.186753168 0.125384608 0.506829333 0.420821931 0.361802801
0.247711055 0.266060022 0.388386469 0.064221385 -0.094802996
0.402391774 0.328671143 0.070337707 0.247983715 -0.05198874
0.371674844 0.554136449 -0.027523451 0.082661238

Queen Anne's Lace, leaves, raw -0.504589402 -0.591675897
0.617969033 0.059172124 0.484631316 -0.530203077 0.401447991
0.576942043 0.499466667 0.378187056 0.645747966 0.541307291
0.478975726 0.509712137 0.683885128 0.668516923 0.607044102
0.306151426 0.640602564 0.489221196 0.510273185 0.376521026
0.594293945 0.284624741 0.089647863 0.674381144

Radish leaves, raw -0.238009825 -0.191129102 0.259932879
0.362216623 0.462102897 -0.501263116 0.616259102 0.328526278
0.501263116 0.300639797 0.44190428 0.240669235 0.036062095
0.367833366 0.33898369 0.360620947 0.097367656 0.166619646
0.387571786 0.057699352 0.469323254 0.27407192 0.565057931
0.361017888 0.057699352 0.259932879

Raisins, golden seedless (Vitis vinifera) -0.266859501 -0.212766359
0.245492164 0.347727958 0.447662181 -0.508475535 0.608966095
0.277983774 0.508475535 0.307884129 0.470881609 0.269841264
0.06491177 0.33898369 0.367833366 0.346196109 0.104580075
0.137642317 0.423793448 0.043274514 0.440441823 0.230797406
0.550569266 0.346577173 0.043274514 0.245492164

Raisins, seedless (Vitis vinifera) -0.152031058 -0.162338248
0.497869344 0.390818109 0.487550808 -0.430325197 0.143307423
0.441117398 0.22418139 0.349406919 0.411523705 0.455978165
0.048959154 0.38394284 0.291178127 0.347867674 0.131416677

0.230349747 0.292466532 0.012883988 0.286339364 0.08503432
0.307995729 0.203791079 0.260256556 0.580417629

Raspberries, red, frozen -0.410431846 -0.507909409 0.549556835
0.10821508 0.616325422 -0.564343788 0.43576871 0.552124857
0.415562244 0.319492142 0.65959668 0.560274056 0.420692642
0.45404023 0.589995779 0.607952172 0.564343788 0.32979834
0.5513816 0.443779433 0.600917287 0.415562244 0.649290482
0.382635366 0.041043185 0.572669038

Rhubarb stalks, cooked -0.234403615 0.562568677 -0.207585286 -
0.039843828 -0.395314587 -0.234403615 -0.510510499 -
0.254517611 0.06491177 -0.086931989 -0.369460955 -0.154976401
-0.494050697 -0.165885636 -0.468807231 -0.421926508 -
0.562568677 -0.416549116 -0.416549116 -0.393076832 -
0.238271806 -0.562568677 -0.557813599 -0.489179238
0.115398703 -0.03249161

Rhubarb, raw (Rheum rhabarbarum) -0.234403615 0.562568677 -
0.207585286 -0.039843828 -0.395314587 -0.234403615 -
0.510510499 -0.254517611 0.06491177 -0.086931989 -0.369460955
-0.154976401 -0.494050697 -0.165885636 -0.468807231 -
0.421926508 -0.562568677 -0.416549116 -0.416549116 -
0.393076832 -0.238271806 -0.562568677 -0.557813599 -
0.489179238 0.115398703 -0.03249161

Rocket, wild, raw (Diplotaxis tenuifolia) -0.088686674 -0.15596622
0.180630113 0.347101301 0.327583426 -0.400619114 0.615373995
0.24186066 0.253827378 0.476112404 0.457682246 0.349433475
0.198780476 0.314990601 0.400619114 0.162082542 -0.039756095
0.353244687 0.371674844 0.070337707 0.33370648 0.045872418
0.457682246 0.52964423 -0.003058161 0.143891785

Rutabagas, raw (Brassica napus var. napobrassica) -0.412939628 -
0.412939628 0.391346468 0.099544033 0.584263742 -0.418445489
0.548387089 0.457489533 0.45148066 0.445183035 0.754875581
0.709841156 0.2587755 0.280798947 0.567103755 0.55058617
0.600138925 0.080188249 0.517075948 0.429457213 0.636627001
0.462492383 0.627680429 0.394102429 0.041293963 0.39685839

Sage, fresh -0.447169974 -0.508475535 0.509035222 -0.105042821
-0.003610179 -0.212766359 0.076576575 0.393509498 0.385864413
0.278906799 0.286151132 0.233376228 0.421926508 0.40750167
0.548143839 0.494050697 0.439957555 0.173863979 0.503481105

0.393076832 0.083034114 0.129823541 0.19197481 -0.057762862
0.144248379 0.610120231

Sauce, pasta, spaghetti/marinara, ready-to-serve -0.238009825 -
0.191129102 0.259932879 0.362216623 0.462102897 -0.501263116
0.616259102 0.328526278 0.501263116 0.300639797 0.44190428
0.240669235 0.036062095 0.367833366 0.33898369 0.360620947
0.097367656 0.166619646 0.387571786 0.057699352 0.469323254
0.27407192 0.565057931 0.361017888 0.057699352 0.259932879

Sauerkraut, canned, solids and liquids -0.438472645 -0.416411129
0.389262229 0.163423617 0.582512981 -0.4770803 0.616259102
0.436194554 0.526718712 0.468111718 0.767260034 0.677606162
0.245434374 0.314376614 0.592903263 0.551537919 0.537749471
0.080326863 0.567827823 0.369530406 0.626684581 0.427441887
0.656464361 0.403065854 0.046880723 0.397544404

Sorghum, grain, white -0.461594812 -0.501263116 0.501814864 -
0.097798488 0.003610179 -0.191129102 0.127627625 0.415170571
0.421926508 0.329617127 0.322372794 0.277134271 0.385864413
0.400289251 0.562568677 0.515687954 0.468807231 0.130397984
0.510725438 0.414714089 0.119135903 0.173098055 0.213707807 -
0.043322147 0.158673217 0.602899873

Soybeans, green, raw (Glycine max) -0.266859501 -0.212766359
0.245492164 0.347727958 0.447662181 -0.508475535 0.608966095
0.277983774 0.508475535 0.307884129 0.470881609 0.269841264
0.06491177 0.33898369 0.367833366 0.346196109 0.104580075
0.137642317 0.423793448 0.043274514 0.440441823 0.230797406
0.550569266 0.346577173 0.043274514 0.245492164

Spices, celery seed (Apium graveolens) -0.447169974 -0.508475535
0.509035222 -0.105042821 -0.003610179 -0.212766359
0.076576575 0.393509498 0.385864413 0.278906799 0.286151132
0.233376228 0.421926508 0.40750167 0.548143839 0.494050697
0.439957555 0.173863979 0.503481105 0.393076832 0.083034114
0.129823541 0.19197481 -0.057762862 0.144248379 0.610120231

Spices, parsley, dried (Petroselinum crispum) -0.247711055 -0.400619114
0.358198699 -0.058362166 -0.101030402 -0.088686674
0.194816893 0.296968152 0.137617253 0.537546263 0.353244687
0.405095444 0.504596594 0.314990601 0.590225107 0.253827378
0.2599437 0.328671143 0.457682246 0.382270147 0.009184582 -
0.070337707 0.12593941 0.205122332 0.076454029 0.401060082

Spinach, raw (Spinacia oleracea) -0.341363425 -0.426704282
0.427173962 0.174202057 0.570483936 -0.61390358 0.395284095
0.350007053 0.360633941 0.204618289 0.564082852 0.414769931
0.393669112 0.382657388 0.495527553 0.443221867 0.346869287
0.32904833 0.514310836 0.233999122 0.468513378 0.22574033
0.5751433 0.355518975 -0.019270516 0.432685884

Strawberries, frozen, unsweetened -0.458484786 -0.519957607
0.515401559 0.151789363 0.617969033 -0.612166837 0.510227447
0.479502942 0.489221196 0.36275085 0.722928998 0.593107032
0.417502906 0.432871111 0.637780513 0.591675897 0.530203077
0.275279013 0.625166358 0.381643761 0.587198791 0.36115282
0.671474977 0.392320589 0.028175043 0.535915053

Sweet potato leaves, cooked, steamed, without salt -0.057811548 -
0.19545809 0.192917273 0.212913626 0.575995858 -0.423951351
0.489929582 0.261816299 0.055058617 0.348404115 0.583438636
0.565089234 0.300069463 0.19270516 0.363386872 0.184446367
0.198211021 0.376055235 0.298632098 0.211975675 0.504340871
0.176187574 0.525371284 0.611823352 -0.09635258 0.121262286

Sweet potato leaves, raw (Ipomoea batatas) -0.412380171 -0.494343931
0.520529932 0.136353156 0.643610901 -0.586553162 0.479147602
0.520529932 0.432871111 0.331878437 0.692056585 0.582747084
0.402134701 0.432871111 0.591675897 0.591675897 0.545571282
0.306151426 0.563421532 0.412380171 0.617969033 0.407257436
0.671474977 0.40770571 0.028175043 0.535915053

Sweet potato, raw, unprepared (Ipomoea batatas) -0.445785982 -
0.559165076 0.544302754 0.023293795 0.575258361 -0.538550695
0.341332226 0.482391541 0.363328459 0.297642931 0.680696442
0.612313536 0.507629124 0.394250031 0.626161813 0.584933052
0.621008218 0.30281933 0.582344865 0.466400363 0.533984219
0.347867674 0.587521264 0.332772774 0.012883988 0.570099093

Taro leaves, cooked, steamed, without salt 0.151141731 -0.064775028
0.108077211 0.108436074 0.410693401 -0.194325083 0.283828105
0.237769864 -0.237508434 0.281933791 0.36868265 0.458491553
0.280691786 0.107958379 0.194325083 0.021591676 0.107958379
0.455431509 0.065061644 0.237508434 0.367462517 0.107958379
0.325308221 0.583616938 -0.107958379 0.021615442

Taro leaves, raw (Colocasia esculenta) -0.445785982 -0.559165076
0.544302754 0.023293795 0.575258361 -0.538550695 0.341332226

0.482391541 0.363328459 0.297642931 0.680696442 0.612313536
0.507629124 0.394250031 0.626161813 0.584933052 0.621008218
0.30281933 0.582344865 0.466400363 0.533984219 0.347867674
0.587521264 0.332772774 0.012883988 0.570099093

Taro, cooked, without salt -0.266859501 -0.212766359 0.245492164
0.347727958 0.447662181 -0.508475535 0.608966095 0.277983774
0.508475535 0.307884129 0.470881609 0.269841264 0.06491177
0.33898369 0.367833366 0.346196109 0.104580075 0.137642317
0.423793448 0.043274514 0.440441823 0.230797406 0.550569266
0.346577173 0.043274514 0.245492164

Tea, black, brewed, prepared with tap water -0.132677613 0.592474169 -
0.036641008 0.082716011 -0.070991953 -0.22646696 -0.017348288
0.052671449 0.251629956 -0.202194695 -0.101097347 -
0.373566459 -0.354569483 -0.098364437 -0.237904686 -
0.146402883 -0.336269123 -0.413580057 -0.119478683 -
0.759464958 -0.083587299 -0.139540248 -0.009190668 -
0.277097622 0.166990789 -0.07671711

Tea, fruit flavored, brewed -0.052305686 0.575362548 -0.035827494 -
0.033181344 -0.272840143 0.041293963 -0.377190105 -0.212209
0.107364303 -0.163141609 -0.284806538 -0.389716713 -
0.536821516 -0.176187574 -0.360633941 -0.231246191 -
0.385410319 -0.367759899 -0.320752994 -0.740538399 -
0.293509851 -0.371645665 -0.334578554 -0.485049144
0.16242292 -0.089568734

Tea, green, brewed, decaffeinated -0.103062294 0.547375294 -
0.002292794 0.062110994 -0.009171177 -0.194673222 -0.091476614
-0.009171177 0.210705134 -0.163328911 -0.050608958 -
0.262850522 -0.398507536 -0.121384479 -0.235898139 -
0.132835845 -0.295445242 -0.381867595 -0.149526468 -
0.737467969 -0.065344633 -0.174060763 -0.032205701 -
0.279720884 0.139706665 -0.05846625

Tea, green, brewed, flavored -0.091501802 0.551298358 -0.004580126
0.050548674 -0.002290063 -0.185291149 -0.107559383 -
0.006870189 0.187578694 -0.174622691 -0.055144008 -0.257911209
-0.395745294 -0.130390068 -0.247054866 -0.137252703 -
0.285943132 -0.376817385 -0.165432023 -0.727439327 -
0.060686669 -0.166990789 -0.036762672 -0.277097622
0.134965158 -0.062976732

Tea, green, large leaf, Quingmao, brewed -0.079834995 0.602891856 -0.10472652 -0.082953361 -0.330715325 0.002752931 -0.424512848 -0.26457226 0.101858441 -0.201853177 -0.334578554 -0.403635167 -0.553339101 -0.209222745 -0.415692558 -0.297316532 -0.445974798 -0.412001691 -0.370525011 -0.718514952 -0.304533695 -0.426704282 -0.395411019 -0.52638856 0.118376027 -0.128152189

Thyme, fresh (Thymus vulgaris) -0.447169974 -0.508475535 0.509035222 -0.105042821 -0.003610179 -0.212766359 0.076576575 0.393509498 0.385864413 0.278906799 0.286151132 0.233376228 0.421926508 0.40750167 0.548143839 0.494050697 0.439957555 0.173863979 0.503481105 0.393076832 0.083034114 0.129823541 0.19197481 -0.057762862 0.144248379 0.610120231

Tomato products, canned, puree, without salt added -0.238009825 -0.191129102 0.259932879 0.362216623 0.462102897 -0.501263116 0.616259102 0.328526278 0.501263116 0.300639797 0.44190428 0.240669235 0.036062095 0.367833366 0.33898369 0.360620947 0.097367656 0.166619646 0.387571786 0.057699352 0.469323254 0.27407192 0.565057931 0.361017888 0.057699352 0.259932879

Tomatoes, cherry, raw 0.033639773 0.021407128 0.180630113 0.605123507 0.43779841 -0.174315187 0.460757414 0.088784293 0.272176345 0.261093899 0.175086497 0.262848189 -0.094802996 0.070337707 0.192664154 0.2599437 0.003058161 -0.039932008 0.132082796 0.266060022 0.639859215 0.535178206 0.322527758 0.603120886 -0.272176345 0.033676801

Tomatoes, red, ripe, canned, packed in tomato juice -0.410431846 -0.507909409 0.549556835 0.10821508 0.616325422 -0.564343788 0.43576871 0.552124857 0.415562244 0.319492142 0.65959668 0.560274056 0.420692642 0.45404023 0.589995779 0.607952172 0.564343788 0.32979834 0.5513816 0.443779433 0.600917287 0.415562244 0.649290482 0.382635366 0.041043185 0.572669038

Tomatoes, red, ripe, cooked -0.293844219 -0.401956714 0.435701153 0.158709736 0.602211148 -0.590460552 0.364401646 0.396848821 0.302160564 0.172631642 0.534601215 0.40644799 0.379779792 0.385324023 0.449082674 0.446310558 0.365919216 0.364753954 0.451069775 0.268895181 0.505080318 0.277211527 0.579151316 0.374647488 -0.019404807 0.435701153

Tomatoes, red, ripe, raw, year round average (Lycopersicon esculentum)
0.071576202 -0.016517585 0.18464939 0.478364379 0.636627001 -
0.19545809 0.361879805 0.107482481 0.074329133 0.179732281 0.298632098
0.464876365 0.022023447 -0.027529309 0.176187574 0.256022569
0.19270516 0.019355784 0.077423137 0.382657388 0.763401209
0.605644787 0.378820347 0.691746222 -0.346869287 0.008267883

Tomatoes, yellow, raw (Lycopersicon esculentum) -0.238009825 -
0.191129102 0.259932879 0.362216623 0.462102897 -0.501263116
0.616259102 0.328526278 0.501263116 0.300639797 0.44190428
0.240669235 0.036062095 0.367833366 0.33898369 0.360620947
0.097367656 0.166619646 0.387571786 0.057699352 0.469323254
0.27407192 0.565057931 0.361017888 0.057699352 0.259932879

Tree Spinach, raw (Cnidoscolus aconitifolius) -0.266859501 -0.212766359
0.245492164 0.347727958 0.447662181 -0.508475535 0.608966095
0.277983774 0.508475535 0.307884129 0.470881609 0.269841264
0.06491177 0.33898369 0.367833366 0.346196109 0.104580075
0.137642317 0.423793448 0.043274514 0.440441823 0.230797406
0.550569266 0.346577173 0.043274514 0.245492164

Tree spinach, cooked -0.266859501 -0.212766359 0.245492164
0.347727958 0.447662181 -0.508475535 0.608966095 0.277983774
0.508475535 0.307884129 0.470881609 0.269841264 0.06491177
0.33898369 0.367833366 0.346196109 0.104580075 0.137642317
0.423793448 0.043274514 0.440441823 0.230797406 0.550569266
0.346577173 0.043274514 0.245492164

Turmeric, steamed (Curcuma longa) -0.003606209 -0.097367656
0.213000554 0.126775818 0.602899873 -0.313740224 0.284427278
0.375458604 0.032455885 0.065198992 0.347727958 0.335478328
0.082942818 0.180310473 0.104580075 0.266859501 0.284890548
0.260795968 0.05795466 0.245222244 0.563187905 0.414714089
0.470881609 0.407950213 -0.036062095 0.166068229

Turnip greens, raw (Brassica rapa (Rapifera Group)) -0.266859501 -
0.212766359 0.245492164 0.347727958 0.447662181 -0.508475535
0.608966095 0.277983774 0.508475535 0.307884129 0.470881609
0.269841264 0.06491177 0.33898369 0.367833366 0.346196109
0.104580075 0.137642317 0.423793448 0.043274514 0.440441823
0.230797406 0.550569266 0.346577173 0.043274514 0.245492164

Water spinach -0.474561822 -0.5335614 0.562396948 0.172628819
0.577805083 -0.615647769 0.526553858 0.539284744 0.543822196

0.38390588 0.685362175 0.534335442 0.407866647 0.50534421
0.654125755 0.633604162 0.513039807 0.304032845 0.641560833
0.392475453 0.575237061 0.377084258 0.675055977 0.369795253
0.066695175 0.598349264

Watercress, raw (Nasturtium officinale) -0.437993846 -0.504589402
0.525658306 0.162080167 0.62822578 -0.607044102 0.515407421
0.515401559 0.484098461 0.357605447 0.70234739 0.572387136
0.397011966 0.453362051 0.617289572 0.601921367 0.525080342
0.295860622 0.599439347 0.391889231 0.607712285 0.391889231
0.681765781 0.402577337 0.038420513 0.546171801

Watercress, steamed -0.238009825 -0.191129102 0.259932879
0.362216623 0.462102897 -0.501263116 0.616259102 0.328526278
0.501263116 0.300639797 0.44190428 0.240669235 0.036062095
0.367833366 0.33898369 0.360620947 0.097367656 0.166619646
0.387571786 0.057699352 0.469323254 0.27407192 0.565057931
0.361017888 0.057699352 0.259932879

Yardlong bean, cooked, boiled, drained, without salt -0.437993846 -
0.504589402 0.525658306 0.162080167 0.62822578 -0.607044102
0.515407421 0.515401559 0.484098461 0.357605447 0.70234739
0.572387136 0.397011966 0.453362051 0.617289572 0.601921367
0.525080342 0.295860622 0.599439347 0.391889231 0.607712285
0.391889231 0.681765781 0.402577337 0.038420513 0.546171801

Yuzu, raw 0.292102967 0.028849676 -0.267153237 0.115909319
0.205780196 0.400289251 0.003646504 -0.299644847 -0.090155237
-0.112287153 -0.19197481 -0.010939511 -0.414714089
0.140642169 -0.259647082 0.057699352 0.119004912 -0.137642317
-0.246307303 0.562568677 0.285204132 0.169491845 -0.036221662
0.061373041 -0.007212419 -0.137186797

## Table S2

Food ATC Number of validated paper correlation PMID1 PMID2
PMID3 PMID4 PMID5 PMID6 PMID7 PMID8
PMID9

Acerola, (west indian cherry), raw (Malpighia emarginata) D02 0
0.616259102

Acerola, (west indian cherry), raw (Malpighia emarginata) P03 0
0.565057931

| | | | | |
|---|---|---|---|---|
| Alcoholic beverage, beer, regular, all | J02 | 0 | 0.552123539 | |
| Alcoholic beverage, wine, berry, colored | C08 | 1 | 0.682720597 | 27167647 |
| Alcoholic beverage, wine, berry, colored | H03 | 0 | 0.568263192 | |
| Alcoholic beverage, wine, berry, colored | J02 | 0 | 0.51023472 | |
| Alcoholic beverage, wine, berry, colored | N05 | 0 | 0.590874777 | |
| Alcoholic beverage, wine, table, red, Cabernet Franc | J02 | 0 | 0.55621344 | |
| Alcoholic beverage, wine, table, red, Cabernet Sauvignon | J02 | 0 | 0.55621344 | |
| Amaranth leaves, cooked, boiled, drained, without salt | C08 | 0 | 0.602899873 | |
| Amaranth leaves, cooked, boiled, drained, without salt | N05 | 0 | 0.563187905 | |
| Annual saw-thistle, leaves | C02 | 0 | 0.575815319 | |
| Annual saw-thistle, leaves | D05 | 0 | 0.556296155 | |
| Annual saw-thistle, leaves | H03 | 0 | 0.702575116 | |
| Annual saw-thistle, leaves | J02 | 0 | 0.601324583 | |
| Annual saw-thistle, leaves | L04 | 0 | 0.733601035 | |
| Annual saw-thistle, leaves | M01 | 0 | 0.575182207 | |
| Annual saw-thistle, leaves | M05 | 0 | 0.665855162 | |
| Annual saw-thistle, leaves | P03 | 0 | 0.631583205 | |
| Annual saw-thistle, leaves | S02 | 0 | 0.602654168 | |
| Apricots, raw (Prunus armeniaca) | C07 | 0 | 0.55597642 | |
| Arctic bramble berries | H02 | 0 | 0.608324644 | |
| Arugula, raw (Eruca sativa) | D02 | 0 | 0.609189332 | |
| Asparagus, raw (Asparagus officinalis) | D02 | 0 | 0.590635342 | |
| Avocados, raw, all commercial varieties (Persea americana) | S01 | 0 | 0.562568677 | |
| Bananas, raw (Musa acuminata Colla) | J02 | 0 | 0.523333872 | |
| Bay leaves, fresh (Laurus nobilis) | D02 | 0 | 0.608966095 | |
| Bayberries, raw | C08 | 0 | 0.602899873 | |
| Bayberries, raw | N05 | 0 | 0.563187905 | |
| Beans, kidney, red, mature seeds, cooked, boiled, without salt | C08 | 1 | 0.676597543 | 32917495 |
| Beans, kidney, red, mature seeds, cooked, boiled, without salt | H03 | 0 | 0.55597642 | |
| Beans, kidney, red, mature seeds, cooked, boiled, without salt | N05 | 0 | 0.603120886 | |

| | | | |
|---|---|---|---|
| Beans, snap, green, canned, regular pack, drained solids | D02 | 0 | 0.616259102 |
| Beans, snap, green, canned, regular pack, drained solids | P03 | 0 | 0.565057931 |
| Beans, snap, green, cooked, boiled, drained, without salt | C08 | 0 | 0.602899873 |
| Beans, snap, green, cooked, boiled, drained, without salt | N05 | 0 | 0.563187905 |
| Beans, snap, green, frozen, all styles, unprepared | D02 | 0 | 0.616259102 |
| Beans, snap, green, frozen, all styles, unprepared | P03 | 0 | 0.565057931 |
| Beans, snap, green, raw (Phaseolus vulgaris) | C08 | 0 | 0.601838056 |
| Beans, snap, green, raw (Phaseolus vulgaris) | H03 | 0 | 0.553978364 |
| Beans, snap, green, raw (Phaseolus vulgaris) | P03 | 0 | 0.592756849 |
| Beans, snap, yellow, cooked, boiled, drained, without salt | D02 | 0 | 0.616259102 |
| Beans, snap, yellow, cooked, boiled, drained, without salt | P03 | 0 | 0.565057931 |
| Bee Pollen | C08 | 0 | 0.575995858 |
| Bee Pollen | H03 | 0 | 0.583438636 |
| Bee Pollen | J02 | 0 | 0.565089234 |
| Bee Pollen | R06 | 0 | 0.611823352 |

| | | | | | | | | | |
|---|---|---|---|---|---|---|---|---|---|
| Bilberry, raw | C08 | 0 | 0.602899873 | | | | | | |
| Bilberry, raw | N05 | 0 | 0.563187905 | | | | | | |
| Blueberries, frozen, unsweetened | C02 | 6 | 0.566685296 | 25578927 | 25150116 | 24004888 | 20660279 | 29882843 | 25263326 |
| Blueberries, frozen, unsweetened | C08 | 6 | 0.587198791 | 25578927 | 25150116 | 24004888 | 20660279 | 29882843 | 25263326 |
| Blueberries, frozen, unsweetened | H03 | 0 | 0.640602564 | | | | | | |
| Blueberries, frozen, unsweetened | J02 | 0 | 0.520587395 | | | | | | |
| Blueberries, frozen, unsweetened | L04 | 1 | 0.607044102 | 26477692 | | | | | |
| Blueberries, frozen, unsweetened | M01 | 2 | 0.612166837 | 26477692 | 30699971 | | | | |
| Blueberries, frozen, unsweetened | M05 | 1 | 0.578857739 | 37269909 | | | | | |
| Blueberries, frozen, unsweetened | N05 | 2 | 0.566685296 | 35564709 | 32731478 | | | | |
| Blueberries, frozen, unsweetened | P03 | 0 | 0.640602564 | | | | | | |
| Blueberries, frozen, unsweetened | S02 | 0 | 0.597455538 | | | | | | |
| Blueberries, rabbiteye, raw (Vaccinium spp.) | J02 | 0 | 0.512199109 | | | | | | |
| Bog whortleberries, wild, frozen | C08 | 0 | 0.602899873 | | | | | | |

| | | | |
|---|---|---|---|
| Bog whortleberries, wild, frozen | N05 | 0 | 0.563187905 |
| Broadbeans (fava beans), mature seeds, canned | D02 | 0 | 0.616259102 |
| Broadbeans (fava beans), mature seeds, canned | P03 | 0 | 0.565057931 |
| Broccoli, cooked, boiled, drained, without salt | D02 | 0 | 0.616259102 |
| Broccoli, cooked, boiled, drained, without salt | P03 | 0 | 0.565057931 |
| Broccoli, frozen, chopped, unprepared | D02 | 0 | 0.608966095 |
| Broccoli, raw (Brassica oleracea var. italica) | C08 | 0 | 0.570483936 |
| Broccoli, raw (Brassica oleracea var. italica) | H03 | 0 | 0.564082852 |
| Brussels sprouts, cooked, boiled, drained, without salt | N05 | 0 | 0.575995858 |
| Brussels sprouts, raw (Brassica oleracea (Gemmifera Group)) | N05 | 0 | 0.570483936 |
| Buckwheat groats, roasted, dry | D02 | 0 | 0.568854556 |
| Buckwheat groats, roasted, dry | D05 | 0 | 0.610120231 |
| Cabbage, chinese (pak-choi), cooked, boiled, drained, without salt | D02 | 0 | 0.579144578 |
| Cabbage, chinese (pak-choi), cooked, boiled, drained, without salt | H02 | 0 | 0.575279276 |
| Cabbage, chinese (pak-choi), cooked, boiled, drained, without salt | H03 | 0 | 0.722159091 |

Cabbage, chinese (pak-choi), cooked, boiled, drained, without salt J02 0
0.640755703

Cabbage, chinese (pak-choi), cooked, boiled, drained, without salt L04 0
0.721414971

Cabbage, chinese (pak-choi), cooked, boiled, drained, without salt M05 0
0.648719184

Cabbage, chinese (pak-choi), cooked, boiled, drained, without salt P03 0
0.587519261

Cabbage, chinese (pak-choi), cooked, boiled, drained, without salt R06 0
0.55385626

Cabbage, chinese (pak-choi), raw (Brassica rapa (Chinensis Group)) C02 0
0.556428548

Cabbage, chinese (pak-choi), raw (Brassica rapa (Chinensis Group)) C08 0
0.556428548

Cabbage, chinese (pak-choi), raw (Brassica rapa (Chinensis Group)) D02 0
0.572387136

Cabbage, chinese (pak-choi), raw (Brassica rapa (Chinensis Group)) G03 0
0.601921367

Cabbage, chinese (pak-choi), raw (Brassica rapa (Chinensis Group)) H03 0
0.733219802

Cabbage, chinese (pak-choi), raw (Brassica rapa (Chinensis Group)) J02 0
0.582747084

Cabbage, chinese (pak-choi), raw (Brassica rapa (Chinensis Group)) L04 0
0.699253333

Cabbage, chinese (pak-choi), raw (Brassica rapa (Chinensis Group)) M01 0
0.653148718

Cabbage, chinese (pak-choi), raw (Brassica rapa (Chinensis Group)) M05 0
0.686911183

Cabbage, chinese (pak-choi), raw (Brassica rapa (Chinensis Group)) N05 0
0.576942043

| | | | |
|---|---|---|---|
| Cabbage, chinese (pak-choi), raw (Brassica rapa (Chinensis Group)) | P03 | 0 | 0.681765781 |
| Cabbage, chinese (pak-choi), raw (Brassica rapa (Chinensis Group)) | S02 | 0 | 0.597455538 |
| Cabbage, chinese (pe-tsai), raw (Brassica rapa (Pekinensis Group)) | C08 | 0 | 0.580373106 |
| Cabbage, chinese (pe-tsai), raw (Brassica rapa (Pekinensis Group)) | H03 | 0 | 0.731740067 |
| Cabbage, chinese (pe-tsai), raw (Brassica rapa (Pekinensis Group)) | J02 | 0 | 0.632902174 |
| Cabbage, chinese (pe-tsai), raw (Brassica rapa (Pekinensis Group)) | L04 | 0 | 0.661821352 |
| Cabbage, chinese (pe-tsai), raw (Brassica rapa (Pekinensis Group)) | M01 | 0 | 0.572039385 |
| Cabbage, chinese (pe-tsai), raw (Brassica rapa (Pekinensis Group)) | M05 | 0 | 0.641560833 |
| Cabbage, chinese (pe-tsai), raw (Brassica rapa (Pekinensis Group)) | P03 | 0 | 0.613218788 |
| Cabbage, Chinese, choi-sum, raw | C08 | 0 | 0.617969033 |
| Cabbage, Chinese, choi-sum, raw | H03 | 0 | 0.722928998 |
| Cabbage, Chinese, choi-sum, raw | J02 | 0 | 0.593107032 |
| Cabbage, Chinese, choi-sum, raw | L04 | 0 | 0.637780513 |
| Cabbage, Chinese, choi-sum, raw | M01 | 0 | 0.591675897 |
| Cabbage, Chinese, choi-sum, raw | M05 | 0 | 0.625166358 |

| | | | | | |
|---|---|---|---|---|---|
| Cabbage, Chinese, choi-sum, raw | N05 | 0 | 0.587198791 | | |
| Cabbage, Chinese, choi-sum, raw | P03 | 0 | 0.671474977 | | |
| Cabbage, cooked, boiled, drained, without salt | C08 | 0 | 0.575258361 | | |
| Cabbage, cooked, boiled, drained, without salt | H03 | 0 | 0.680696442 | | |
| Cabbage, cooked, boiled, drained, without salt | J02 | 0 | 0.612313536 | | |
| Cabbage, cooked, boiled, drained, without salt | L04 | 0 | 0.626161813 | | |
| Cabbage, cooked, boiled, drained, without salt | M01 | 0 | 0.584933052 | | |
| Cabbage, cooked, boiled, drained, without salt | M02 | 2 | 0.621008218 | 35719197 | 26889617 |
| Cabbage, cooked, boiled, drained, without salt | M05 | 0 | 0.582344865 | | |
| Cabbage, cooked, boiled, drained, without salt | S02 | 0 | 0.570099093 | | |
| Cabbage, napa, raw | D02 | 0 | 0.616259102 | | |
| Cabbage, napa, raw | P03 | 0 | 0.565057931 | | |
| Cabbage, raw (Brassica oleracea (Capitata Group)) | C02 | 0 | 0.559460092 | | |
| Cabbage, raw (Brassica oleracea (Capitata Group)) | D02 | 0 | 0.58457507 | | |
| Cabbage, raw (Brassica oleracea (Capitata Group)) | G03 | 0 | 0.688232713 | | |

| | | | |
|---|---|---|---|
| Cabbage, raw (Brassica oleracea (Capitata Group)) | H03 | 0 | 0.572378188 |
| Cabbage, raw (Brassica oleracea (Capitata Group)) | L02 | 1 | 0.58637427125040978 |
| Cabbage, raw (Brassica oleracea (Capitata Group)) | L04 | 0 | 0.655197542 |
| Cabbage, raw (Brassica oleracea (Capitata Group)) | M01 | 0 | 0.635927026 |
| Cabbage, raw (Brassica oleracea (Capitata Group)) | M05 | 0 | 0.669157109 |
| Cabbage, raw (Brassica oleracea (Capitata Group)) | P03 | 0 | 0.61661998 |
| Cabbage, raw (Brassica oleracea (Capitata Group)) | S02 | 0 | 0.625603157 |
| Cabbage, red, pickled | C08 | 0 | 0.602899873 |
| Cabbage, red, pickled | N05 | 0 | 0.563187905 |
| Cabbage, red, raw (Brassica oleracea (Capitata Group)) | D05 | 0 | 0.581507781 |
| Cabbage, red, raw (Brassica oleracea (Capitata Group)) | M01 | 0 | 0.553339101 |
| Cabbage, savoy, raw (Brassica oleracea (Capitata Group)) | C02 | 0 | 0.561556922 |
| Cabbage, savoy, raw (Brassica oleracea (Capitata Group)) | D02 | 0 | 0.562027188 |
| Cabbage, savoy, raw (Brassica oleracea (Capitata Group)) | G03 | 0 | 0.612166837 |
| Cabbage, savoy, raw (Brassica oleracea (Capitata Group)) | H03 | 0 | 0.748656009 |

| | | | | |
|---|---|---|---|---|
| Cabbage, savoy, raw (Brassica oleracea (Capitata Group)) | J02 | 0 | 0.608646954 | |
| Cabbage, savoy, raw (Brassica oleracea (Capitata Group)) | L04 | 0 | 0.724867008 | |
| Cabbage, savoy, raw (Brassica oleracea (Capitata Group)) | M01 | 0 | 0.663394188 | |
| Cabbage, savoy, raw (Brassica oleracea (Capitata Group)) | M05 | 0 | 0.707492792 | |
| Cabbage, savoy, raw (Brassica oleracea (Capitata Group)) | N05 | 0 | 0.561556922 | |
| Cabbage, savoy, raw (Brassica oleracea (Capitata Group)) | P03 | 0 | 0.666329575 | |
| Cabbage, savoy, raw (Brassica oleracea (Capitata Group)) | S02 | 0 | 0.607712285 | |
| Capers, canned (Capparis spinosa) | D02 | 1 | 0.616259102 | 36221990 |
| Capers, canned (Capparis spinosa) | P03 | 0 | 0.565057931 | |
| Capers, raw | D02 | 1 | 0.608966095 | 36221990 |
| Carob fiber (Caromax) | C08 | 0 | 0.676597543 | |
| Carob fiber (Caromax) | H03 | 0 | 0.55597642 | |
| Carob fiber (Caromax) | N05 | 1 | 0.603120886 | 36606510 |
| Carob kibbles | C08 | 0 | 0.682720597 | |
| Carob kibbles | H03 | 0 | 0.568263192 | |
| Carob kibbles | J02 | 0 | 0.51023472 | |
| Carob kibbles | N05 | 1 | 0.590874777 | 36606510 |
| Carrots, raw (Daucus carota) | C08 | 0 | 0.570483936 | |
| Carrots, raw (Daucus carota) | H03 | 0 | 0.564082852 | |
| Cashew apple, raw | C08 | 0 | 0.682720597 | |
| Cashew apple, raw | H03 | 0 | 0.568263192 | |
| Cashew apple, raw | J02 | 0 | 0.51023472 | |
| Cashew apple, raw | N05 | 0 | 0.590874777 | |
| Catsup | D02 | 0 | 0.616259102 | |
| Catsup | P03 | 0 | 0.565057931 | |
| Cauliflower, frozen, cooked, boiled, drained, without salt | M05 | 0 | 0.568263192 | |
| Cauliflower, raw (Brassica oleracea (Botrytis Group)) | C02 | 0 | 0.559460092 | |
| Cauliflower, raw (Brassica oleracea (Botrytis Group)) | D02 | 0 | 0.58457507 | |
| Cauliflower, raw (Brassica oleracea (Botrytis Group)) | G03 | 0 | 0.688232713 | |
| Cauliflower, raw (Brassica oleracea (Botrytis Group)) | H03 | 0 | 0.572378188 | |

| | | | |
|---|---|---|---|
| Cauliflower, raw (Brassica oleracea (Botrytis Group)) | L02 | 1 | 0.586374271 17652276 |
| Cauliflower, raw (Brassica oleracea (Botrytis Group)) | L04 | 0 | 0.655197542 |
| Cauliflower, raw (Brassica oleracea (Botrytis Group)) | M01 | 0 | 0.635927026 |
| Cauliflower, raw (Brassica oleracea (Botrytis Group)) | M05 | 0 | 0.669157109 |
| Cauliflower, raw (Brassica oleracea (Botrytis Group)) | P03 | 0 | 0.61661998 |
| Cauliflower, raw (Brassica oleracea (Botrytis Group)) | S02 | 0 | 0.625603157 |
| Celeriac, raw (Apium graveolens) | D02 | 0 | 0.554268542 |
| Celeriac, raw (Apium graveolens) | D05 | 0 | 0.602899873 |
| Celeriac, raw (Apium graveolens) | G03 | 0 | 0.562568677 |
| Celeriac, raw (Apium graveolens) | M01 | 0 | 0.562568677 |
| Celery hearts, green | L04 | 0 | 0.562568677 |
| Celery hearts, green | S02 | 0 | 0.602899873 |
| Celery hearts, white | L04 | 0 | 0.562568677 |
| Celery hearts, white | S02 | 0 | 0.602899873 |
| Celery, Chinese, raw | S02 | 0 | 0.610120231 |

Celery, raw (Apium graveolens) C02 0 0.609067391

Celery, raw (Apium graveolens) D05 0 0.578751819

Celery, raw (Apium graveolens) G03 0 0.693738574

Celery, raw (Apium graveolens) H03 0 0.566847964

Celery, raw (Apium graveolens) L02 0 0.591880133

Celery, raw (Apium graveolens) L04 0 0.710256159

Celery, raw (Apium graveolens) M01 0 0.685479782

Celery, raw (Apium graveolens) M05 0 0.696808229

Celery, raw (Apium graveolens) P03 0 0.555787516

Celery, raw (Apium graveolens) S02 0 0.697258144

Chard, swiss, red and white stems, raw (Beta vulgaris subsp. Vulagaris) C08 1 0.645982269 31029925

Chard, swiss, red and white stems, raw (Beta vulgaris subsp. Vulagaris) H03 0 0.592836735

Chard, swiss, red and white stems, raw (Beta vulgaris subsp. Vulagaris) N05 0 0.566382558

Chard, swiss, red and white stems, raw (Beta vulgaris subsp. Vulagaris) P03 0 0.592836735

Chard, swiss, red leaf, raw (Beta vulgaris subsp. Vulagaris) C08 1 0.645982269 31029925

Chard, swiss, red leaf, raw (Beta vulgaris subsp. Vulagaris) H03 0
0.592836735

Chard, swiss, red leaf, raw (Beta vulgaris subsp. Vulagaris) N05 0
0.566382558

Chard, swiss, white leaf, raw (Beta vulgaris subsp. vulagaris) C08 1
0.652105324 31029925

Chard, swiss, white leaf, raw (Beta vulgaris subsp. vulagaris) H03 0
0.605123507

Chard, swiss, white leaf, raw (Beta vulgaris subsp. vulagaris) J02 0
0.522604046

Chard, swiss, white leaf, raw (Beta vulgaris subsp. vulagaris) N05 0
0.554136449

Chard, swiss, white leaf, raw (Beta vulgaris subsp. vulagaris) P03 0
0.568263192

Cherries, sour, dry, sweetened D02 0 0.553527363

Cherries, sour, dry, unsweetened D02 0 0.553527363

Cherries, sour, powder D02 0 0.590635342

Cherries, sour, red, frozen, unsweetened D02 0 0.553527363

Cherries, sour, red, raw (Prunus cerasus) H02 4 0.619385092
24566440 27455316 19883392 27231439

Chia seeds, raw D02 1 0.616259102 20548903

Chia seeds, raw P03 0 0.565057931

Chicory greens, raw (Cichorium intybus) C02 1 0.559460092 26872721

| | | | | |
|---|---|---|---|---|
| Chicory greens, raw (Cichorium intybus) | D02 | 1 | 0.58457507 | 27161285 |
| Chicory greens, raw (Cichorium intybus) | G03 | 0 | 0.688232713 | |
| Chicory greens, raw (Cichorium intybus) | H03 | 0 | 0.572378188 | |
| Chicory greens, raw (Cichorium intybus) | L02 | 0 | 0.586374271 | |
| Chicory greens, raw (Cichorium intybus) | L04 | 0 | 0.655197542 | |
| Chicory greens, raw (Cichorium intybus) | M01 | 1 | 0.635927026 | 20618964 |
| Chicory greens, raw (Cichorium intybus) | M05 | 0 | 0.669157109 | |
| Chicory greens, raw (Cichorium intybus) | P03 | 0 | 0.61661998 | |
| Chicory greens, raw (Cichorium intybus) | S02 | 0 | 0.625603157 | |
| Chokeberry, raw | D02 | 0 | 0.616259102 | |
| Chokeberry, raw | P03 | 0 | 0.565057931 | |
| Corn poppy, leaves | D02 | 0 | 0.58486436 | |
| Corn poppy, leaves | H03 | 0 | 0.740296147 | |
| Corn poppy, leaves | J02 | 0 | 0.681107862 | |
| Corn poppy, leaves | L04 | 0 | 0.668700552 | |

Corn poppy, leaves M05 0 0.585863507

Corn poppy, leaves P03 0 0.622633183

Corn poppy, leaves R06 0 0.596140623

Cowpeas, black seed cultivar, mature seeds, raw (Vigna unguiculata Subsp. Sinensis) C08 0 0.676597543

Cowpeas, black seed cultivar, mature seeds, raw (Vigna unguiculata Subsp. Sinensis) H03 0 0.55597642

Cowpeas, black seed cultivar, mature seeds, raw (Vigna unguiculata Subsp. Sinensis) N05 0 0.603120886

Cranberries, dried, sweetened (Includes foods for USDA's Food Distribution Program) C08 3 0.616325422 17761017 25904733 34444779

Cranberries, dried, sweetened (Includes foods for USDA's Food Distribution Program) D05 0 0.552124857

Cranberries, dried, sweetened (Includes foods for USDA's Food Distribution Program) H03 0 0.65959668

Cranberries, dried, sweetened (Includes foods for USDA's Food Distribution Program) J02 0 0.560274056

Cranberries, dried, sweetened (Includes foods for USDA's Food Distribution Program) L04 2 0.589995779 30553231 33636418

Cranberries, dried, sweetened (Includes foods for USDA's Food Distribution Program) M01 2 0.607952172 30553231 33636418

Cranberries, dried, sweetened (Includes foods for USDA's Food Distribution Program) N05 0 0.600917287

Cranberries, dried, sweetened (Includes foods for USDA's Food Distribution Program) P03 0 0.649290482

Cranberries, dried, sweetened (Includes foods for USDA's Food Distribution Program) S02 0 0.572669038

Cranberry juice cocktail, bottled C08 3 0.616325422 17761017 25904733 34444779

Cranberry juice cocktail, bottled D05 0 0.552124857

Cranberry juice cocktail, bottled H03 0 0.65959668

Cranberry juice cocktail, bottled J02 0 0.560274056

Cranberry juice cocktail, bottled L04 2 0.589995779 30553231 33636418

Cranberry juice cocktail, bottled M01 2 0.607952172 30553231 33636418

Cranberry juice cocktail, bottled N05 0 0.600917287

Cranberry juice cocktail, bottled P03 0 0.649290482

Cranberry juice cocktail, bottled S02 0 0.572669038

Cranberry sauce, canned, sweetened C08 3 0.621461467 17761017 25904733 34444779

Cranberry sauce, canned, sweetened H03 0 0.669902878

Cranberry sauce, canned, sweetened J02 0 0.591400392

Cranberry sauce, canned, sweetened L04 2 0.584865381 30553231 33636418

Cranberry sauce, canned, sweetened M01 2 0.58743058 30553231 33636418

Cranberry sauce, canned, sweetened N05 0 0.590645196

Cranberry sauce, canned, sweetened P03 0 0.628678086

Cranberry sauce, jellied, canned, OCEAN SPRAY C08 3 0.602899873 17761017 25904733 34444779

Cranberry sauce, jellied, canned, OCEAN SPRAY N05 0 0.563187905

Cranberry Sauce, whole, canned, OCEAN SPRAY C08 3 0.602899873 17761017 25904733 34444779

Cranberry Sauce, whole, canned, OCEAN SPRAY N05 0 0.563187905

Crowberries, raw C08 0 0.602899873

Crowberries, raw N05 0 0.563187905

Cucumber, with peel, raw (Cucumis sativus) D02 0 0.608966095

Currants, european black, raw (Ribes nigrum) J02 0 0.655266721

Currants, white, raw C08 0 0.676597543

Currants, white, raw H03 0 0.55597642

| | | | | |
|---|---|---|---|---|
| Currants, white, raw | N05 | 0 | 0.603120886 | |
| Dill weed, fresh (Anethum graveolens) | J02 | 0 | 0.531684945 | |
| Dill weed, fresh (Anethum graveolens) | R06 | 0 | 0.63387104 | |
| Dock, raw (Rumex spp.) | C08 | 0 | 0.658228379 | |
| Dock, raw (Rumex spp.) | H03 | 0 | 0.568263192 | |
| Dock, raw (Rumex spp.) | N05 | 0 | 0.590874777 | |
| Dock, raw (Rumex spp.) | P03 | 0 | 0.605123507 | |
| Drumstick (horseradish tree) leaves, raw (Moringa oleifera) | D02 | 1 | 0.615373995 | 25097471 |
| Eggplant, long, cooked | H03 | 0 | 0.631535335 | |
| Eggplant, long, cooked | J02 | 0 | 0.621836108 | |
| Eggplant, long, cooked | L01 | 1 | 0.565326367 | 34322510 |
| Eggplant, long, cooked | L04 | 0 | 0.606691711 | |
| Eggplant, long, cooked | M02 | 0 | 0.625995538 | |
| Eggplant, long, cooked | M05 | 0 | 0.562288039 | |
| Elderberries, raw (Sambucus spp.) | D02 | 0 | 0.590635342 | |

Fennel, leaves, raw H03 0 0.568566934

Fennel, leaves, raw J02 0 0.546487266

Fennel, leaves, raw M04 0 0.522258315

Fennel, leaves, raw R06 0 0.587198791

Garlic chives, raw D02 0 0.608966095

Garlic, raw (Allium sativum) C08 8 0.676597543 28956671 23169470 20594781 35276764 32349742 34496450 33974725 36558397

Garlic, raw (Allium sativum) H03 0 0.55597642

Garlic, raw (Allium sativum) N05 0 0.603120886

Goji berry (wolfberry), dried C08 0 0.676597543

Goji berry (wolfberry), dried H03 0 0.55597642

Goji berry (wolfberry), dried N05 2 0.603120886 19857084 18447631

Gooseberries, raw (Ribes spp.) D02 0 0.616259102

Gooseberries, raw (Ribes spp.) P03 0 0.565057931

Grapefruit, raw (not specified as to color) (Citrus paradisi) N05 0 0.587019703

Grapefruit, raw, pink and red, all areas (Citrus paradisi) C08 2 0.558736051 32520418 19153985

Grapefruit, raw, pink and red, all areas (Citrus paradisi) N03 0 0.670233503

Grapefruit, raw, pink and red, all areas (Citrus paradisi) N05 0 0.588014796

Greek greens pie (prepared from wild greens) M04 0 0.537694522

Hartwort, leaves H03 0 0.563421532

Hartwort, leaves J02 0 0.520587395

Hartwort, leaves M04 0 0.53254912

Hartwort, leaves R06 0 0.592327164

Hawthorn leaves, raw D02 0 0.568854556

Hawthorn leaves, raw D05 0 0.610120231

Honey, mixed varieties (samples obtained in Argentina, Australia, Italy, Portugaul and Spain) H03 0 0.687887134

Honey, mixed varieties (samples obtained in Argentina, Australia, Italy, Portugaul and Spain) J02 1 0.665400153 28760590

Honey, mixed varieties (samples obtained in Argentina, Australia, Italy, Portugaul and Spain) L01 0 0.575182207

Honey, mixed varieties (samples obtained in Argentina, Australia, Italy, Portugaul and Spain) L04 1 0.626363674 28474502

Horseradish, root, whole M05 0 0.568263192

Jams and preserves, grape C08 5 0.575258361 22810991 25942487 15630270 26337448 35107113

Jams and preserves, grape H03 0 0.680696442

Jams and preserves, grape J02 0 0.612313536

Jams and preserves, grape L04 0 0.626161813

Jams and preserves, grape M01 0 0.584933052

Jams and preserves, grape M02 0 0.621008218

Jams and preserves, grape M05 0 0.582344865

Jams and preserves, grape S02 0 0.570099093

Jams and preserves, guava C08 2 0.575258361 1332463 27488183

Jams and preserves, guava H03 0 0.680696442

Jams and preserves, guava J02 0 0.612313536

Jams and preserves, guava L04 0 0.626161813

Jams and preserves, guava M01 0 0.584933052

| | | | | | | | | | | | |
|---|---|---|---|---|---|---|---|---|---|---|---|
| Jams and preserves, guava | M02 | 0 | 0.621008218 | | | | | | | | |
| Jams and preserves, guava | M05 | 0 | 0.582344865 | | | | | | | | |
| Jams and preserves, guava | S02 | 0 | 0.570099093 | | | | | | | | |
| Jams and preserves, peach | D02 | 0 | 0.616259102 | | | | | | | | |
| Jams and preserves, peach | P03 | 0 | 0.565057931 | | | | | | | | |
| Jams and preserves, raspberry | D02 | 0 | 0.616259102 | | | | | | | | |
| Jams and preserves, raspberry | P03 | 0 | 0.565057931 | | | | | | | | |
| Jams and preserves, sour orange | C10 | 8 | 0.57187614 | 23919812 | 36184225 | 26471075 | 20729016 | 30638420 | 21056284 | 28526377 | 33652807 |
| Jams and preserves, strawberry | D02 | 0 | 0.608966095 | | | | | | | | |
| Juice concentrate, black currant | C08 | 3 | 0.602899873 | 28555052 | 32396017 | 31902311 | | | | | |
| Juice concentrate, black currant | N05 | 1 | 0.563187905 | 29343201 | | | | | | | |
| Juice concentrate, sour cherry | D02 | 0 | 0.553527363 | | | | | | | | |
| Juice, black Currant | C08 | 3 | 0.610120231 | 28555052 | 32396017 | 31902311 | | | | | |
| Juice, blood orange | N03 | 0 | 0.562568677 | | | | | | | | |
| Juice, Cranberry cocktail, bottled | C02 | 3 | 0.605094064 | 17761017 | 25904733 | 34444779 | | | | | |

Juice, Cranberry cocktail, bottled D05 0 0.583135005

Juice, Cranberry cocktail, bottled J02 0 0.598860138

Juice, Cranberry cocktail, bottled S02 0 0.707569671

Juice, cranberry, raw C08 3 0.602899873 17761017 25904733 34444779

Juice, cranberry, raw N05 0 0.563187905

Juice, crowberry C08 0 0.602899873

Juice, crowberry N05 0 0.563187905

Juice, grape, canned or bottled, unsweetened, without added ascorbic acid C02 5 0.58551354 22810991 25942487 15630270 26337448 35107113

Juice, grape, canned or bottled, unsweetened, without added ascorbic acid C08 5 0.56591476 22810991 25942487 15630270 26337448 35107113

Juice, grape, canned or bottled, unsweetened, without added ascorbic acid D05 0 0.580613845

Juice, grape, canned or bottled, unsweetened, without added ascorbic acid H03 0 0.587457696

Juice, grape, canned or bottled, unsweetened, without added ascorbic acid J02 0 0.64831826

Juice, grape, canned or bottled, unsweetened, without added ascorbic acid S02 0 0.6712582

Juice, grape, red	C08	1	0.602899873	25666417

Juice, grape, red	N05	0	0.563187905

Juice, grapefruit, white, raw	N03	0	0.555816752

Juice, orange, frozen concentrate, unsweetened, diluted with 3 volume water	N03	0	0.562568677

Juice, orange, raw	N03	0	0.581430427

Juice, sour orange	C10	8	0.584108785	23919812	36184225	26471075	20729016	30638420	21056284	28526377	33652807

Juice, tangelo	C10	0	0.565759818

Juice, tangerine, frozen concentrate, sweetened, diluted with 3 volume water	N03	0	0.639155686

Juice, tangor (e.g., murcot or temple)	C10	0	0.565759818

Juice, tangor, diluted from frozen concentrate (ex. Murcot or temple)	N03	0	0.562568677

Juice, tomato, canned, without salt added	C08	5	0.658228379	33407609	30223563	21138408	29131285	16368299

Juice, tomato, canned, without salt added	H03	0	0.568263192

Juice, tomato, canned, without salt added	N05	3	0.590874777	37475550	25880734	30792141

Juice, tomato, canned, without salt added	P03	0	0.605123507

| | | | |
|---|---|---|---|
| Juniper berries, green, unripe (Juniperus communis) | C02 | 0 | 0.584751722 |
| Juniper berries, green, unripe (Juniperus communis) | D05 | 0 | 0.603120886 |
| Juniper berries, green, unripe (Juniperus communis) | L02 | 0 | 0.584108785 |
| Juniper berries, green, unripe (Juniperus communis) | M01 | 0 | 0.602457752 |
| Juniper berries, green, unripe (Juniperus communis) | S02 | 0 | 0.670474488 |
| Juniper berries, ripe (Juniperus communis) | C02 | 0 | 0.584751722 |
| Juniper berries, ripe (Juniperus communis) | D05 | 0 | 0.603120886 |
| Juniper berries, ripe (Juniperus communis) | L02 | 0 | 0.584108785 |
| Juniper berries, ripe (Juniperus communis) | M01 | 0 | 0.602457752 |
| Juniper berries, ripe (Juniperus communis) | S02 | 0 | 0.670474488 |
| Kale, canned | D02 | 0 | 0.608966095 |
| Kale, Chinese, raw | D05 | 0 | 0.558387301 |
| Kale, Chinese, raw | J02 | 0 | 0.50822487 |
| Kale, Chinese, raw | N05 | 0 | 0.638156915 |
| Kale, raw (Brassica oleracea (Acephala Group)) | D02 | 0 | 0.578266016 |

Kiwifruit, green, raw (Actinidia deliciosa) C02 3 0.556764227 22258209 35807858 25483553

Kiwifruit, green, raw (Actinidia deliciosa) S02 0 0.577475369

Kohlrabi, raw (Brassica oleracea (Gongylodes Group)) M05 0 0.568263192

Kumquats, raw (Fortunella spp.) N06 0 0.562568677

Leeks, (bulb and lower leaf-portion), raw (Allium ampeloprasum) C08 0 0.652105324

Leeks, (bulb and lower leaf-portion), raw (Allium ampeloprasum) H03 0 0.605123507

Leeks, (bulb and lower leaf-portion), raw (Allium ampeloprasum) J02 0 0.522604046

Leeks, (bulb and lower leaf-portion), raw (Allium ampeloprasum) N05 0 0.554136449

Lemons, raw, without peel (Citrus limon) N03 0 0.584395955

Lettuce, butterhead (includes boston and bibb types), raw (Lactuca sativa var. capitata) D02 0 0.616259102

Lettuce, butterhead (includes boston and bibb types), raw (Lactuca sativa var. capitata) P03 0 0.565057931

Lettuce, green leaf, raw (Lactuca sativa var. crispa) C02 0 0.602583912

Lettuce, green leaf, raw (Lactuca sativa var. crispa) D05 0 0.602583912

Lettuce, green leaf, raw (Lactuca sativa var. crispa) H03 0 0.640602564

Lettuce, green leaf, raw (Lactuca sativa var. crispa) J02 0 0.520587395

Lettuce, green leaf, raw (Lactuca sativa var. crispa) L04 0 0.642903248

Lettuce, green leaf, raw (Lactuca sativa var. crispa) M01 0 0.663394188

Lettuce, green leaf, raw (Lactuca sativa var. crispa) M05 1 0.604584749
9925126

Lettuce, green leaf, raw (Lactuca sativa var. crispa) N05 0 0.566685296

Lettuce, green leaf, raw (Lactuca sativa var. crispa) P03 0 0.640602564

Lettuce, green leaf, raw (Lactuca sativa var. crispa) S02 0 0.648739275

Lettuce, iceberg (includes crisphead types), raw (Lactuca sativa var. capitata)
C08 0 0.602583912

Lettuce, iceberg (includes crisphead types), raw (Lactuca sativa var. capitata)
D02 0 0.582747084

Lettuce, iceberg (includes crisphead types), raw (Lactuca sativa var. capitata)
D05 0 0.556428548

Lettuce, iceberg (includes crisphead types), raw (Lactuca sativa var. capitata)
G03 0 0.566062222

Lettuce, iceberg (includes crisphead types), raw (Lactuca sativa var. capitata)
H03 0 0.733219802

Lettuce, iceberg (includes crisphead types), raw (Lactuca sativa var. capitata)
J02 0 0.598287006

| | | | | |
|---|---|---|---|---|
| Lettuce, iceberg (includes crisphead types), raw (Lactuca sativa var. capitata) | L04 | 0 | 0.658271453 | |
| Lettuce, iceberg (includes crisphead types), raw (Lactuca sativa var. capitata) | M01 | 0 | 0.648025983 | |
| Lettuce, iceberg (includes crisphead types), raw (Lactuca sativa var. capitata) | M05 | 1 | 0.635457162 | 9925126 |
| Lettuce, iceberg (includes crisphead types), raw (Lactuca sativa var. capitata) | N05 | 0 | 0.62822578 | |
| Lettuce, iceberg (includes crisphead types), raw (Lactuca sativa var. capitata) | P03 | 0 | 0.70234739 | |
| Lettuce, iceberg (includes crisphead types), raw (Lactuca sativa var. capitata) | S02 | 0 | 0.571813669 | |
| Limes, raw (Citrus latifolia) | N03 | 0 | 0.565759818 | |
| Lingonberries (cowberries), raw | D02 | 0 | 0.616259102 | |
| Lingonberries (cowberries), raw | P03 | 0 | 0.565057931 | |
| Locust bean powder | D02 | 0 | 0.616259102 | |
| Locust bean powder | P03 | 0 | 0.565057931 | |
| Lotus root, raw (Nelumbo nucifera) | C08 | 0 | 0.614579313 | |
| Lotus root, raw (Nelumbo nucifera) | H03 | 0 | 0.597264196 | |
| Lovage, leaves, raw | D02 | 0 | 0.616259102 | |
| Lovage, leaves, raw | P03 | 0 | 0.565057931 | |
| Mangos, raw (Mangifera indica) | H03 | 0 | 0.666391997 | |
| Mangos, raw (Mangifera indica) | J02 | 0 | 0.645816268 | |
| Mangos, raw (Mangifera indica) | L04 | 2 | 0.608397718 | 32109839 24344049 |
| Mangos, raw (Mangifera indica) | M02 | 0 | 0.605644787 | |
| Mangos, raw (Mangifera indica) | M05 | 0 | 0.5751433 | |
| Medlar | C01 | 0 | 0.590225107 | |
| Melons, cantaloupe, raw (Cucumis melo) | M05 | 0 | 0.55597642 | |
| Mizuna (Japanese mustard) | D02 | 0 | 0.615373995 | |
| Mung beans, mature seeds, sprouted, raw (Vigna radiata) | D02 | 0 | 0.608966095 | |
| Mustard greens, raw (Brassica juncea) | D02 | 0 | 0.578266016 | |
| Nalta jute, raw | C08 | 0 | 0.658228379 | |
| Nalta jute, raw | H03 | 0 | 0.568263192 | |
| Nalta jute, raw | N05 | 0 | 0.590874777 | |
| Nalta jute, raw | P03 | 0 | 0.605123507 | |
| Nectarines, without skin, raw (Prunus persica var. nucipersica) | D02 | 0 | 0.616259102 | |
| Nectarines, without skin, raw (Prunus persica var. nucipersica) | P03 | 0 | 0.565057931 | |
| New Zealand spinach, raw (Tetragonia tetragonioides) | D02 | 0 | 0.608966095 | |
| Nuts, almonds (Prunus dulcis) | A02 | 0 | 0.645417219 | |
| Oil, olive, salad or cooking | S02 | 0 | 0.610120231 | |
| Onion, spring, red, leaves | D02 | 0 | 0.616259102 | |
| Onion, spring, red, leaves | P03 | 0 | 0.565057931 | |
| Onions, cooked, boiled, drained, without salt | D02 | 0 | 0.616259102 | |
| Onions, cooked, boiled, drained, without salt | P03 | 0 | 0.565057931 | |
| Onions, raw (Allium cepa) | D02 | 0 | 0.553752319 | |
| Onions, raw (Allium cepa) | H02 | 0 | 0.55494588 | |
| Onions, raw (Allium cepa) | H03 | 0 | 0.694293524 | |
| Onions, raw (Allium cepa) | J02 | 0 | 0.652197175 | |
| Onions, raw (Allium cepa) | L04 | 1 | 0.652292133 | 6364880 |
| Onions, raw (Allium cepa) | M05 | 1 | 0.55494588 | 26686359 |
| Onions, raw (Allium cepa) | R06 | 1 | 0.609151233 | 6364880 |
| Onions, red, raw | D02 | 0 | 0.553752319 | |
| Onions, red, raw | H02 | 0 | 0.564724662 | |
| Onions, red, raw | H03 | 0 | 0.708961697 | |
| Onions, red, raw | J02 | 0 | 0.691575118 | |
| Onions, red, raw | L04 | 0 | 0.637688578 | |
| Onions, red, raw | R06 | 0 | 0.623770862 | |
| Onions, spring or scallions (includes tops and bulb), raw (Allium cepa or Allium fistulosum) | D02 | 0 | 0.616259102 | |
| Onions, spring or scallions (includes tops and bulb), raw (Allium cepa or Allium fistulosum) | P03 | 0 | 0.565057931 | |
| Onions, sweet, raw (Allium cepa) | C02 | 1 | 0.580373106 | 36364848 |
| Onions, sweet, raw (Allium cepa) | C08 | 1 | 0.765270733 | 36364848 |
| Onions, sweet, raw (Allium cepa) | H03 | 0 | 0.662173229 | |
| Onions, sweet, raw (Allium cepa) | M01 | 0 | 0.582300181 | |

Onions, sweet, raw (Allium cepa) N05 1 0.577805083 33282442

Onions, sweet, raw (Allium cepa) P03 0 0.754929013

Onions, young green, tops only (Allium cepa) H03 0 0.677452445

Onions, young green, tops only (Allium cepa) J02 0 0.634681505

Onions, young green, tops only (Allium cepa) L04 0 0.635927026

Onions, young green, tops only (Allium cepa) M05 1 0.613854868 26686359

Oranges, raw, all commercial varieties (Citrus sinensis) C08 2 0.614024443 21068346 26471075

Oranges, raw, all commercial varieties (Citrus sinensis) N03 0 0.703404577

Oranges, raw, all commercial varieties (Citrus sinensis) N05 0 0.616461048

Oranges, raw, navels (Citrus sinensis) C08 2 0.578255214 21068346 26471075

Oranges, raw, navels (Citrus sinensis) N03 0 0.699480056

Oranges, raw, navels (Citrus sinensis) N05 0 0.573375423

Oregano, fresh D05 0 0.625603157

Oregano, fresh M01 0 0.602891856

Oregano, fresh N05 0 0.564972014

| | | | |
|---|---|---|---|
| Oregano, fresh | S02 | 0 | 0.570483936 |
| Pako fern, steamed (Athyrium esculentum) | D02 | 0 | 0.616259102 |
| Pako fern, steamed (Athyrium esculentum) | P03 | 0 | 0.565057931 |
| Papayas, raw (Carica papaya) | H03 | 0 | 0.621532014 |
| Papayas, raw (Carica papaya) | J02 | 0 | 0.603361371 |
| Papayas, raw (Carica papaya) | L01 | 0 | 0.574594405 |
| Papayas, raw (Carica papaya) | L04 | 0 | 0.618793975 |
| Papayas, raw (Carica papaya) | M02 | 0 | 0.618793975 |
| Papayas, raw (Carica papaya) | M05 | 0 | 0.57713687 |
| Parsley, fresh (Petroselinum crispum) | H03 | 0 | 0.753801411 |
| Parsley, fresh (Petroselinum crispum) | J02 | 0 | 0.686346566 |
| Parsley, fresh (Petroselinum crispum) | L04 | 0 | 0.694130598 |
| Parsley, fresh (Petroselinum crispum) | M01 | 0 | 0.632657778 |
| Parsley, fresh (Petroselinum crispum) | M02 | 0 | 0.668516923 |
| Parsley, fresh (Petroselinum crispum) | M05 | 0 | 0.645747966 |
| Parsley, fresh (Petroselinum crispum) | N05 | 0 | 0.556428548 |
| Parsley, fresh (Petroselinum crispum) | P03 | 0 | 0.609730151 |
| Parsley, fresh (Petroselinum crispum) | S02 | 0 | 0.576942043 |
| Peas, green, frozen, cooked, boiled, drained, without salt | C02 | 0 | 0.582070417 |
| Peas, green, frozen, cooked, boiled, drained, without salt | H03 | 0 | 0.640602564 |
| Peas, green, frozen, cooked, boiled, drained, without salt | L04 | 0 | 0.648025983 |
| Peas, green, frozen, cooked, boiled, drained, without salt | M01 | 0 | 0.617289572 |
| Peas, green, frozen, cooked, boiled, drained, without salt | M05 | 0 | 0.63031176 |
| Peas, green, frozen, cooked, boiled, drained, without salt | P03 | 0 | 0.620020956 |
| Peas, green, frozen, cooked, boiled, drained, without salt | S02 | 0 | 0.623097407 |
| Peppers, hot chili, green, raw (Capsicum frutescens) | C02 | 0 | 0.567727975 |
| Peppers, hot chili, green, raw (Capsicum frutescens) | D05 | 0 | 0.620091235 |
| Peppers, hot chili, green, raw (Capsicum frutescens) | M01 | 0 | 0.602891856 |
| Peppers, hot chili, green, raw (Capsicum frutescens) | S02 | 0 | 0.614579313 |
| Peppers, sweet, yellow, raw (Capsicum annuum) | C08 | 0 | 0.575995858 |

| | | | | | | | | | |
|---|---|---|---|---|---|---|---|---|---|
| Perilla leaves, raw | D05 | 0 | 0.581507781 | | | | | | |
| Perilla leaves, raw | M01 | 0 | 0.553339101 | | | | | | |
| Pitanga, (surinam-cherry), raw (Eugenia uniflora) | C08 | 0 | 0.676597543 | | | | | | |
| Pitanga, (surinam-cherry), raw (Eugenia uniflora) | H03 | 0 | 0.55597642 | | | | | | |
| Pitanga, (surinam-cherry), raw (Eugenia uniflora) | N05 | 0 | 0.603120886 | | | | | | |
| Plum, red, whole, raw | C02 | 1 | 0.570130806 | 29241576 | | | | | |
| Plum, red, whole, raw | C08 | 1 | 0.596164176 | 29241576 | | | | | |
| Plum, red, whole, raw | D05 | 0 | 0.57533748 | | | | | | |
| Plum, red, whole, raw | H03 | 0 | 0.694787002 | | | | | | |
| Plum, red, whole, raw | J02 | 0 | 0.570608252 | | | | | | |
| Plum, red, whole, raw | L04 | 0 | 0.644917708 | | | | | | |
| Plum, red, whole, raw | M01 | 0 | 0.650118657 | | | | | | |
| Plum, red, whole, raw | M05 | 6 | 0.613815585 | 35798020 | 35057457 | 21736808 | 34714130 | 26902092 | 11860726 |
| Plum, red, whole, raw | N05 | 0 | 0.609180861 | | | | | | |
| Plum, red, whole, raw | P03 | 0 | 0.679115115 | | | | | | |

| | | | | |
|---|---|---|---|---|
| Plum, red, whole, raw | S02 | 1 | 0.603974187 | 29281192 |
| Plum, yellow, whole, raw (Prunus domestica) | C08 | 1 | 0.668837536 | 29241576 |
| Plum, yellow, whole, raw (Prunus domestica) | H03 | 0 | 0.563319636 | |
| Plum, yellow, whole, raw (Prunus domestica) | N05 | 0 | 0.598272108 | |
| Plum, yellow, whole, raw (Prunus domestica) | P03 | 0 | 0.600258629 | |
| Plums, dried (prunes), uncooked | C08 | 1 | 0.602211148 | 29241576 |
| Potato, flesh and skin, raw (Solanum tuberosum) | D02 | 0 | 0.608966095 | |
| Prickly pears, raw (Opuntia spp.) | D02 | 0 | 0.590635342 | |
| Purslane, raw (Portulaca oleracea) | D02 | 1 | 0.559712026 | 37872023 |
| Purslane, raw (Portulaca oleracea) | R06 | 0 | 0.554136449 | |
| Queen Anne's Lace, leaves, raw | C02 | 0 | 0.617969033 | |
| Queen Anne's Lace, leaves, raw | D05 | 0 | 0.576942043 | |
| Queen Anne's Lace, leaves, raw | H03 | 0 | 0.645747966 | |
| Queen Anne's Lace, leaves, raw | J02 | 0 | 0.541307291 | |
| Queen Anne's Lace, leaves, raw | L04 | 0 | 0.683885128 | |

| | | | |
|---|---|---|---|
| Queen Anne's Lace, leaves, raw | M01 | 0 | 0.668516923 |
| Queen Anne's Lace, leaves, raw | M02 | 0 | 0.607044102 |
| Queen Anne's Lace, leaves, raw | M05 | 0 | 0.640602564 |
| Queen Anne's Lace, leaves, raw | P03 | 0 | 0.594293945 |
| Queen Anne's Lace, leaves, raw | S02 | 0 | 0.674381144 |
| Radish leaves, raw | D02 | 0 | 0.616259102 |
| Radish leaves, raw | P03 | 0 | 0.565057931 |
| Raisins, golden seedless (Vitis vinifera) | D02 | 0 | 0.608966095 |
| Raisins, seedless (Vitis vinifera) | S02 | 0 | 0.580417629 |
| Raspberries, red, frozen | C08 | 0 | 0.616325422 |
| Raspberries, red, frozen | D05 | 0 | 0.552124857 |
| Raspberries, red, frozen | H03 | 0 | 0.65959668 |
| Raspberries, red, frozen | J02 | 0 | 0.560274056 |
| Raspberries, red, frozen | L04 | 0 | 0.589995779 |
| Raspberries, red, frozen | M01 | 0 | 0.607952172 |

| | | | | |
|---|---|---|---|---|
| Raspberries, red, frozen | N05 | 0 | 0.600917287 | |
| Raspberries, red, frozen | P03 | 0 | 0.649290482 | |
| Raspberries, red, frozen | S02 | 0 | 0.572669038 | |
| Rhubarb stalks, cooked | C01 | 1 | 0.562568677 | 17708625 |
| Rhubarb, raw (Rheum rhabarbarum) | C01 | 1 | 0.562568677 | 17708625 |
| Rocket, wild, raw (Diplotaxis tenuifolia) | D02 | 0 | 0.615373995 | |
| Rutabagas, raw (Brassica napus var. napobrassica) | C08 | 0 | 0.584263742 | |
| Rutabagas, raw (Brassica napus var. napobrassica) | H03 | 0 | 0.754875581 | |
| Rutabagas, raw (Brassica napus var. napobrassica) | J02 | 0 | 0.709841156 | |
| Rutabagas, raw (Brassica napus var. napobrassica) | L04 | 0 | 0.567103755 | |
| Rutabagas, raw (Brassica napus var. napobrassica) | M02 | 0 | 0.600138925 | |
| Rutabagas, raw (Brassica napus var. napobrassica) | N05 | 0 | 0.636627001 | |
| Rutabagas, raw (Brassica napus var. napobrassica) | P03 | 0 | 0.627680429 | |
| Sage, fresh | S02 | 0 | 0.610120231 | |
| Sauce, pasta, spaghetti/marinara, ready-to-serve | D02 | 0 | 0.616259102 | |
| Sauce, pasta, spaghetti/marinara, ready-to-serve | P03 | 0 | 0.565057931 | |
| Sauerkraut, canned, solids and liquids | C08 | 0 | 0.582512981 | |
| Sauerkraut, canned, solids and liquids | D02 | 0 | 0.616259102 | |
| Sauerkraut, canned, solids and liquids | H03 | 0 | 0.767260034 | |
| Sauerkraut, canned, solids and liquids | J02 | 0 | 0.677606162 | |
| Sauerkraut, canned, solids and liquids | L04 | 0 | 0.592903263 | |
| Sauerkraut, canned, solids and liquids | M05 | 0 | 0.567827823 | |
| Sauerkraut, canned, solids and liquids | N05 | 0 | 0.626684581 | |
| Sauerkraut, canned, solids and liquids | P03 | 0 | 0.656464361 | |
| Sorghum, grain, white | L04 | 0 | 0.562568677 | |
| Sorghum, grain, white | S02 | 0 | 0.602899873 | |
| Soybeans, green, raw (Glycine max) | D02 | 1 | 0.608966095 | 29369914 |
| Spices, celery seed (Apium graveolens) | S02 | 0 | 0.610120231 | |
| Spices, parsley, dried (Petroselinum crispum) | L04 | 0 | 0.590225107 | |
| Spinach, raw (Spinacia oleracea) | C08 | 2 | 0.570483936 | 22019438 27075914 |
| Spinach, raw (Spinacia oleracea) | H03 | 0 | 0.564082852 | |
| Strawberries, frozen, unsweetened | C08 | 3 | 0.617969033 | 33758944 25512803 29099521 |
| Strawberries, frozen, unsweetened | H03 | 0 | 0.722928998 | |
| Strawberries, frozen, unsweetened | J02 | 0 | 0.593107032 | |
| Strawberries, frozen, unsweetened | L04 | 0 | 0.637780513 | |
| Strawberries, frozen, unsweetened | M01 | 2 | 0.591675897 | 28846633 30382270 |
| Strawberries, frozen, unsweetened | M05 | 1 | 0.625166358 | 34812466 |
| Strawberries, frozen, unsweetened | N05 | 0 | 0.587198791 | |
| Strawberries, frozen, unsweetened | P03 | 0 | 0.671474977 | |
| Sweet potato leaves, cooked, steamed, without salt | C08 | 0 | 0.575995858 | |
| Sweet potato leaves, cooked, steamed, without salt | H03 | 0 | 0.583438636 | |
| Sweet potato leaves, cooked, steamed, without salt | J02 | 0 | 0.565089234 | |
| Sweet potato leaves, cooked, steamed, without salt | R06 | 0 | 0.611823352 | |
| Sweet potato leaves, raw (Ipomoea batatas) | C08 | 0 | 0.643610901 | |
| Sweet potato leaves, raw (Ipomoea batatas) | H03 | 0 | 0.692056585 | |

| | | | |
|---|---|---|---|
| Sweet potato leaves, raw (Ipomoea batatas) | J02 | 0 | 0.582747084 |
| Sweet potato leaves, raw (Ipomoea batatas) | L04 | 0 | 0.591675897 |
| Sweet potato leaves, raw (Ipomoea batatas) | M01 | 0 | 0.591675897 |
| Sweet potato leaves, raw (Ipomoea batatas) | M05 | 0 | 0.563421532 |
| Sweet potato leaves, raw (Ipomoea batatas) | N05 | 0 | 0.617969033 |
| Sweet potato leaves, raw (Ipomoea batatas) | P03 | 0 | 0.671474977 |
| Sweet potato, raw, unprepared (Ipomoea batatas) | C08 | 0 | 0.575258361 |
| Sweet potato, raw, unprepared (Ipomoea batatas) | H03 | 0 | 0.680696442 |
| Sweet potato, raw, unprepared (Ipomoea batatas) | J02 | 0 | 0.612313536 |
| Sweet potato, raw, unprepared (Ipomoea batatas) | L04 | 0 | 0.626161813 |
| Sweet potato, raw, unprepared (Ipomoea batatas) | M01 | 0 | 0.584933052 |
| Sweet potato, raw, unprepared (Ipomoea batatas) | M02 | 0 | 0.621008218 |
| Sweet potato, raw, unprepared (Ipomoea batatas) | M05 | 0 | 0.582344865 |
| Sweet potato, raw, unprepared (Ipomoea batatas) | S02 | 0 | 0.570099093 |
| Taro leaves, cooked, steamed, without salt | R06 | 0 | 0.583616938 |

Taro leaves, raw (Colocasia esculenta) C08 0 0.575258361

Taro leaves, raw (Colocasia esculenta) H03 0 0.680696442

Taro leaves, raw (Colocasia esculenta) J02 0 0.612313536

Taro leaves, raw (Colocasia esculenta) L04 0 0.626161813

Taro leaves, raw (Colocasia esculenta) M01 0 0.584933052

Taro leaves, raw (Colocasia esculenta) M02 0 0.621008218

Taro leaves, raw (Colocasia esculenta) M05 0 0.582344865

Taro leaves, raw (Colocasia esculenta) S02 0 0.570099093

Taro, cooked, without salt D02 0 0.608966095

Tea, black, brewed, prepared with tap water C01 9 0.592474169
23038021 11447078 19516176 25658240 28034564
27854314 15165919 17010979 15643124

Tea, fruit flavored, brewed C01 0 0.575362548

Tea, green, brewed, decaffeinated C01 7 0.547375294 31084381
33491287 30912296 22749178 27797683 27165772
18525384

Tea, green, brewed, flavored C01 7 0.551298358 31084381
33491287 30912296 22749178 27797683 27165772
18525384

Tea, green, large leaf, Quingmao, brewed C01 7 0.602891856
31084381 33491287 30912296 22749178 27797683
27165772 18525384

Thyme, fresh (Thymus vulgaris)	S02	0	0.610120231

Tomato products, canned, puree, without salt added	D02	3	0.616259102
11340098	15830922	27662341

Tomato products, canned, puree, without salt added	P03	0	0.565057931

Tomatoes, cherry, raw	C07	0	0.605123507

Tomatoes, cherry, raw	N05	0	0.639859215

Tomatoes, cherry, raw	R06	0	0.603120886

Tomatoes, red, ripe, canned, packed in tomato juice	C08	5	0.616325422
33407609	30223563	21138408	29131285	16368299

Tomatoes, red, ripe, canned, packed in tomato juice	D05	0	0.552124857

Tomatoes, red, ripe, canned, packed in tomato juice	H03	0	0.65959668

Tomatoes, red, ripe, canned, packed in tomato juice	J02	0	0.560274056

Tomatoes, red, ripe, canned, packed in tomato juice	L04	0	0.589995779

Tomatoes, red, ripe, canned, packed in tomato juice	M01	0	0.607952172

Tomatoes, red, ripe, canned, packed in tomato juice	N05	3	0.600917287
37475550	30792141	25880734

Tomatoes, red, ripe, canned, packed in tomato juice	P03	0	0.649290482

Tomatoes, red, ripe, canned, packed in tomato juice	S02	0	0.572669038

Tomatoes, red, ripe, cooked C08 5 0.602211148 33407609 30223563 21138408 29131285 16368299

Tomatoes, red, ripe, raw, year round average (Lycopersicon esculentum) C08 5 0.636627001 33407609 30223563 21138408 29131285 16368299

Tomatoes, red, ripe, raw, year round average (Lycopersicon esculentum) N05 3 0.763401209 37475550 30792141 25880734

Tomatoes, red, ripe, raw, year round average (Lycopersicon esculentum) N06 3 0.605644787 32281554 35611327 22840609

Tomatoes, red, ripe, raw, year round average (Lycopersicon esculentum) R06 2 0.691746222 17519582 18324527

Tomatoes, yellow, raw (Lycopersicon esculentum) D02 3 0.616259102 11340098 15830922 27662341

Tomatoes, yellow, raw (Lycopersicon esculentum) P03 0 0.565057931

Tree spinach, cooked D02 0 0.608966095

Tree Spinach, raw (Cnidoscolus aconitifolius) D02 0 0.608966095

Turmeric, steamed (Curcuma longa) C08 1 0.602899873 21742514

Turmeric, steamed (Curcuma longa) N05 0 0.563187905

Turnip greens, raw (Brassica rapa (Rapifera Group)) D02 0 0.608966095

Water spinach C02 0 0.562396948

Water spinach C08 0 0.577805083

| | | | |
|---|---|---|---|
| Water spinach | H03 | 0 | 0.685362175 |
| Water spinach | J02 | 0 | 0.534335442 |
| Water spinach | L04 | 0 | 0.654125755 |
| Water spinach | M01 | 0 | 0.633604162 |
| Water spinach | M05 | 0 | 0.641560833 |
| Water spinach | N05 | 0 | 0.575237061 |
| Water spinach | P03 | 0 | 0.675055977 |
| Water spinach | S02 | 0 | 0.598349264 |
| Watercress, raw (Nasturtium officinale) | C08 | 0 | 0.62822578 |
| Watercress, raw (Nasturtium officinale) | H03 | 0 | 0.70234739 |
| Watercress, raw (Nasturtium officinale) | J02 | 0 | 0.572387136 |
| Watercress, raw (Nasturtium officinale) | L04 | 0 | 0.617289572 |
| Watercress, raw (Nasturtium officinale) | M01 | 0 | 0.601921367 |
| Watercress, raw (Nasturtium officinale) | M05 | 0 | 0.599439347 |
| Watercress, raw (Nasturtium officinale) | N05 | 0 | 0.607712285 |
| Watercress, raw (Nasturtium officinale) | P03 | 0 | 0.681765781 |
| Watercress, steamed | D02 | 0 | 0.616259102 |
| Watercress, steamed | P03 | 0 | 0.565057931 |
| Yardlong bean, cooked, boiled, drained, without salt | C08 | 0 | 0.62822578 |
| Yardlong bean, cooked, boiled, drained, without salt | H03 | 0 | 0.70234739 |
| Yardlong bean, cooked, boiled, drained, without salt | J02 | 0 | 0.572387136 |
| Yardlong bean, cooked, boiled, drained, without salt | L04 | 0 | 0.617289572 |
| Yardlong bean, cooked, boiled, drained, without salt | M01 | 0 | 0.601921367 |
| Yardlong bean, cooked, boiled, drained, without salt | M05 | 0 | 0.599439347 |
| Yardlong bean, cooked, boiled, drained, without salt | N05 | 0 | 0.607712285 |
| Yardlong bean, cooked, boiled, drained, without salt | P03 | 0 | 0.681765781 |
| Yuzu, raw | N03 | 0 | 0.562568677 |